\documentclass{jfm}
\usepackage{graphicx}
\usepackage{epstopdf,epsfig}
\usepackage{newtxtext}
\usepackage{newtxmath}
\usepackage{natbib}
\usepackage{hyperref}
\hypersetup{
    colorlinks = true,
    urlcolor   = blue,
    citecolor  = black,
}

\newcommand{\RomanNumeralCaps}[1]
\linenumbers

\usepackage{calc}

\title{Spatio-temporal fluctuations of interscale and interspace energy transfer dynamics in homogeneous turbulence}

\author{H. S. Larssen\aff{1}
  \corresp{\email{h.larssen18@imperial.ac.uk}} \and J. C. Vassilicos\aff{2,1}\corresp{\email{john-christos.vassilicos@cnrs.fr}}}

\affiliation{\aff{1}Department of Mathematics, Imperial College London, London SW7 2AZ, UK
\aff{2}Univ. Lille, CNRS, ONERA, Arts et Métiers ParisTech, Centrale Lille, UMR 9014 - LMFL - Laboratoire de Mécanique des fluides de Lille - Kampé de Feriet, F-59000 Lille, France}

\begin{document}

\maketitle 				

\begin{abstract}
	We study fluctuations of all co-existing
          energy exchange/transfer/transport processes in stationary
          periodic turbulence including those which average to zero
          and are not present in average cascade theories. We use a
          Helmholtz decomposition of accelerations which leads to a
          decomposition of all terms in the Kármán-Howarth-Monin-Hill
          (KHMH) equation (scale-by-scale two-point energy balance)
          causing it to break into two energy balances, one resulting
          from the integrated two-point vorticity equation and the
          other from the integrated two-point pressure equation. The
          various two-point acceleration terms in the Navier-Stokes
          difference (NSD) equation for the dynamics of two-point
          velocity differences have similar alignment tendencies with
          the two-point velocity difference, implying similar
          characteristics for the NSD and KHMH equations. We introduce
          the two-point sweeping concept and show how it articulates
          with the fluctuating interscale energy transfer as the
          solenoidal part of the interscale transfer rate does not
          fluctuate with turbulence dissipation at any scale above the
          Taylor length but with the sum of the time-derivative and
          the solenoidal interspace transport rate terms. The pressure
          fluctuations play an important role in the interscale and
          interspace turbulence transfer/transport dynamics as the
          irrotational part of the interscale transfer rate is equal
          to the irrotational part of the interspace transfer rate and
          is balanced by two-point fluctuating pressure work. We also
          study the homogeneous/inhomogeneous decomposition of
          interscale transfer. The statistics of the latter are skewed
          towards forward cascade events whereas the statistics of the
          former are not. We also report statistics conditioned on
          intense forward/backward interscale transfer events.
\end{abstract}

\section{Introduction}\label{sec:intro}

Modeling of turbulence dissipation is a cornerstone of one-point
turbulent flow prediction methods based on the Reynolds Averaged
Navier Stokes (RANS) equations such as the widely used $k-\varepsilon$
and the $k-\omega$ models (see \citet{Pope2000},
\citet{Leschziner2016}) and also of two-point turbulence flow
prediction methods based on filtered Navier Stokes equations, namely
Large Eddy Simulations (LES) (see \citet{Pope2000},
\citet{Sagaut2000}). The mechanism of turbulence dissipation away from
walls is the turbulence cascade
\citep{Pope2000,Vassilicos2015}. Indeed, without
  cascade, the larger turbulent eddies which constitute the most
  energetic part of the turbulence, would take much longer to lose
  their energy than just one or a few eddy turnover times. The
physical understanding of this cascade which, to this day, has
underpinned these prediction methods is based on Kolmogorov's average
cascade in statistically homogeneous and stationary
turbulence. Notwithstanding recent advances which have shown that the
turbulence dissipation and cascade are different from Kolmogorov's
both in non-stationary (see
e.g. \citet{Vassilicos2015,Goto2016,Steiros2022}) and in
non-homogeneous turbulence \citep{Chen2021,Chen2022}, Kolmogorov's
cascade is in fact valid only as an {\it average} cascade even in
homogeneous stationary turbulence. Turbulence has been known to be
intermittent since the late 1940s (see \citet{Frisch1995} and
references therein), and
this intermittency has mainly been taken into account as structure
function exponent corrections to Kolmogorov's average picture.
However, \citet{Yasuda2018} studied intermittent
  fluctuations without reference to structure function exponents which
  require high Reynolds numbers to be well defined and to be predicted
  from Kolmogorov's theory or various intermittency-accounting
  variants of this theory (see \citet{Frisch1995} and references
  therein). They concentrated their attention on the actual
  fundamental basis of Kolmogorov's theory which is scale-by-scale
  equilibrium for statistically homogeneous and stationary turbulence,
  and not on the theory's structure function and energy spectrum
  scaling consequences. The scale-by-scale equilibrium implied by
  statistical homogeneity and stationarity is that the average
  interscale turbulence energy transfer rate is balanced by nothing
  more than average scale-by-scale viscous diffusion rate, average
  turbulence dissipation rate and average energy input rate by a
  stirring force, irrespective of Reynolds number (except that the
  Reynolds number needs to be large enough for the presence of random
  fluctuations). It is most natural for a study of intermittency to
  start with the fluctuations around this balance, which means that
  along with the fluctuations of interscale transfer, dissipation,
  diffusion and energy input, all other fluctuating turbulent energy
  change rates need to be taken into account as well even if their
  spatio-temporal average is zero in statistically stationary
  homogeneous turbulence.
The intermittency corrections to Kolmogorov's average cascade theory
which have been developed since the 1960s (e.g. see
\citet{Frisch1995,Sreenivasan1997}) are often based on the
intermittent fluctuations of the local (in space and time) turbulence
dissipation rate, yet \citet{Yasuda2018} demonstrated that these
dissipation fluctuations are much less intense than the fluctuations
of other turbulent energy change rates such as the
  non-linear interspace energy transfer rate (which is a
  scale-by-scale rate of turbulent transport in physical space), the
fluctuating work resulting from the correlation of the fluctuating
pressure gradient with the fluctuating velocity and
  the time-derivative of the scale-by-scale turbulent kinetic
  energy. \citet{Yasuda2018} made these observations using Direct
  Numerical Simulations (DNS) of statistically stationary periodic
  turbulence at low to moderate Taylor length-based Reynolds numbers
  from about 80 to 170. Even though their Reynolds numbers were not
  high enough to test the high Reynolds number scaling consequences of
  Kolmogorov's theory, they observed an energy spectrum with a
  near-decade power law range where the power law exponent was not too
  far from Kolmogorov's $-5/3$. However, they did not observe a
  significant range of scales where the scale-by-scale equilibrium
  reduces to a scale-by-scale balance between average interscale
  turbulence energy transfer rate and average turbulence dissipation
  as predicted by the Kolmogorov theory for statistically stationary
  homogeneous turbulence at asymptotically high Reynolds numbers. This
  high Reynolds number scale-by-scale equilibrium is the trademark of
  the Kolmgorov average cascade and is typically not put in question
  by existing intermittency corrections to Kolmogorov's theory
  (e.g. see \citet{Frisch1995}).

Given the low to moderate Reynolds numbers of the DNS
  used by \citet{Yasuda2018}, their observations concern interscale
  turbulence energy transfers more than the turbulence cascade per se
  if the concept of turbulence cascade is taken to have meaning only
  at very large Reynolds numbers. They demonstrated that
an interscale transfer picture appears that is
  radically different from Kolmogorov's if the average is lifted and
all spatio-temporal intermittent fluctuations are
taken into account. This different picture involves
  highly fluctuating processes which vanish on average in
  statistically stationary and homogeneous turbulence and are not
  taken into account by the Kolmogorov theory for that very reason. We
  stress once more that \citet{Yasuda2018} made this demonstration in
  statistically homogeneous and stationary turbulence, the very type
  of turbulence where Kolmogorov's theory has been designed for.


It is hard to imagine that the complex turbulence
  energy transfer picture educed by the DNS of \citet{Yasuda2018} does
  not survive at asympotically high Reynolds numbers because it is
  known that the small-scale turbulence becomes increasingly
  intermittent with increasing Reynolds number (e.g. see
  \citet{Frisch1995,Sreenivasan1997}). A DNS study at higher Reynolds
  numbers is nevertheless needed to ascertain this point. However,
  this is not the study proposed in this paper. In this paper our aim
  is to gain deeper insight into the fluctuating energy transfer
  picture revealed by the DNS of \citet{Yasuda2018} and we do this in
  terms of Helmholtz decomposed solenoidal and irrotational
  acceleration fields. Given that the computational cost involved in
  this Helmholtz decomposition is high (see following two sections) it
  is not possible for us to carry out our study for a variety of
  increasing Reynolds numbers and thereby combine it with a Reynolds
  number dependence study. We therefore limit ourselves to Reynolds
  numbers comparable to those of \citet{Yasuda2018}.

The radically different turbulence energy transfer
picture which appears when all intermittent turbulence fluctuations
are taken into account exhibits correlations between fluctuations of
different processes: in particular, the fluctuating pressure-velocity
term mentioned above is correlated with the interscale energy transfer
rate, and the time derivative of the turbulent kinetic energy below a
certain two-point length $r$ is correlated with the inter-space energy
transport rate at the same length $r$. \citet{Yasuda2018} explained
the former correlation as resulting from the link between
non-linearity and non-locality (via the pressure field) and the latter
correlation as reflecting the passive sweeping of small turbulent
eddies by large ones \citep{Tennekes1975}. However, this sweeping
(also termed ``random Taylor hypothesis'') has been studied by
reference to the one-point incompressible Navier-Stokes equation
(e.g. \citet{Tennekes1975}, \citet{Tsinober2001}) rather than the
two-point Kármán-Howarth-Monin-Hill (KHMH) equation, used by
\citet{Yasuda2018} in their study of the fluctuating turbulence
cascade. The KHMH equation is a scale-by-scale energy budget local in
space and time, directly derived from the incompressible Navier-Stokes
equations for the instantaneous velocity field (see \citet{HILL2002})
without decomposition (e.g. Reynolds decomposition), without averages
(e.g. Reynolds averages), and without any assumption made about the
turbulent flow (e.g. homogeneity, isotropy, etc.). The
  initial trigger of the present paper is to substantiate the claim
of \citet{Yasuda2018} concerning correlations being caused by random
sweeping by translating the sweeping analysis of \citet{Tsinober2001}
to the KHMH equation. It is in doing so that we
  espouse the Helmholtz decomposition which \citet{Tsinober2001}
introduced for the analysis of the acceleration
field. We apply it to the two-point Navier-Stokes
difference (NSD) equation (which is the equation
  governing the dynamics of two-point velocity differences) and the
KHMH equation which derives from it. This decomposition into
solenoidal and irrotational terms breaks the Navier-Stokes equation
into two equations, one being the irrotational balance between
non-linearity and non-locality (pressure) and the other being the
solenoidal balance between local unsteadiness and advection which
encapsulates the sweeping. With this decomposition we substantiate all
the correlations observed by \citet{Yasuda2018}
  between different KHMH terms representing different energy change
  processes, not only the ones caused by sweeping. In fact, we educe
  the relation between interspace turbulence energy transfer/transport
  and two-point sweeping (i.e. the random Taylor hypothesis that we
  generalise to two-point statistics), and we extend the correlation
  study to solenoidal and irrotational sub-terms of the KHMH equation
  which leads to even stronger correlations than those found by
  \citet{Yasuda2018}. This approach sheds some light on the way that
  two-point sweeping and interscale energy transfer relate to each
  other. We then ask whether the scale-by-scale equilibrium which is
  at the basis of Kolmogorov's theory and which disappears when the
  average is lifted does nevertheless exist locally at relatively
  high energy transfer events, a question which leads
  us to consider whether two-point sweeping also holds at such
  events. Finally, we study the recently introduced decomposition
\citep{AlvesPortela2020} of the interscale transfer rate into a
homogeneous and an inhomogeneous interscale transfer component. We
analyse their fluctuations and the correlations of these fluctuations,
both unconditionally and conditionally on relatively rare
intense interscale transfer events.


In the following section we introduce our direct numerical simulations
(DNSs) of forced periodic turbulence. Subsection \ref{subsec:ns} is a
reminder of the application of this decomposition to the one-point
Navier-Stokes equation by \citet{Tsinober2001}. In this sub-section we
also validate our DNS by retrieving the conclusions of
\citet{Tsinober2001} on sweeping and by comparing our DNS results on
one-point acceleration dynamics to theirs. In subsections
\ref{subsec:nsd}-\ref{subsec:nsdTwo} we apply the Helmholtz
decomposition to the two-point NSD equation for the case of
homogeneous/periodic turbulence and in subsection \ref{subsec:khmh1}
we derive from the Helmholtz decomposed Navier-Stokes difference
equations corresponding KHMH equations. Subsection \ref{subsec:khmh1}
formalises the connection between the NS and KHMH dynamics, clarifies
under which conditions a link exists between NS and KHMH dynamics and
provides results on scale and Reynolds number dependencies of the KHMH
dynamics. By considering the NSD dynamics in terms of solenoidal and
irrotational dynamics, we derive two new KHMH equations. In section
\ref{sec:khmh2} we use these two new KHMH equations to obtain new
results on the fluctuating cascade dynamics across scales both
unconditionally and conditionally on rare extreme interscale energy
transfer events. In section \ref{sec:khmh3} we analyse the
inhomogeneous and homogeneous contributions to the interscale energy
transfer rate. Finally, section \ref{sec:conc} summarises our results.
\section{DNS of body-forced period turbulence} \label{sec:dns}

\begin{table}
\setlength{\tabcolsep}{6.5pt}
	\begin{center}
	\begin{tabular}{lccccccc}
		$N$	 &  $\langle \Rey_{\lambda}\rangle_t$ & $\nu/10^{3}$ & $k_{\text{max}}\langle \eta \rangle_t$ & $2\upi/\langle L \rangle_t$ & $\langle \lambda\rangle/\langle L \rangle_t$ & $T_s/T$ & $\Delta T/T$  \\ [3pt]
		256 &	112 & 1.80  & 1.88 & 5.6 & 3.5 & 21 & 0.01 \\ [2pt]
		512 &	174 & 0.72 & 1.89  & 5.4 & 5.2 & 27 & 0.12
	\end{tabular}
	\caption{Specifications of the numerial
            simulations. $N$ denotes the number of grid points in each
            Cartesian coordinate, $Re_{\lambda}$ the Taylor-scale
            Reynolds number, $\nu$ the kinematic viscosity,
            $k_{\text{max}}=\sqrt{2}/3N$ is the highest resolved
            wavenumber, $\eta$ and $\lambda$ are, respectively, the
            Kolmogorov and Taylor lengths and $\langle \ldots
            \rangle_t$ denotes a time-average. $L$ is the integral
            lengths calculated from the three-dimensional energy
            spectrum $E(k,t)$:
            $L(t)=(3\pi/4)\int_{0}^{\infty}k^{-1}E(k,t)dk/K(t)$ where
            $K(t)$ is the kinetic energy per unit mass. $T_s$ denotes
            the total sampling time over which converged statistics
            are calculated by sampling randomly in space-time, $\Delta
            T$ denotes the time between samples and $T \equiv \langle
            L \rangle_t/\sqrt{2/3 \langle K \rangle_t}$ is the
            turnover time.}
	\label{tab:DNS}
	\end{center}
\end{table}

Our study requires turbulence data from a turbulent flow where the
Kolmogorov equilibrium theory for statistically homogeneous and
stationary turbulence is applicable. We therefore follow
\citet{Yasuda2018} and perform Direct Numerical Simulations of
body-forced periodic Navier-Stokes turbulence with the same
pseudo-spectral code that they used. This code solves numerically the
vorticity equation
\begin{equation}
	\frac{\partial \boldsymbol{\omega}}{\partial t} = \nabla_{\boldsymbol{x}} \times (\boldsymbol{u} \times \boldsymbol{\omega}) + \nu \nabla_{\boldsymbol{x}}^2 \boldsymbol{\omega}+\nabla_{\boldsymbol{x}} \times \boldsymbol{f},
\end{equation}
\noindent subjected to the continuity equation
\begin{equation}
	\nabla_{\boldsymbol{x}} \cdot \boldsymbol{u} = 0  ,
\end{equation}
\noindent where $\boldsymbol{u} (\boldsymbol{x},t)$, $\boldsymbol{f}
(\boldsymbol{x}, t)$ and $\boldsymbol{\omega} (\boldsymbol{x},t)$ are
the velocity, force and vorticity fields respectively and $\nu$ is the
kinematic viscosity. All fields are $2\pi$-periodic in each one of the
three orthogonal spatial coordinates $x_1$, $x_2$ and $x_3$, and
$\boldsymbol{x}=(x_{1}, x_{2}, x_{3})$. The
  pseudo-spectral method is fully dealised with a combination of
  phase-shifting and spherical truncation \citep{Patterson1971}.
\noindent
The forcing method is a negative damping forcing \citep{Linkmann2015,
  McComb2015}
\begin{align}
	\widehat{\boldsymbol{f}}(\boldsymbol{k},t) &=
        (\epsilon_W/2K_f)\widehat{\boldsymbol{u}}(\boldsymbol{k},t)
        &&\text{for} \ 0<|\boldsymbol{k}|<k_f, \\ &=0
        &&\text{otherwise},
\end{align}

\noindent where $\widehat{\boldsymbol{f}}(\boldsymbol{k},t)$ and
$\widehat{\boldsymbol{u}}(\boldsymbol{k},t)$ are the Fourier
transforms of $\boldsymbol{f}(\boldsymbol{x},t)$ and
$\boldsymbol{u}(\boldsymbol{x},t)$ respectively, $k_f$ is the cutoff
wavenumber, $\epsilon_W$ is the energy input rate per unit mass and
$K_f$ is the kinetic energy per unit mass in the wavenumber band
$0<|\boldsymbol{k}|<k_f$. Note that this forcing is incompressible and
has therefore no irrotational part. The addition of a potential, i.e.
irrotational, term to the forcing would effectively just be subsumed
into the pressure required to keep the flow incompressible.

We perform two DNS of forced periodic/homogeneous
  turbulence with forcing parameters $\epsilon_W=0.1$ and $k_f=2.5$ at
  both simulation sizes $256^3$ and $512^3$. Average statistics are
  given in table \ref{tab:DNS}. For these two simulation sizes
  respectively, deviations around these averages are as follows: the
  standard deviations of $L$ are $0.007L_b$ and $0.006L_b$ (where
  $L_{b}=2\pi$) and the maximum $L$ values are $0.188L_b$ and
  $0.202L_b$; the standard deviations of $\lambda$ are $2.5\%$ and
  $3.7\%$ of $\langle \lambda \rangle_t$; and the standard deviation
  of $k_{\text{max}} \eta $ are $0.025$ and $0.035$.

\citet{McComb2015a} performed DNS with identical
  combinations of $N$, $\nu$ and forcing as in our simulations. The
  time-averaged Taylor-scale Reynolds numbers $\langle
  \Rey_{\lambda}\rangle_t$, the ratios of the box size to the
  time-averaged integral length $2\pi/\langle L\rangle_t$ and the
  time-averaged Kolmogorov microscales $\langle \eta \rangle_t$ are
  all very similar (and $\langle \ldots \rangle_t$ denotes a
  time-average). This study reports slightly poorer small-scale
  resolution $k_{\text{max}}\langle \eta \rangle_t$ than ours due to
  their more severe spherical truncation.

We have also verified that the results do not significantly change
when the flow is forced at small wavenumbers with an ABC forcing with
$A=B=C$ \citep{Podvigina1994}. In contrast to the negative damping
forcing, this forcing is independent of time and of the velocity field
and is also maximally helical as $\nabla_{\boldsymbol{x}}\times
\boldsymbol{f}$ is parallel to $\boldsymbol{f}$
\citep{Galanti2000}. The helicity input of the ABC
  forcing has been studied in the context of the energy cascade in
  terms of its effect on the dissipation coefficient in
  \citet{Linkmann2018a}.

Our Reynolds numbers are relatively limited due to the high
computational expense of our NSD and KHMH post-processing (which is
typically at least one order of magnitude more expensive than the
DNS). We detail the computational expense of the post-processing once
the relevant terms have been introduced in section \ref{subsec:nsdTwo}.

In the following section we show how we apply the Helmholtz
decomposition to the KHMH equation. We start in subsection
\ref{subsec:ns} by applying this decomposition to the one-point
Navier-Stokes equation following \citet{Tsinober2001}. In this
sub-section we also validate our DNS by retrieving the conclusions of
\citet{Tsinober2001}, in particular on sweeping, and by comparing our
DNS results on one-point acceleration dynamics to theirs. In
subsections \ref{subsec:nsd} and \ref{subsec:nsdTwo} we apply the
Helmholtz decomposition to the two-point Navier-Stokes difference
equation for the case of homogeneous/periodic turbulence and in
subsection \ref{subsec:khmh1} we derive from the Helmholtz-decomposed
Navier-Stokes difference equations corresponding KHMH equations.

\section{Helmholtz decomposition of two-point Navier-Stokes dynamics and corresponding turbulent energy exchanges} \label{sec:nsTokhmh}

\subsection{Solenoidal and irrotational acceleration fluctuations} \label{subsec:ns}

The Helmholtz decomposition states that a twice continously
differentiable $3$D vector field $\boldsymbol{q}(\boldsymbol{x}, t)$
defined on a domain $V \subseteq \mathbb{R}^3$ can be expressed as the
sum of an irrotational vector field
$\boldsymbol{q}_I(\boldsymbol{x},t)$ and a solenoidal vector field
$\boldsymbol{q}_{S}(\boldsymbol{x},t)$ \citep{Helmholtz1867,Stewart2012,Bhatia2013}
\begin{equation}
	\boldsymbol{q}_I(\boldsymbol{x},t) = -\nabla_{\boldsymbol{x}} \phi(\boldsymbol{x},t)  , \quad \boldsymbol{q}_S(\boldsymbol{x},t) = \nabla_{\boldsymbol{x}} \times \boldsymbol{B}(\boldsymbol{x},t) , \label{helmholtz0}
\end{equation}
\noindent where $\phi(\boldsymbol{x},t)$ is a scalar potential and
$\boldsymbol{B}(\boldsymbol{x},t)$ is a vector potential.
The Helmholtz decomposition and its interpretation can be applied to
any vector field $\boldsymbol{q}(\boldsymbol{x},t)$ satisfying the
above conditions, and \citet{Tsinober2001} applied it to fluid
accelerations and the incompressible Navier-Stokes equation.\par

The solenoidal and irrotational Navier-Stokes equations in
homogeneous/periodic turbulence can be derived from the incompressible
Navier-Stokes equation in Fourier space (see appendix
\ref{sec:appA}). After transformation back to physical space, one
obtains
\begin{align}
	\frac{\partial \boldsymbol{u}}{\partial t}+(\boldsymbol{u}\cdot \nabla_{\boldsymbol{x}}\boldsymbol{u})^{T} &= \nu\nabla_{\boldsymbol{x}}^2 \boldsymbol{u} + \boldsymbol{f}^{T}  ,  \label{rotHIT0} \\
	(\boldsymbol{u}\cdot \nabla_{\boldsymbol{x}}\boldsymbol{u})^{L} &= -\frac{1}{\rho}\nabla_{\boldsymbol{x}} p+\boldsymbol{f}^{L}   , \label{poissonHIT0}
\end{align}

\noindent where superscripts $L$ and $T$ denote fields obtained from
longitudinal and transverse parts of respective Fourier vector fields
(see appendix \ref{sec:appA} for precise definitions and \citep{Pope2000,Stewart2012}), $p=p(\boldsymbol{x},t)$ is the pressure field and $\rho$ is
the density. For any periodic vector field $\boldsymbol{q}$, $\boldsymbol{q}^{L}$
equals the irrotational field $\boldsymbol{q}_I$ and $\boldsymbol{q}^{T}$ equals the
solenoidal field $\boldsymbol{q}_S$ (see appendix \ref{sec:appA} and
\citet{Stewart2012}). From equations
\eqref{rotHIT0}-\eqref{poissonHIT0}, one arrives at
\citep{Tsinober2001}
\begin{align}
	\frac{\partial \boldsymbol{u}}{\partial t}+(\boldsymbol{u}\cdot \nabla_{\boldsymbol{x}}\boldsymbol{u})_S &= \nu\nabla_{\boldsymbol{x}}^2 \boldsymbol{u} + \boldsymbol{f}_S  ,  \label{rotHIT1} \\
	(\boldsymbol{u}\cdot \nabla_{\boldsymbol{x}}\boldsymbol{u})_I &= -\frac{1}{\rho}\nabla_{\boldsymbol{x}} p+\boldsymbol{f}_I   , \label{poissonHIT1}
\end{align}

\noindent which we refer to as Tsinober equations. \eqref{rotHIT1}
contains only solenoidal vector fields and \eqref{poissonHIT1}
contains only irrotational vector fields. Note that in the case of an
incompressible periodic velocity field, the velocity field is
solenoidal, i.e. $\boldsymbol{u}=\boldsymbol{u}_S$. This follows
immediately from the scalar potential $\phi$ being the solution to
$\nabla_{\boldsymbol{x}}^2\phi=0$ with periodic boundary conditions
for $\nabla_{\boldsymbol{x}}\phi$, yielding $\phi=const$.\par

In appendix \ref{sec:appC} we show that \eqref{rotHIT1} is the
integrated vorticity equation and that \eqref{poissonHIT1} is the
integrated Poisson equation for pressure. The procedure presented in
appendix \ref{sec:appC} for obtaining the Tsinober equations is also
used in this same appendix to obtain generalised Tsinober equations
for non-homogeneous/non-periodic turbulence with arbitrary boundary
conditions.\par

\begin{table}
	\begin{center}
	\setlength{\tabcolsep}{6.5pt}
	\begin{tabular}{rccccccccc}
	 & $\boldsymbol{a}_c$ & $\boldsymbol{a}_l$ & $\boldsymbol{a}_{c_S}$ & $\boldsymbol{a}_{c_I}$  & $\boldsymbol{a}_p$ & $\boldsymbol{a}$ & $\boldsymbol{a}_\nu$ & $\boldsymbol{f}$ & $\langle \Rey_\lambda \rangle_t$ \\ [5pt]
	$\langle \boldsymbol{q}^2 \rangle/ (3\langle \epsilon \rangle^{3/2} \nu^{-1/2})$	 & 8.47  & 5.87 & 5.93 & 2.55 &  2.55 & 2.60  & 0.05 &  0.007 & 112  \\ [3pt]
	$\langle \boldsymbol{q}^2 \rangle/ (3\langle \epsilon \rangle^{3/2} \nu^{-1/2})$	 & 14.28 & 11.21 & 11.26 & 3.03 & 3.03 & 3.09 & 0.05 & 0.005 & 174 \\ [3pt]
	$ \langle \boldsymbol{q}^2 \rangle/\langle \boldsymbol{a}_{c}^2 \rangle$  & 1 & 0.69 &  0.70 & 0.30 & 0.30 & 0.31 & 0.0062 & 0.00081 & 112 \\ [3pt]
	 $ \langle \boldsymbol{q}^2 \rangle/\langle \boldsymbol{a}_{c}^2 \rangle$ & 1 & 0.78 & 0.79 & 0.21 & 0.21 & 0.22 & 0.0038 & 0.00032 & 174\\ [3pt]
	\end{tabular}
	\caption{Normalised average magnitudes $\langle
          \boldsymbol{q}^2 \rangle/ (3\langle \epsilon \rangle^{3/2}
          \nu^{-1/2})$ and $ \langle \boldsymbol{q}^2 \rangle/\langle
          \boldsymbol{a}_{c}^2 \rangle$ for Navier-Stokes
          accelerations and forces $\boldsymbol{q}$ defined in the
          fourth paragraph of \ref{subsec:ns} for our two $\langle
          \Rey_\lambda \rangle_t$. The accelerations and forces
          $\boldsymbol{q}$ are listed on the top row,
          $\boldsymbol{q}^2 \equiv q_i q_i$, $\epsilon$ denotes the
          viscous dissipation rate and $\langle \ldots \rangle$
          denotes a spatio-temporal average.}
     \label{tab:nsTab}
     \end{center}
\end{table}

\begin{figure}
	\centerline{\includegraphics{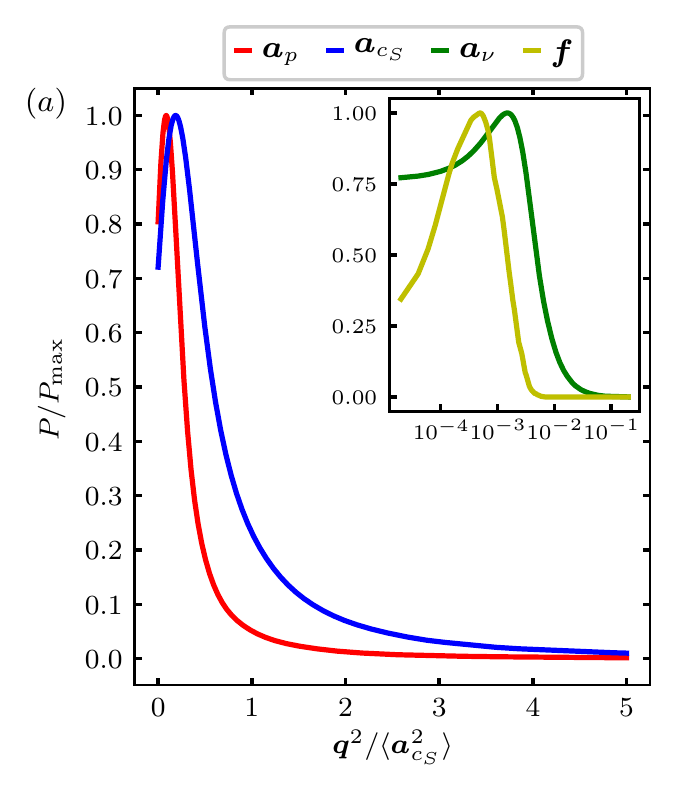} \hfill 	\includegraphics{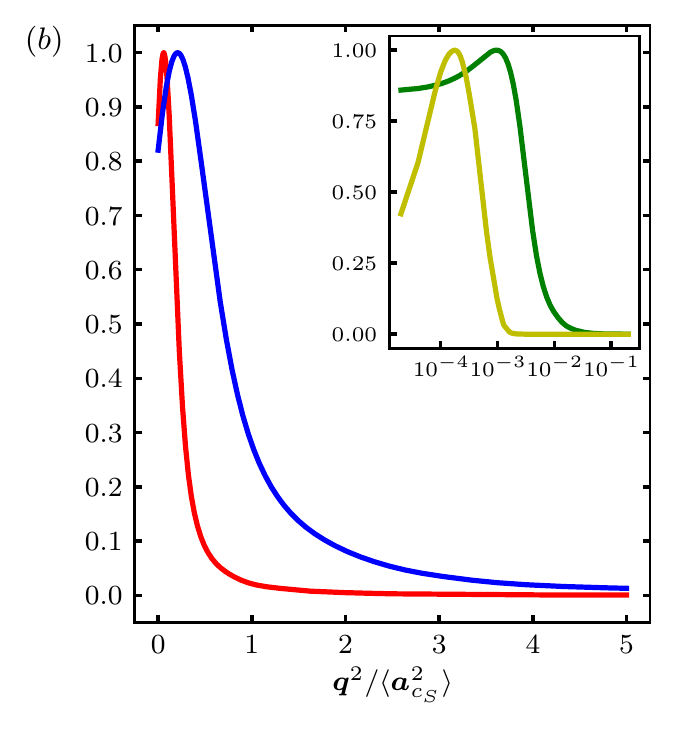}}
	\caption{Probability density functions (PDFs) $P$ of Navier-Stokes acceleration and force magnitudes $\boldsymbol{q}^2$ for terms $\boldsymbol{q}$ listed at the top of $(a)$. $P_{\text{max}}$ for the PDF of $\boldsymbol{q}^2$ denotes its maximum value. $(a)$ $\langle \Rey_\lambda\rangle_t=112$, $(b)$ $\langle \Rey_\lambda\rangle_t=174$. }
	\label{fig:nsPDF}
\end{figure}

Following the notation of \citet{Tsinober2001}, we define
$\boldsymbol{a}_l \equiv \partial \boldsymbol{u}/\partial t$, $
\ \boldsymbol{a}_c \equiv \boldsymbol{u}\cdot
\nabla_{\boldsymbol{x}}\boldsymbol{u}$, $ \ \boldsymbol{a} \equiv
\boldsymbol{a}_l+\boldsymbol{a}_c$, $ \ \boldsymbol{a}_p \equiv
-1/\rho \nabla_{\boldsymbol{x}}p$ and $ \ \boldsymbol{a}_{\nu} \equiv
\nu \nabla_{\boldsymbol{x}}^2 \boldsymbol{u}$. In such notation,
equations \eqref{rotHIT1}-\eqref{poissonHIT1} are $\boldsymbol{a}_l +
\boldsymbol{a}_{c_S} = \boldsymbol{a}_{\nu} + \boldsymbol{f}_S$ and
$\boldsymbol{a}_{c_I} = \boldsymbol{a}_p +
\boldsymbol{f}_I$. \citet{Tsinober2001} in fact wrote these equations
for statistically homogeneous/periodic Navier-Stokes turbulence
without body forces, i.e. with $\boldsymbol{f}=0$. In general,
however, the body forcing can be considered, as in the present work,
to be non-zero and typically incompressible, i.e. $\boldsymbol{f}_I =
\boldsymbol{0}$ but $\boldsymbol{f}_{s}\not\equiv \boldsymbol{0}$,
given that a compressible component of the forcing can be subsumed
into the pressure field in incompressible turbulence. In body-forced
statistically stationary homogeneous/periodic turbulence, the average
forcing magnitude $\langle \boldsymbol{f}^2 \rangle$, where the
brackets denote spatio-temporal averaging, tends to be small compared
to $\langle \boldsymbol{a}_{\nu}^2 \rangle$ when the forcing is
applied only to the largest scales \citep{Vedula1999}. Given that
$\langle \boldsymbol{f} \cdot \boldsymbol{u}\rangle=\langle \epsilon
\rangle$, where $\epsilon$ is the local turbulence dissipation rate,
$\boldsymbol{f}^2$ can be quite small if $\boldsymbol{f}$ is not close
to orthogonal to the velocity field. This is indeed the case with the
negative damping and ABC forcings used in this study. In cases where
$\boldsymbol{f}$ is close to orthogonal to the velocity field, which
is conceivable in electromagnetic situations (Lorentz force),
$\boldsymbol{f}^2$ needs to be large enough for $\langle
\boldsymbol{f} \cdot \boldsymbol{u}\rangle$ to balance $\langle
\epsilon \rangle$. In this study we have not considered such forcings
and some of our results might not be applicable to such
situations. Our results for the forcings we used indicate that
$\langle \boldsymbol{f}^2 \rangle$ is indeed much smaller than
$\langle \boldsymbol{a}_{\nu}^2 \rangle$ (see results from our DNS in
table \ref{tab:nsTab}) and the probability to find values of
$\boldsymbol{f}^{2}$ large enough to be comparable to the other terms
in the Tsinober equations is extremely small (see results from our DNS
in figure \ref{fig:nsPDF} and table \ref{tab:nsExtreme} where we see,
in particular, that $|\boldsymbol{f}|>0.1 |\boldsymbol{a}_{c_S}|$ in
$15.3\%$ and $6.3\%$ of the spatio-temporal domain for the two
Reynolds numbers respectively, the percentage being
  smaller for the higher Reynolds number. If we consider
$|\boldsymbol{f}|>\sqrt{0.1} |\boldsymbol{a}_{c_S}| \approx 0.32
|\boldsymbol{a}_{c_S}|$, we see that this is only satisfied in $0.8\%$
and $0.3\%$ of the spatio-temporal domain respectively. Furthermore,
figure \ref{fig:nsPDF} and table \ref{tab:nsExtreme} show that
$\boldsymbol{f}$ is also typically much smaller than
$\boldsymbol{a}_{\nu}$. We can therefore write $\boldsymbol{a}_l +
\boldsymbol{a}_{c_S} \approx \boldsymbol{a}_{\nu}$, this being a good
approximation in the majority of the flow for the majority of the
time. With $\boldsymbol{a}_{c_I} = \boldsymbol{a}_p$ given that
$\boldsymbol{f}_I = \boldsymbol{0}$, these two equations are very
close to the way that \citet{Tsinober2001} originally wrote them
($\boldsymbol{a}_l + \boldsymbol{a}_{c_S} = \boldsymbol{a}_{\nu}$ and
$\boldsymbol{a}_{c_I} = \boldsymbol{a}_p$ for the ${\bf f}\equiv 0$
case) and we can therefore expect our DNS to retrieve the DNS results
and conclusions of \citet{Tsinober2001}.

\begin{table}
\setlength{\tabcolsep}{6.5pt}
	\begin{center}
	\def~{\hphantom{0}}
	\begin{tabular}{rcccc}
		$\alpha$	 &  0.001 & 0.01 & 0.1 & 1   \\ [3pt] 
		Prob($\boldsymbol{a}_\nu^2>\alpha\boldsymbol{a}_{c_S}^2$) &	(0.893, 0.808) & (0.441, 0.308)  & (0.068, 0.037) & (0.004, 0.002) \\ [3pt]
		Prob($\boldsymbol{f}^2>\alpha\boldsymbol{a}_{c_S}^2$)	& (0.707, 0.476) 	 & (0.155, 0.063)  & (0.008, 0.003) & ($3*10^{-4}$, $9*10^{-5}$) \\
	\end{tabular}
	\caption{Probabilities of events $\boldsymbol{q}^2
          > \alpha \boldsymbol{p}^2$ for NS terms
          $(\boldsymbol{q},\boldsymbol{p})$ with $\alpha$ specified on
          the top row. The two probability values in the brackets for
          each $(\boldsymbol{q},\boldsymbol{p},\alpha)$ combination
          refer to $\langle \Rey_\lambda\rangle_t=112$ and $\langle
          \Rey_\lambda\rangle_t=174$ respectively.}
	\label{tab:nsExtreme}
	\end{center}
\end{table}

The DNS of \citet{Tsinober2001} showed that $\boldsymbol{a}_{\nu}$ is
typically negligible (i.e. in a statistical sense, not
  everywhere at any time in the flow) compared to all the other
acceleration terms in the Tsinober equations, namely
$\boldsymbol{a}_l$, $\boldsymbol{a}_{c_S}$, $\boldsymbol{a}_{c_I}$ and
$\boldsymbol{a}_p$. This is confirmed by our DNS results in tables
\ref{tab:nsTab}-\ref{tab:nsExtreme} and in figure \ref{fig:nsPDF}
which are for similar Reynolds numbers to those of
\citet{Tsinober2001} and where we report rms values,
  and probabilities of various acceleration terms. It is worth noting
that $\boldsymbol{a}_{\nu}$ is not everywhere always negligible, at
these Reynolds numbers at least. For example,
$|\boldsymbol{a}_{\nu}|>0.1 |\boldsymbol{a}_{c_S}|$ in $44.1\%$ and
$30.8\%$ of the space-time domain for our lower and higher Reynolds
number respectively; and if we consider $|\boldsymbol{a}_{\nu}|>0.32
|\boldsymbol{a}_{c_S}|$, this is satisfied in $6.8\%$ and $3.7\%$ of
cases. Note that the DNS results of
  \citet{Tsinober2001} suggest that the viscous force {\it typically}
  decreases in magnitude compared to $\boldsymbol{a}_{c_S}$ as the
  Reynolds number increases and our results for our two Reynolds
  numbers agree with this trend. One may therefore expect the
the first of the two Tsinober equations for
homogeneous/periodic turbulence with the kind of forcing we consider
here to {\it typically} reduce to
\begin{equation}
	\boldsymbol{a}_l + \boldsymbol{a}_{c_S} \approx 0
        , \label{rotHIT2}
\end{equation}
at high enough Reynold numbers, the approximation
being valid in the sense that the neglected terms are significantly
smaller than the retained ones in the majority of the flow for the
majority of the time. There exist, however, some relatively rare
spacio-temporal events where the neglegted viscous force and/or body
force are significant (for example, as stated a few
  lines above, $|\boldsymbol{a}_{\nu}|$ is larger than $0.32
  |\boldsymbol{a}_{c_S}|$ in $6.8\%$ and $3.7\%$ of all
  spatio-temporal events for our lower and higher Reynolds numbers
  respectively) and where the right hand side of \eqref{rotHIT2} is
  therefore not zero. In fact, many of these relatively rare events
  can be expected to account for some or even much of the average
  turbulence dissipation which is a sum of squares of fluctuating
  velocity gradients. More generally, one cannot use equation
  \eqref{rotHIT2} to derive statistics of fluctuating velocity
  gradients, as in \citet{Tang2022} for example.

The second of the two Tsinober equations, namely
\begin{equation}
	\boldsymbol{a}_{c_I} = \boldsymbol{a}_p , \label{poissonHIT2}
\end{equation}
is exact everywhere and at any time and we keep it as
  it is.

\begin{table}
	\begin{center}
	\setlength{\tabcolsep}{6.5pt}
	\begin{tabular}{lcccccc}
	$\langle \Rey_{\lambda}\rangle_t$ &$\langle\cos$($\boldsymbol{a}_{c_I}$, $\boldsymbol{a}_p) \rangle$ & $\langle\cos$($\boldsymbol{a}$, $\boldsymbol{a}_p)\rangle$  & $\langle\cos$($\boldsymbol{a}_l$, $\boldsymbol{a}_{c_S})\rangle$ & $\langle\cos$($\boldsymbol{a}_l$, $\boldsymbol{a}_c)\rangle$ & $\langle\cos$($\boldsymbol{a}_c$, $\boldsymbol{a}_p)\rangle$ \\ [3pt]
	112 & 0.9999 & 0.972 & -0.985 & -0.726 & 0.388 \\ [2pt]
	174 & 0.9999 & 0.975 & -0.990 & -0.796 & 0.308 \\ [2pt]
	
	\end{tabular}
	\caption{NS average alignments $\langle
          \cos(\boldsymbol{q},\boldsymbol{p})\rangle$ for NS
          acceleration pairs $(\boldsymbol{q},\boldsymbol{p})$.}
     \label{tab:nsTab2}
     \end{center}
\end{table}

Equations \eqref{rotHIT2}-\eqref{poissonHIT2} suggest similar
magnitudes and strong alignment between $\boldsymbol{a}_l$ and
$-\boldsymbol{a}_{c_S}$ and equal magnitudes as well as perfect
alignment between $\boldsymbol{a}_{c_I}$ and $\boldsymbol{a}_p$. Such
magnitudes and alignments were observed in the DNS of
\citet{Tsinober2001} and are also strongly confirmed by our own DNS
      in table \ref{tab:nsTab2} ($\boldsymbol{a}_{c_S}$
      and $\boldsymbol{a}_{c_I}$ are calculated on the basis of
      equation \eqref{decompQ} in appendix \ref{sec:appA} and aliasing
      errors associated with non-linear terms are removed by
      phase-shifting and truncation \citep{Patterson1971}). As
      suggested by previous DNS and experimental results
      (e.g. \citet{Tsinober2001,Chevillard2005,Yeung2006}), and as
      also supported by our own DNS results in tables
        \ref{tab:nsTab} and \ref{tab:nsTab2}, $\boldsymbol{a}\approx
      \boldsymbol{a}_p$ and $\langle \boldsymbol{a}_{l}^2
      \rangle/\langle \boldsymbol{a}^2 \rangle \sim \langle
      \Rey_{\lambda} \rangle_t^{1/2}$
In fact, the scaling $\langle \boldsymbol{a}_{l}^2 \rangle/\langle
\boldsymbol{a}^2 \rangle \sim \langle \Rey_{\lambda} \rangle_t^{1/2}$
follows from the analysis of \citet{Tennekes1975} who expressed the
concept of passive sweeping by pointing out that "at high Reynolds
number the dissipative eddies flow past an Eulerian observer in a time
much shorter than the time scale which characterizes their own
dynamics". It then follows from equations
\eqref{rotHIT2}-\eqref{poissonHIT2}, from $\langle
\boldsymbol{a}_{l}^2 \rangle/\langle \boldsymbol{a}^2 \rangle \sim
\langle \Rey_{\lambda} \rangle_t^{1/2}$ and from $\langle
\boldsymbol{a}_{p}^2 \rangle \approx \langle \boldsymbol{a}^2 \rangle$
that $\langle \boldsymbol{a}_{c_S}^2 \rangle/\langle
\boldsymbol{a}_{c_I}^2 \rangle \sim \langle \Rey_{\lambda}
\rangle_t^{1/2}$ with increasing $\langle Re_\lambda\rangle_t$, i.e.,
$\boldsymbol{a}_c$ becomes increasingly solenoidal with increasing
$\langle Re_\lambda\rangle_t$.  In this way, the anti-alignment in
\eqref{rotHIT2} leads to an increasing anti-alignment tendency between
$\boldsymbol{a}_l$ and $\boldsymbol{a}_c$ with increasing Reynolds
number, which is consistent with the notion of passive sweeping of
small eddies by large ones, i.e. the random Taylor hypothesis of
\citet{Tennekes1975}. These observations and conclusions were
all made by \citet{Tsinober2001} and their DNS who
  showed, in particular, that the Taylor length-based Reynolds number
  does not need to be so large to make them, and are now confirmed by
our DNS in table \ref{tab:nsTab}.

As a final point, it is a general property of isotropic random vector
fields $\boldsymbol{q}$ that $\langle
\boldsymbol{q}_I(\boldsymbol{x},t) \cdot
\boldsymbol{q}_S(\boldsymbol{x} + \boldsymbol{r},t)\rangle_x=0$ for
any $\boldsymbol{r}$ (including $\boldsymbol{r}=0$), where $\langle
... \rangle_{\boldsymbol{x}}$ signifies a spatial average
\citep{Monin1975}. Thus, $\langle \boldsymbol{a}_{c}^2\rangle =
\langle \boldsymbol{a}_{c_I}^2\rangle + \langle
\boldsymbol{a}_{c_S}^2\rangle$ if the small-scale turbulence is
isotropic. Both our DNS and the DNS of \citet{Tsinober2001} confirm
this equality. From this equality and from \eqref{rotHIT2}, $\langle
\boldsymbol{a}_{c_S}^2 \rangle/\langle \boldsymbol{a}_{c_I}^2 \rangle
\sim \langle \Rey_{\lambda} \rangle_t^{1/2}$, \eqref{poissonHIT2},
$\boldsymbol{a} \approx \boldsymbol{a}_{p}$ and $\langle
\boldsymbol{a}^{2}\rangle \gg \langle \boldsymbol{a}_{\nu}^{2}\rangle
\gg \langle \boldsymbol{f}^{2}\rangle$, we have all in all
\begin{equation}
	\langle \boldsymbol{a}_{c}^2 \rangle \ge \langle \boldsymbol{a}_{c_S}^2
        \rangle \approx \langle \boldsymbol{a}_{l}^2 \rangle \gg \langle
        \boldsymbol{a}_{c_I}^2 \rangle = \langle\boldsymbol{a}_p^2 \rangle
        \approx \langle\boldsymbol{a}^2 \rangle \gg \langle \boldsymbol{a}_{\nu}^2
        \rangle \gg \langle \boldsymbol{f}^2 \rangle \label{magNS},
\end{equation}
for large enough $\langle \Rey_{\lambda} \rangle_t$. The average
magnitude ordering in \eqref{magNS} is confirmed in our DNS
(see table \ref{tab:nsTab}) and the DNS of
\citet{Tsinober2001} even though the Reynolds numbers of these DNS are
moderate and so the difference between $\langle \boldsymbol{a}_{c_I}^2
\rangle$ and $\langle\boldsymbol{a}_l^2\rangle$ is not so
large. Tsinober's way to formulate sweeping is
  encapsulated in $\langle \boldsymbol{a}_{c_S}^2 \rangle \approx
  \langle \boldsymbol{a}_{l}^2 \rangle \gg \langle
  \boldsymbol{a}_{c_I}^2 \rangle = \langle\boldsymbol{a}_p^2 \rangle
  \approx \langle\boldsymbol{a}^2 \rangle$ and in the alignments
  implied by equations \eqref{rotHIT2}-\eqref{poissonHIT2} which are
  also statistically confirmed by our DNS in table \ref{tab:nsTab2}.

\subsection{From one-point to two-point Navier-Stokes dynamics in periodic/homogeneous turbulence} \label{subsec:nsd}

The Navier-Stokes difference (NSD) equation at centroid
$\boldsymbol{x}$ and separation vector $\boldsymbol{r}$ is derived by
subtracting the Navier-Stokes (NS) equation at location
$\boldsymbol{x}^{+}=\boldsymbol{x}+\boldsymbol{r}/2$ from the NS
equation at location
$\boldsymbol{x}^{-}=\boldsymbol{x}-\boldsymbol{r}/2$. Defining $\delta
\boldsymbol{q}(\boldsymbol{x},\boldsymbol{r},t)\equiv
\boldsymbol{q}(\boldsymbol{x}+\boldsymbol{r}/2,t)-\boldsymbol{q}(\boldsymbol{x}-\boldsymbol{r}/2,t)$
for any NS term $\boldsymbol{q}(\boldsymbol{x},t)$, the NSD equation
\citep{HILL2002} reads
\begin{equation}
	\frac{\partial \delta \boldsymbol{u}}{\partial t}+\delta
        \boldsymbol{a}_c =
        -\frac{1}{\rho}\nabla_{\boldsymbol{x}}\delta p +\delta
        \boldsymbol{a}_\nu +\delta \boldsymbol{f} , \label{NSD1}
\end{equation}

\noindent The NSD equation governs the evolution of $\delta
\boldsymbol{u}$, which can be thought of as pertaining to the momentum
at scales smaller or comparable to $|\boldsymbol{r}|$. We derive the
solenoidal NSD equation by subtracting equation \eqref{rotHIT1} at
$\boldsymbol{x}-\boldsymbol{r}/2$ from the same equation at
$\boldsymbol{x}+\boldsymbol{r}/2$. The same operation is used to
derive the irrotational NSD equation. The resulting equations read
\begin{align}
	\frac{\partial \delta \boldsymbol{u}}{\partial t}+\delta
        \boldsymbol{a}_{c_S} &= \delta \boldsymbol{a}_\nu +\delta
        \boldsymbol{f}_S , \label{NSD2} \\ \delta \boldsymbol{a}_{c_I}
        &= -\frac{1}{\rho}\nabla_{\boldsymbol{x}}\delta p + \delta
        \boldsymbol{f}_I , \label{NSD3}
\end{align}

\noindent where $\delta
\boldsymbol{a}_{c_S}(\boldsymbol{x},\boldsymbol{r},t) \equiv
\boldsymbol{a}_{c_S}(\boldsymbol{x}+\boldsymbol{r}/2,t)-
\boldsymbol{a}_{c_S}(\boldsymbol{x}-\boldsymbol{r}/2,t)$ and $\delta
\boldsymbol{a}_{c_I}(\boldsymbol{x},\boldsymbol{r},t) \equiv
\boldsymbol{a}_{c_I}(\boldsymbol{x}+\boldsymbol{r}/2,t)-
\boldsymbol{a}_{c_I}(\boldsymbol{x}-\boldsymbol{r}/2,t)$ and note that
these terms refer to solenoidal and irrotational terms in
$\boldsymbol{x}$-space rather than $\boldsymbol{r}$-space. The
forcings we consider have no irrotational part and so $\delta
\boldsymbol{f}_I = 0$.  At the moderate $\langle
  \Rey_\lambda \rangle_t$ of our DNS, the approximate equation
\eqref{rotHIT2} is valid in the sense explained in the
  text which accompanies it in the previous sub-section, i.e. for a
  majority of spacio-temporal events. If the magnitude of the
separation vector $\boldsymbol{r}$ is not too small for viscosity to
matter directly nor too large for the forcing to be directly present,
we may safely subtract equation \eqref{rotHIT2} at
$\boldsymbol{x}-\boldsymbol{r}/2$ from equation \eqref{rotHIT2} at
$\boldsymbol{x}+\boldsymbol{r}/2$ to obtain an approximation to
\eqref{NSD2} for sufficiently high Reynolds number:
  this is the first of the two equations below where $\delta
\boldsymbol{a}_l \equiv \partial \delta \boldsymbol{u}/\partial
  t$:
\begin{align}
	\delta \boldsymbol{a}_{l} + \delta \boldsymbol{a}_{c_S}&\approx 0    , \label{NSD51}  \\
	\delta \boldsymbol{a}_{c_I} &= -\frac{1}{\rho}\nabla_{\boldsymbol{x}}\delta p   . \label{NSD61}
\end{align}
The second equation, equation \eqref{NSD61}, follows directly from
\eqref{NSD3} with $\delta \boldsymbol{f}_I = 0$ without any
restriction on either $\boldsymbol{r}$ or Reynolds
  number and is exact.

\begin{figure}
	\centerline{\includegraphics{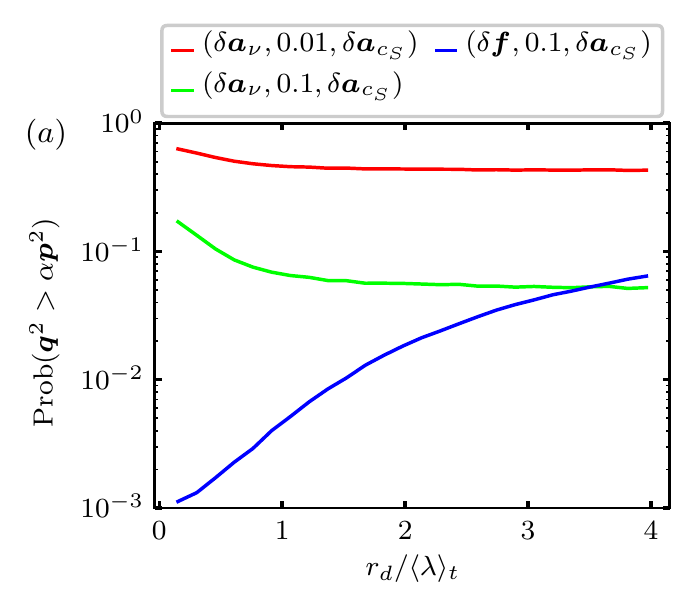}	\hfill			  \includegraphics{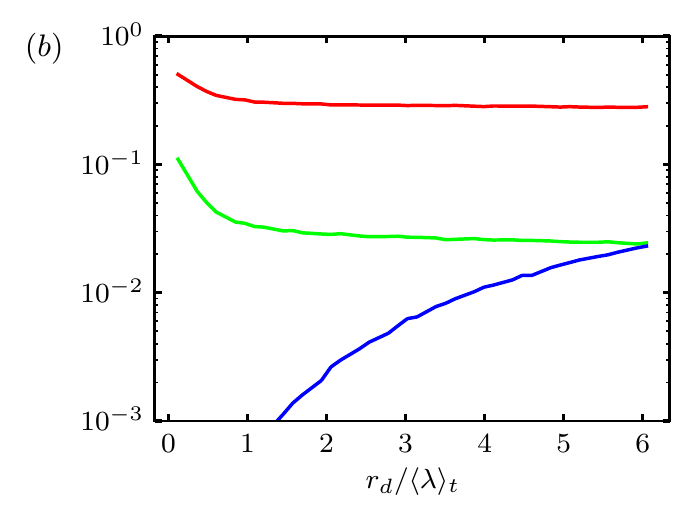}}
	\caption{Navier-Stokes difference (NSD) exceedance
          probabilities $\text{Prob}(\boldsymbol{q}^2>\alpha
          \boldsymbol{p}^2)$ for the NSD terms on top of ($a$) as a
          function of separation length $r_d=|\boldsymbol{r}|$. The
          legend entries read $(\boldsymbol{q},\alpha,
          \boldsymbol{p})$ for the NSD terms introduced in the first
          paragraph of \ref{subsec:nsd}. ($a$) $\langle
            \Rey_\lambda \rangle_t = 112$, $\langle L\rangle_{t} = 3.5
            \langle \lambda \rangle_{t}$. $(b)$ $\langle \Rey_\lambda
            \rangle_t = 174$, $\langle L\rangle_{t} = 5.2 \langle
            \lambda \rangle_{t}$. NSD terms are sampled at scale
          $r_d=|\boldsymbol{r}|$ at random orientations
          $\boldsymbol{r}$.}
	\label{fig:nsdExtreme}
\end{figure}

Like equation \eqref{rotHIT2}, \eqref{NSD51} can be expected to be
valid broadly except where and when $\delta \boldsymbol{a}_\nu +\delta
\boldsymbol{f}_S$ is large enough not to be negligible. Figure
\ref{fig:nsdExtreme} shows statistically converged
estimations of exceedance probabilities of NSD viscous
  and external force terms which suggest that \eqref{NSD51} is indeed
a good approximation for most of space and time at the
  Reynolds numbers of our two DNS, at the very least for separation
distances larger than $\langle \lambda \rangle_{t}$ and smaller than
$\langle L\rangle_t$. With regards to the forcing,
$\text{Prob}(|\delta \boldsymbol{f}|>0.32 |\delta
\boldsymbol{a}_{c_S}|)$ is typically of the order of $1\%$, in
particular for our higher Reynolds number. With regards to the viscous
force, $\text{Prob}(|\delta \boldsymbol{a}_{\nu}|>0.32 |\delta
\boldsymbol{a}_{c_S}|)$ is typically of the order of $5\%$ for $r\ge
\langle \lambda \rangle_{t}$ and even less for our higher Reynolds
number.

The link between non-linearity and non-locality (via the pressure
field) invoked in the two-point analysis of \citet{Yasuda2018} has its
root in equation \eqref{NSD61} which parallels \eqref{poissonHIT2} and
states that $\delta \boldsymbol{a}_{c_I}$ and $\delta
\boldsymbol{a}_p$ are perfectly aligned and have the same
magnitudes. Furthermore, similarly to the way that equation
\eqref{rotHIT2} supports the concept of sweeping of small turbulent
eddies by large ones in the usual one-point sense, \eqref{NSD51}
suggests similar magnitudes for and strong alignment between $\delta
\boldsymbol{a}_l$ and $-\delta \boldsymbol{a}_{c_S}$. A two-point
concept of sweeping similar to the one of \citet{Tennekes1975}
which relies on alignment between $\delta
  \boldsymbol{a}_l$ and $-\delta \boldsymbol{a}_{c}$ should also
require that $\delta \boldsymbol{a}_{c}$ tends towards $\delta
\boldsymbol{a}_{c_s}$ with increasing Reynolds number, i.e. that
$\delta \boldsymbol{a}_c$ becomes increasingly solenoidal. We
therefore seek to obtain inequalities and approximate equalities
similar to \eqref{magNS}. Note that equations
\eqref{NSD51}-\eqref{NSD61} immediately imply $\langle
\delta\boldsymbol{a}_{c_S}^2 \rangle \approx \langle \delta
\boldsymbol{a}_{l}^2 \rangle$, $\langle \delta\boldsymbol{a}_{c_I}^2
\rangle = \langle\delta\boldsymbol{a}_p^2 \rangle$ and
$\langle\delta\boldsymbol{a}_p^2 \rangle \approx \langle
\delta\boldsymbol{a}^2 \rangle$.  It therefore remains to argue that
$\langle \delta \boldsymbol{a}_{c}^2 \rangle \ge \langle
\delta\boldsymbol{a}_{c_S}^2 \rangle \gg
\langle\delta\boldsymbol{a}_{c_I}^2 \rangle$ which is exactly what we
need to complete the new concept of two-point
sweeping.

We start from
\begin{align}
	\langle \delta \boldsymbol{q} \cdot \delta \boldsymbol{q} \rangle (\boldsymbol{r}) &=  \langle  \boldsymbol{q}^{+} \cdot \boldsymbol{q}^{+} \rangle- \langle \boldsymbol{q}^{+} \cdot \boldsymbol{q}^{-} \rangle + \langle  \boldsymbol{q}^{-} \cdot \boldsymbol{q}^{-} \rangle- \langle \boldsymbol{q}^{-} \cdot \boldsymbol{q}^{+} \rangle , \label{mag0}  \\
	 &= 2 \big[ \langle  \boldsymbol{q} \cdot \boldsymbol{q} \rangle -  \langle \boldsymbol{q}^{+} \cdot \boldsymbol{q}^{-} \rangle(\boldsymbol{r}) \big], \label{mag1}
\end{align}
\noindent where $\boldsymbol{q}^{+} \equiv
\boldsymbol{q}(\boldsymbol{x}+\boldsymbol{r}/2)$ and
$\boldsymbol{q}^{-} \equiv
\boldsymbol{q}(\boldsymbol{x}-\boldsymbol{r}/2)$ and where we used
$\langle \boldsymbol{q}^{+} \cdot \boldsymbol{q}^{+} \rangle =\langle
\boldsymbol{q}^{-} \cdot \boldsymbol{q}^{-} \rangle = \langle
\boldsymbol{q} \cdot \boldsymbol{q} \rangle$ because of statistical
homogeneity/periodicity. Previous studies \citep{Hill1997a,
  Vedula1999, Xu2007, G2007} demonstrated that fluid accelerations,
pressure-gradients and viscous forces have limited spatial
correlations in terms of alignments at scales larger than $\langle
\lambda\rangle_{t}$ for moderate and high $\langle
\Rey_\lambda \rangle_t$. Thus, if we assume the two-point term to be
negligible compared to the one-point term in Eq. \eqref{mag1} for
scales $|\boldsymbol{r}|$ larger than $\langle \lambda\rangle_{t}$,
we have that $\langle \delta \boldsymbol{q} \cdot \delta
\boldsymbol{q} \rangle (\boldsymbol{r})$ is approximately equal to $2
\langle \boldsymbol{q}\cdot \boldsymbol{q} \rangle$ for $\vert
\boldsymbol{r}\vert$ larger than $\langle \lambda\rangle_{t}$.  From
\eqref{magNS} we therefore obtain
\begin{equation}
	\langle \delta \boldsymbol{a}_{c}^2 \rangle \ge \langle
        \delta\boldsymbol{a}_{c_S}^2 \rangle \approx \langle \delta
        \boldsymbol{a}_{l}^2 \rangle \gg \langle \delta\boldsymbol{a}_{c_I}^2 \rangle
        = \langle\delta\boldsymbol{a}_p^2 \rangle \approx \langle
        \delta\boldsymbol{a}^2 \rangle \gg \langle\delta \boldsymbol{a}_{\nu}^2
        \rangle \gg \langle \delta\boldsymbol{f}^2 \rangle \label{magNSDD} ,
\end{equation}
for $\vert \boldsymbol{r}\vert$ larger than $\langle
\lambda\rangle_{t}$, but $\langle \delta \boldsymbol{a}_{c}^2 \rangle
\ge \langle \delta\boldsymbol{a}_{c_S}^2 \rangle$ and $\langle
\delta\boldsymbol{a}_{c_I}^2 \rangle = \langle\delta\boldsymbol{a}_p^2
\rangle$ are in fact valid for any $\boldsymbol{r}$. Inequality
$\langle \delta \boldsymbol{a}_{c}^2 \rangle \ge \langle
\delta\boldsymbol{a}_{c_S}^2 \rangle$ follows from $\langle \delta
\boldsymbol{a}_c^2 \rangle = \langle \delta \boldsymbol{a}_{c_I}^2
\rangle + \langle \delta \boldsymbol{a}_{c_S}^2 \rangle$ which itself
follows from $\langle \boldsymbol{a}_{c_I}(\boldsymbol{x},t) \cdot
\boldsymbol{a}_{c_S}(\boldsymbol{x}+\boldsymbol{r},t)\rangle_x=0$ for
any $\boldsymbol{r}$ if the turbulence is isotropic
\citep{Monin1975}. Equality $\langle \delta\boldsymbol{a}_{c_I}^2
\rangle = \langle\delta\boldsymbol{a}_p^2\rangle$ follows directly
from \eqref{NSD61} which is exact and holds for any
$\boldsymbol{r}$ and any Reynolds number. Of
equalities/inequalities \eqref{magNSDD}, the ones that we did not
already directly derive from/with equations
\eqref{NSD51}-\eqref{NSD61} are $\langle \delta \boldsymbol{a}_{c}^2
\rangle \ge \langle \delta\boldsymbol{a}_{c_S}^2 \rangle \gg
\langle\delta\boldsymbol{a}_{c_I}^2 \rangle$ and $\langle\delta
\boldsymbol{a}_{\nu}^2 \rangle \gg \langle \delta\boldsymbol{f}^2
\rangle$. The present way to formulate the new
  concept of two-point sweeping follows from Tsinober's way to
  formulate sweeping and is encapsulated in
  $\langle\delta\boldsymbol{a}_{c_S}^2 \rangle \approx \langle \delta
  \boldsymbol{a}_{l}^2 \rangle \gg \langle
  \delta\boldsymbol{a}_{c_I}^2 \rangle =
  \langle\delta\boldsymbol{a}_p^2 \rangle \approx \langle
  \delta\boldsymbol{a}^2 \rangle$ and in the alignments implied by
  equations \eqref{NSD51}-\eqref{NSD61}. We confirm equations
  \eqref{NSD51}-\eqref{NSD61}-\eqref{magNSDD} with our DNS in the
  remainder of this subsection.

\begin{figure}
	\centerline{\includegraphics{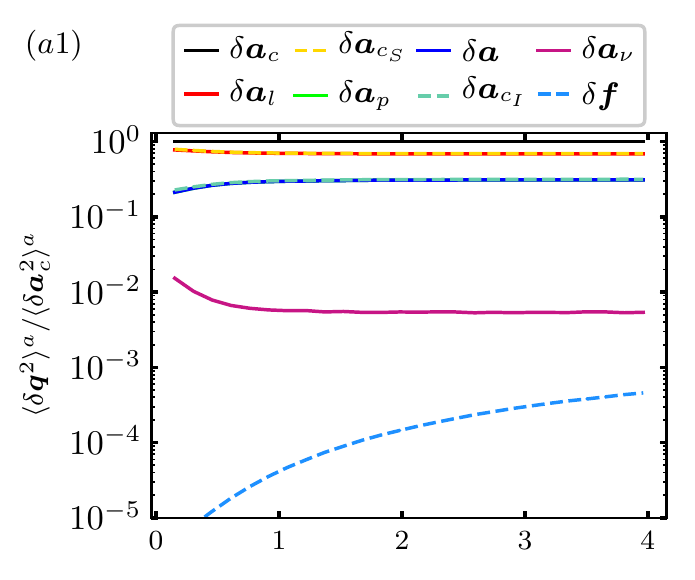} \hfill
          \includegraphics{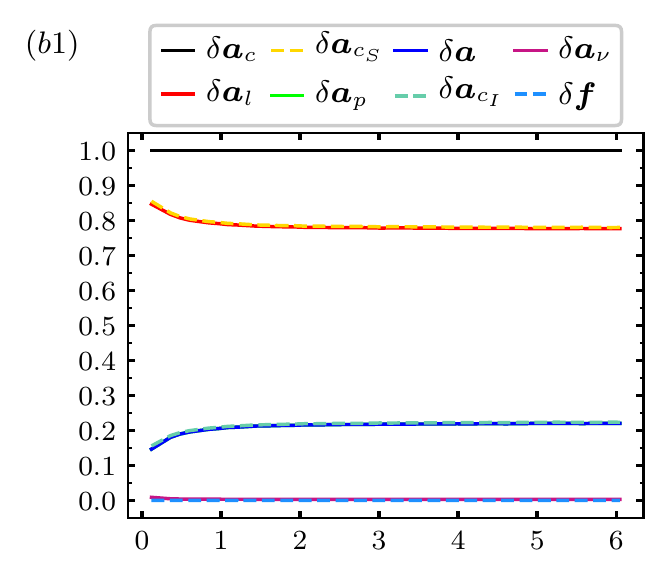}}%
        \centerline{\includegraphics{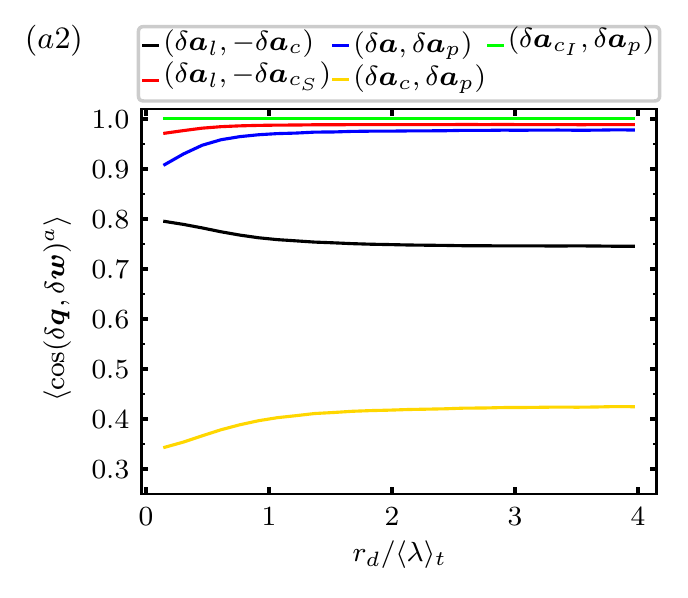} \hfill
          \includegraphics{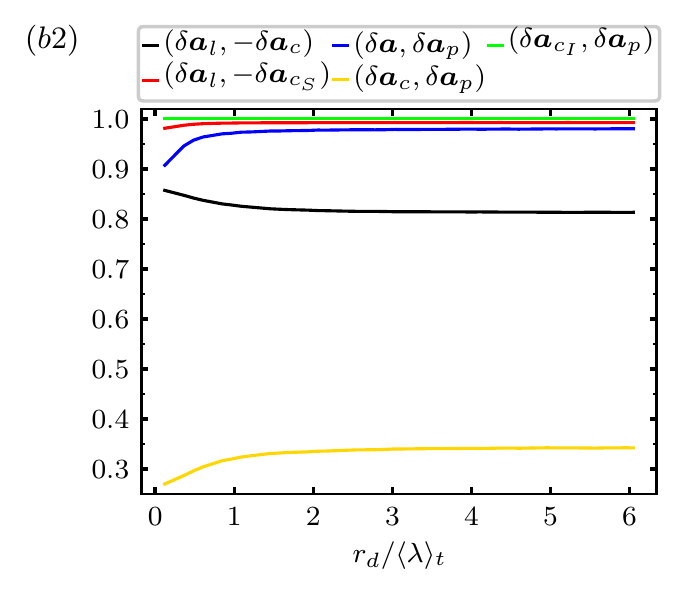}}%
	\caption{($a1$,$b1$) spatio-temporal averages of spherically
          averaged NSD magntiudes $(\delta \boldsymbol{q}^2)^a \equiv
          (\upi r_d^2)^{-1} \iiint_{|\boldsymbol{r}|=r_d} \delta
          \boldsymbol{q}(\boldsymbol{x},\boldsymbol{r},t) \cdot \delta
          \boldsymbol{q}(\boldsymbol{x},\boldsymbol{r},t),
          \ d\boldsymbol{r}$ for NSD terms $\delta \boldsymbol{q}$
          listed on top of the figures as a function of $r_d$: ($a1$)
          $\langle \Rey_\lambda\rangle_t=112$, ($b1$) $\langle
          \Rey_\lambda\rangle_t=174$. The magnitudes of
            the terms $\delta \boldsymbol{a}_l$ and $\delta
            \boldsymbol{a}_{c_S})$ overlap and the magntiudes of the
            terms $(\delta \boldsymbol{a}_p$, $\delta \boldsymbol{a}$
            and $\delta \boldsymbol{a}_{c_I})$ also overlap. ($a2$,
          $b2$) average NSD alignments between NSD terms $(\delta
          \boldsymbol{q},\delta \boldsymbol{w})$ listed on top of the
          figures as a function of $r_d$: ($a2$) $\langle
          \Rey_\lambda\rangle_t=112$, ($b2$) $\langle
          \Rey_\lambda\rangle_t=174$.}
	\label{fig:nsdFig1}
\end{figure}

To test \eqref{magNSDD} with our DNS data in a manageable way, we
calculate spatio-temporal averages of $\boldsymbol{r}$-orientation-averaged
quantities
\begin{equation}
	(\delta \boldsymbol{q} \cdot \delta \boldsymbol{q} )^a(\boldsymbol{x},r_d,t) \equiv
  \frac{1}{\upi r_d^2} \iiint_{|\boldsymbol{r}|=r_d} \delta
  \boldsymbol{q}(\boldsymbol{x},\boldsymbol{r},t) \cdot \delta \boldsymbol{q}(\boldsymbol{x},\boldsymbol{r},t),
  \ d\boldsymbol{r},
  \label{ORIENTAVG}
\end{equation}
which we plot in figure \ref{fig:nsdFig1}($a1$,$a2$) as ratios of such
quantities versus two-point length $r_d$. In figure
  \ref{fig:nsdFig1}($a1$,$a2$) we plot spatio-temporal averages of
  $\boldsymbol{r}$-orientation-averaged quantities \eqref{ORIENTAVG}
  for various acceleration/force terms in the NSD and the Helmholtz
  decomposed NSD equations. A comparison of relative magnitudes in the
  plots of figure \ref{fig:nsdFig1}($a1$,$a2$) with relative
  magnitudes in table \ref{tab:nsTab} makes it clear that the results
are consistent with \eqref{magNSDD} and $\langle \delta \boldsymbol{q}
\cdot \delta \boldsymbol{q} \rangle (\boldsymbol{r}) /\langle
\boldsymbol{q}\cdot \boldsymbol{q} \rangle$ close to $2$ for $r_{d}
\geq \langle \lambda \rangle_t$ at both $\langle Re_\lambda \rangle_t$
to a good degree of accuracy ($\langle \delta \boldsymbol{q} \cdot
\delta \boldsymbol{q} \rangle (\boldsymbol{r}) /\langle
\boldsymbol{q}\cdot \boldsymbol{q} \rangle$ increases from 1.8 to 2.0
as $r_d$ grows from $\langle \lambda\rangle_t$ to $\langle
L\rangle_t$). Note, in particular, that in Figure
\ref{fig:nsdFig1}($a1$,$b1$) the average quantities corresponding to $\delta \boldsymbol{a}_l$ and $\delta \boldsymbol{a}_{c_S}$ overlap
and those corresponding to $\delta \boldsymbol{a}_p$, $\delta
\boldsymbol{a}$ and $\delta \boldsymbol{a}_{c_I}$ also overlap. At
scales below $\langle \lambda \rangle_t$, the average relative
magnitudes change slightly, but the NSD magnitude separations still
abide by \eqref{magNSDD}, the NSD analogue to \eqref{magNS}, at all
scales.

In figure \ref{fig:nsdFig1}($b1$,$b2$) we use our DNS data to plot
spatio-temporal averages of $\boldsymbol{r}$-orientation-averaged
cosines of angles between various NSD terms $\delta \boldsymbol{q}$
and $\delta \boldsymbol{w}$ to test for average alignments as a
function of $r_d$. These alignment results are of course in perfect
agreement with \eqref{NSD61} but they are also in good agreement with
\eqref{NSD51} and acceptable agreement with $\delta \boldsymbol{a}
\approx \delta \boldsymbol{a}_{p}$ {(the cosine of the
  angle between these two acceleration vectors is higher than $0.9$
  for all $r_d$)}. They also show that we should not expect $\delta
\boldsymbol{a}_l$ to be extremely well aligned with $-\delta
\boldsymbol{a}_c$ at our moderate Reynolds numbers. This demonstrates
the pertinence of the solenoidal-irrotational decomposition which has
revealed very good alignments at our moderate Reynolds numbers for
which there are significantly weaker alignments without
{this decomposition}.

In conclusion, figure \ref{fig:nsdFig1} provides strong support for
equations \eqref{NSD51}-\eqref{NSD61}-\eqref{magNSDD} which establish
the two-point link between non-linearity and non-locality, and also a
concept of two-point sweeping.

\subsection{Interscale transfer and physical space transport accelerations} \label{subsec:nsdTwo}

The convective non-linearity is responsible for non-linear turbulence
transport through space and non-linear transfer through scales. We
want to separate these two effects and therefore decompose the
two-point non-linear acceleration term $\delta \boldsymbol{a}_c$ into an
interscale transfer acceleration $\boldsymbol{a}_{\mathit{\Pi}}$ and a
physical space transport acceleration $\boldsymbol{a}_{\mathcal{T}}$
\citep{HILL2002}, i.e $\delta \boldsymbol{a}_c =
\boldsymbol{a}_\mathit{\Pi}+\boldsymbol{a}_\mathcal{T}$ with
\begin{equation}
	\boldsymbol{a}_\mathcal{T}(\boldsymbol{x},\boldsymbol{r},t) =
        \frac{1}{2}(\boldsymbol{u}^{+}+\boldsymbol{u}^-)\cdot \nabla_{\boldsymbol{x}} \delta
        \boldsymbol{u} , \quad \boldsymbol{a}_\mathit{\Pi}(\boldsymbol{x},\boldsymbol{r},t)=\delta
        \boldsymbol{u} \cdot \nabla_{\boldsymbol{r}} \delta \boldsymbol{u}
        . \label{nsdTransTerms}
\end{equation}
\noindent With this decomposition of the non-linear term, the NSD
equation \eqref{NSD1} reads
\begin{equation}
	\frac{\partial \delta \boldsymbol{u}}{\partial t}+\boldsymbol{a}_{\mathit{\Pi}}+\boldsymbol{a}_{\mathcal{T}} = -\frac{1}{\rho}\nabla_{\boldsymbol{x}}\delta p  + \delta \boldsymbol{a}_{\nu} + \delta \boldsymbol{f} . \label{NSDTransport}
\end{equation}

\noindent
We note relations $\boldsymbol{a}_{\mathit{\Pi}} = \delta
\boldsymbol{a}_{C} + u^{+}_{j} \partial \boldsymbol{u}^{-}  / \partial x^{-}_{j} -u^{-}_{j} \partial \boldsymbol{u}^{+} / \partial x^{+}_{j}
$ and $\boldsymbol{a}_{\mathcal{T}}=\delta
\boldsymbol{a}_{C} - u^{+}_{j} \partial \boldsymbol{u}^{-}/ \partial x^{-}_{j}
 + u^{-}_{j} \partial \boldsymbol{u}^{+}/  \partial x^{+}_{j}
$ which can be easily used to show that $\langle
\boldsymbol{a}_{\mathit{\Pi}}^{2}\rangle$ and $\langle
\boldsymbol{a}_{\mathcal{T}}^{2}\rangle$ tend
towards each other as the amplitude of the separation vector $\boldsymbol{r}$
grows above the integral length scale. We report DNS evidence of this
tendency, below in this paper.

We want to consider the effects of the interscale transfer and
interspace transport terms in the solenoidal and irrotational NSD
dynamics and we therefore need to break down the NSD equation
\eqref{NSDTransport} into two equations, one irrotational and one
solenoidal. We therefore perform Helmholtz decompositions in centroid
space $\boldsymbol{x}$ for a given separation $\boldsymbol{r}$ at time
$t$, for example $\delta
\boldsymbol{q}(\boldsymbol{x},\boldsymbol{r},t)= \delta
\boldsymbol{q}_{\overline{I}}(\boldsymbol{x},\boldsymbol{r},t)+\delta
\boldsymbol{q}_{\overline{S}}(\boldsymbol{x},\boldsymbol{r},t)$ where
$\delta
\boldsymbol{q}_{\overline{I}}(\boldsymbol{x},\boldsymbol{r},t)$ and
$\delta
\boldsymbol{q}_{\overline{S}}(\boldsymbol{x},\boldsymbol{r},t)$ are,
respectively, the irrotational and solenoidal parts in centroid space
of $\delta \boldsymbol{q}(\boldsymbol{x},\boldsymbol{r},t)$. This
decomposition in centroid space differs in general from the difference
of the Helmholtz decomposed terms in the NS equations which gives
equations \eqref{NSD2}-\eqref{NSD3}, but in periodic/homogeneous
turbulence $\delta \boldsymbol{q}_{I} = \delta
\boldsymbol{q}_{{\overline{I}}}$ and $\delta \boldsymbol{q}_{S} =
\delta \boldsymbol{q}_{{\overline{S}}}$ (see appendix
\ref{sec:appB}). Furthermore, from $\delta \boldsymbol{a}_c =
\boldsymbol{a}_\mathit{\Pi}+\boldsymbol{a}_\mathcal{T}$ immediately
follow $\delta
\boldsymbol{a}_{c_{\overline{S}}}=\boldsymbol{a}_{\mathit{\Pi}_{\overline{S}}}+\boldsymbol{a}_{\mathcal{T}_{\overline{S}}}$
and $\delta
\boldsymbol{a}_{c_{\overline{I}}}=\boldsymbol{a}_{\mathit{\Pi}_{\overline{I}}}+\boldsymbol{a}_{\mathcal{T}_{\overline{I}}}$. Thus,
we can rewrite the NSD solenoidal and irrotational equations
\eqref{NSD2}-\eqref{NSD3} as
\begin{align}
	 \boldsymbol{a}_{\mathit{\Pi}_{\overline{I}}}+\boldsymbol{a}_{\mathcal{T}_{\overline{I}}}
         &= \delta \boldsymbol{a}_p, \label{NSDirrTr} \\ \delta \boldsymbol{a}_l +
         \boldsymbol{a}_{\mathit{\Pi}_{\overline{S}}}+\boldsymbol{a}_{\mathcal{T}_{\overline{S}}}
         &= \delta \boldsymbol{a}_{\nu}+ \delta \boldsymbol{f} \label{NSDsolTr},
\end{align}
in periodic/homogeneous turbulence.

We emphasize that the interscale transfer term
$\boldsymbol{a}_{\mathit{\Pi}_{\overline{S}}}$ is related non-locally
in space to two-point vortex stretching and compression terms
governing the evolution of vorticity difference $\delta
\boldsymbol{\omega}$. This follows from the fact that, as for the
Tsinober equations, the NSD solenoidal equation is an integrated
vorticity difference equation. We provide mathematical detail on the
connection between $\boldsymbol{a}_{\mathit{\Pi}_{\overline{S}}}$ and
$\delta \boldsymbol{\omega}$ in appendix \ref{sec:appC}. This relation
between $\boldsymbol{a}_{\mathit{\Pi}_{\overline{S}}}$ and the
vorticity difference dynamics provides an instantaneous connection
between the interscale momentum dynamics and two-point vorticity
stretching and compression dynamics.

\begin{figure}
	\centerline{\includegraphics{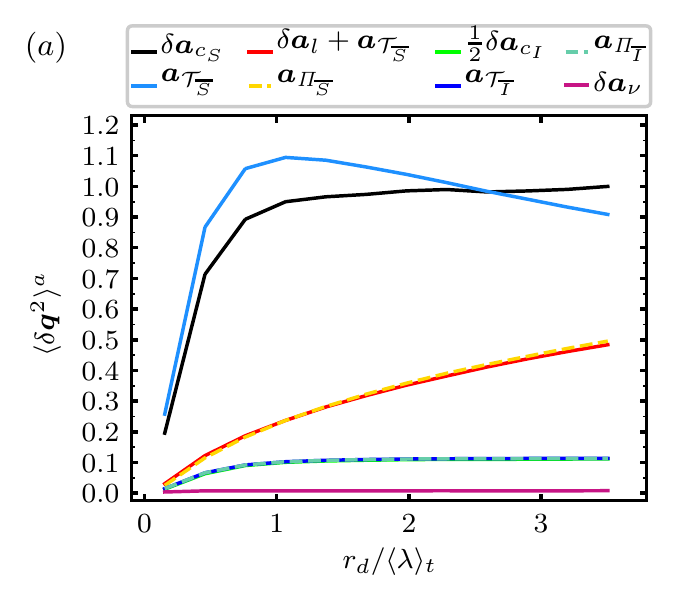} \hfill
          \includegraphics{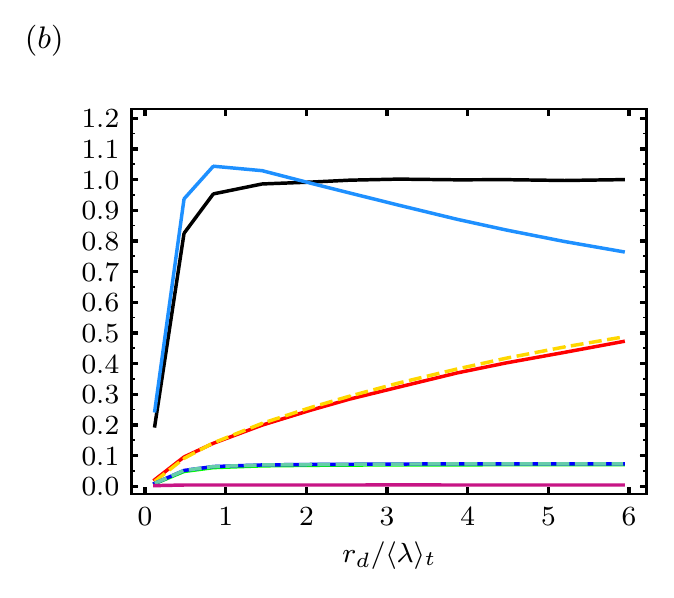}}%
	\caption{Average magnitudes $\langle \delta \boldsymbol{q}^2
          \rangle^a$ of NSD terms present in the irrotational and
          solenoidal NSD equations
          \eqref{NSDsolTr}-\eqref{transportPressure} listed on top of
          $(a)$. All values have been normalised with $\langle \delta
          \boldsymbol{a}_{c_S}^2 \rangle^a$ at the largest considered
          separation $r_d$. {The magnitudes of the terms
            $(\delta
            \boldsymbol{a}_l+\boldsymbol{a}_{\mathcal{T}_{\overline{S}}}$
            and $\boldsymbol{a}_{\mathit{\Pi}_{\overline{S}}})$
            overlap and the magnitudes of the terms $(1/2\delta
            \boldsymbol{a}_{c_I}$,
            $\boldsymbol{a}_{\mathcal{T}_{\overline{I}}}$ and
            $\boldsymbol{a}_{\mathit{\Pi}_{\overline{I}}})$ also
            overlap.}  ($a$) $\langle \Rey_\lambda\rangle_t=112$,
          ($b$) $\langle \Rey_\lambda\rangle_t=174$.}
	\label{fig:decomposedNSD}
\end{figure}

\begin{figure}
	\centerline{\includegraphics{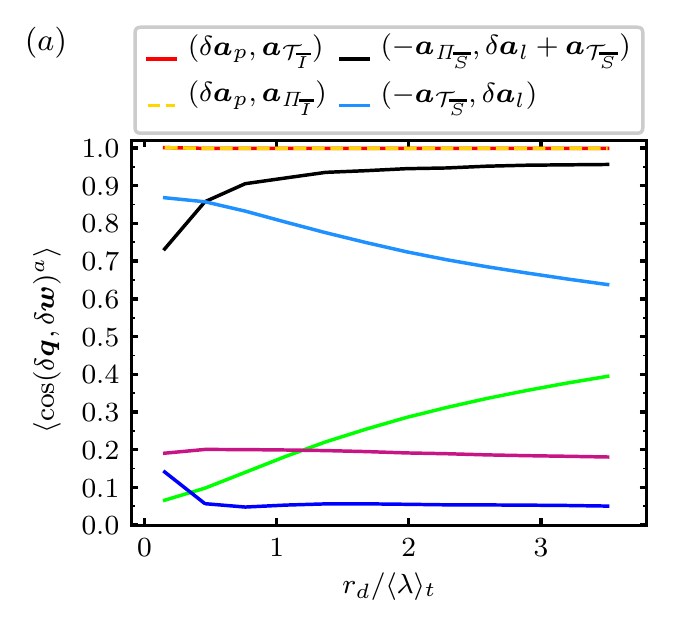}	\hfill \includegraphics{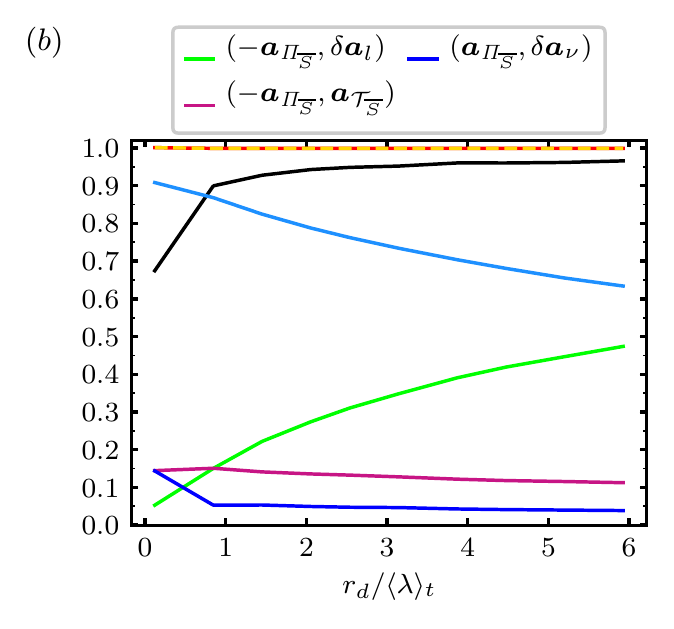}}%
	\caption{Average alignments of NSD terms $(\delta
          \boldsymbol{q},\delta \boldsymbol{w})$ listed on top of
                    {($a$) and $(b)$. The average
                       alignments of $(\delta
                       \boldsymbol{a}_p,\boldsymbol{a}_{\mathcal{T}_{\overline{I}}})$
                       and $(\delta
                       \boldsymbol{a}_p,\boldsymbol{a}_{\mathit{\Pi}_{\overline{I}}})$
                       overlap}: ($a$) $\langle
                     \Rey_\lambda\rangle_t=112$, ($b$) $\langle
                     \Rey_\lambda\rangle_t=174$.}
	\label{fig:nsdDecompAlign}
\end{figure}

Equation \eqref{NSDirrTr} can also be obtained by integrating the
Poisson equation for $\delta p$ in centroid space similarly to
equation \eqref{NSDsolTr} which, as already mentioned, can be obtained
by integrating the vorticity difference equation in that same
space. We use this approach in appendix \ref{sec:appC} to derive these
equations for periodic/homogeneous turbulence but also their
generalised form for non-homogeneous turbulence. By deriving the exact
equations for
$\boldsymbol{a}_{\mathcal{T}_{\overline{I}}}(\boldsymbol{x},\boldsymbol{r},t)$
and
$\boldsymbol{a}_{\mathit{\Pi}_{\overline{I}}}(\boldsymbol{x},\boldsymbol{r},t)$
in Fourier centroid space we show in appendix \ref{sec:appB} that we
have
$\boldsymbol{a}_{\mathcal{T}_{\overline{I}}}(\boldsymbol{x},\boldsymbol{r},t)=\boldsymbol{a}_{\mathit{\Pi}_{\overline{I}}}(\boldsymbol{x},\boldsymbol{r},t)$
in periodic/homogeneous turbulence. This result combined with
\eqref{NSDirrTr} yields
\begin{equation}
	\boldsymbol{a}_{\mathit{\Pi}_{\overline{I}}} = \boldsymbol{a}_{\mathcal{T}_{\overline{I}}}  = \frac{1}{2}\delta \boldsymbol{a}_p = \frac{1}{2}\delta \boldsymbol{a}_{c_I}  \label{transportPressure},
\end{equation}
in periodic/homogeneous turbulence. {In figure
  \ref{fig:decomposedNSD} we plot spatio-temporal averages of
  $\boldsymbol{r}$-orientation-averaged quantities \eqref{ORIENTAVG}
  for various acceleration/force terms in the NSD and the Helmholtz
  decomposed NSD equations and in figure \ref{fig:nsdDecompAlign} we
  plot spatio-temporal averages of
  $\boldsymbol{r}$-orientation-averaged cosines of angles between
  various two-point acceleration terms in these equations.} The
overlapping magnitudes in figure \ref{fig:decomposedNSD} and the
average alignments in figure \ref{fig:nsdDecompAlign} confirm
\eqref{transportPressure}, {or rather validate our DNS
  given that \eqref{transportPressure} is exact}.


{The computational procedure to calculate the various
  $\boldsymbol{r}$-orientation-averaged terms in these figures is
  computationally expensive. To calculate the NSD irrotational and
  solenoidal parts of the interscale and interspace transport terms at
  a given time $t$ and separation $\boldsymbol{r}$, we use the
  pseudo-spectral algorithm of \citet{Patterson1971} with one
  phase-shift and spherical truncation. We apply this algorithm to
  $\delta u_j$ and $\partial \delta u_i/\partial r_j$ for the
  interscale transfer and for $(u_j^{+}+u_j^{-})/2$ and $\partial
  \delta u_i/\partial x_j$ for the interspace transfer. Hence, we
  express these vectors/tensors in Fourier-space (see equations
  \eqref{FourTransport1}-\eqref{FourTransport4} in appendix
  \ref{sec:appB}) and apply the pseudo-spectral method of
  \citet{Patterson1971} to calculate
  $\widehat{\boldsymbol{a}_{\mathcal{T}}}(\boldsymbol{k},\boldsymbol{r},t)$
  and
  $\widehat{\boldsymbol{a}_{\mathit{\Pi}}}(\boldsymbol{k},\boldsymbol{r},t)$
  without aliasing errors. We next decompose these fields to
  irrotational and solenoidal fields with the projection operator and
  inverse these fields to physical space to obtain
  $\boldsymbol{a}_{\mathit{\Pi}_{\overline{S}}}(\boldsymbol{x},\boldsymbol{r},t)$,
  $\boldsymbol{a}_{\mathit{\Pi}_{\overline{I}}}(\boldsymbol{x},\boldsymbol{r},t)$,
  $\boldsymbol{a}_{\mathcal{T}_{\overline{S}}}(\boldsymbol{x},\boldsymbol{r},t)$
  and
  $\boldsymbol{a}_{\mathcal{T}_{\overline{I}}}(\boldsymbol{x},\boldsymbol{r},t)$. These
  fields can then be sampled over $\boldsymbol{x}$ to calculate
  e.g. $\boldsymbol{a}_{\mathit{\Pi}_{\overline{S}}}^2(\boldsymbol{x},\boldsymbol{r},t)$
  or KHMH terms such as $2\delta \boldsymbol{u} \cdot
  \boldsymbol{a}_{\mathit{\Pi}_{\overline{S}}}(\boldsymbol{x},\boldsymbol{r},t)$
  (see section \ref{subsec:khmh1}). If we assume that the cost of a
  DNS time-step is similar to the cost of the pseudo-spectral method
  to calculate the NS non-linear term, the calculation of solenoidal
  and irrotational interspace and interscale transfers for one $t$ and
  one $\boldsymbol{r}$ has similar cost to one DNS time-step. The
  total cost of the pseudo-spectral post-processing method is
  proportional to the total number $N_r$ of separation vectors
  $\boldsymbol{r}$ that we use in our spherical averaging across
  scales $r_d$ and to the total number $T_s/\Delta T$ of samples in
  time (see table \ref{tab:DNS}). With a total number of separation
  vectors $N_r \sim 10^{3}-10^4$ and our $T_s/\Delta T$ values, the
  total cost of the pseudo-spectral post-processing method in terms of
  DNS time-steps is at least one order of magnitude larger than the
  cost of the DNS itself. This high post-processing cost limits the
  $\langle\Rey_{\lambda}\rangle_t$ values of this study.}

The NSD solenoidal equation \eqref{NSDsolTr} describes a balance
between the time-derivative, solenoidal interscale transfer,
solenoidal interspace transport, viscous and forcing terms. From the
point we made in the sentence directly following equation
\eqref{NSDTransport}, we expect
$\langle\boldsymbol{a}_{\mathcal{T}_{\overline{S}}}^{2}\rangle$ and
$\langle \boldsymbol{a}_{\mathit{\Pi}_{\overline{S}}}^{2}\rangle$ to
tend to become equal to each other as the amplitude of
$\boldsymbol{r}$ tends to values significantly larger than $\langle L
\rangle_t$. Figure \ref{fig:decomposedNSD} confirms this trend for the
orientation-averaged {fluctuation} magnitudes of
$\boldsymbol{a}_{\mathcal{T}_{\overline{S}}}$ and
$\boldsymbol{a}_{\mathit{\Pi}_{\overline{S}}}$. With
decreasing $r_d$, {$\langle \boldsymbol{a}_{\mathit{\Pi}_{\overline{S}}}^2\rangle^a$} decreases relative to {$\langle \boldsymbol{a}_{\mathcal{T}_{\overline{S}}}^2 \rangle$}. At all scales
$r_{d}\geq \langle \lambda\rangle_t$ the {fluctuation magnitudes} of
$\boldsymbol{a}_{\mathcal{T}_{\overline{S}}}$ and
$\boldsymbol{a}_{\mathit{\Pi}_{\overline{S}}}$ are one order of
magnitude larger than those of the viscous term $\delta
\boldsymbol{a}_{\nu}$ and this separation is greater for the larger
$\langle \Rey_\lambda\rangle_t$. The {fluctuation magnitudes} of $\delta
\boldsymbol{a}_{\nu}$ are themselves much larger than those of $\delta
\boldsymbol{f}$ (not shown in figure \ref{fig:decomposedNSD} for not
overloading the figure but see figure \ref{fig:nsdFig1}($a1$)). These
observations suggest that the solenoidal NSD equation \eqref{NSDsolTr}
reduces to the approximate
\begin{equation}
	\delta \boldsymbol{a}_l + \boldsymbol{a}_{\mathcal{T}_{\overline{S}}} \approx -\boldsymbol{a}_{\mathit{\Pi}_{\overline{S}}} \label{vortNew} ,
\end{equation}
where this equation is understood as typical in terms of {fluctuation magnitudes}: i.e. in most regions of the flow for the majority of the
time, the removed terms are at least one order of magnitude smaller
than the retained terms. (As for the NS dynamics, we do expect
dynamically important regions localised in space and time where the
dynamics differ from \eqref{vortNew}.) Figure \ref{fig:decomposedNSD}
confirms equation \eqref{vortNew} in this sense and shows
that the {relatively rare spatio-temporal events} which
are neglected when writing equation \eqref{vortNew} are indeed present
as the fluctuation magnitudes do show a very small deviation from
equation \eqref{vortNew}. An additional important observation to be
made from figure \ref{fig:decomposedNSD} is that $\delta
\boldsymbol{a}_{c_S}$ tends to become increasingly dominated by
$\boldsymbol{a}_{\mathcal{T}_{\overline{S}}}$ rather than
$\boldsymbol{a}_{\mathit{\Pi}_{\overline{S}}}$ as $r_d$ decreases.

Equation \eqref{vortNew} is the same as equation \eqref{NSD51}, and
similarly to figure \ref{fig:nsdFig1} which provides support for
equation \eqref{NSD51}, figures \ref{fig:decomposedNSD} and
\ref{fig:nsdDecompAlign} provide strong support for equation
\eqref{vortNew}, in particular for $r_{d} > \langle \lambda
\rangle_{t}$. It is interesting to note that the average alignment
between the left and the right hand side of equation \eqref{vortNew}
lies between 90\% and 100\% (typically 95\%) for $r_{d} > \langle
\lambda \rangle_{t}$. Whilst this is strong support for approximate
equation \eqref{vortNew}, the fact that the alignment is not 100\% is
a reminder of the nature of the approximation, i.e. that
{relatively rare} spatio-temporal events do exist where
the viscous and/or driving forces are not negligible.

At length-scales $r_{d} \le \langle \lambda \rangle_{t}$, the
alignment between $\delta \boldsymbol{a}_l$ and
$-\boldsymbol{a}_{\mathcal{T}_{\overline{S}}}$ improves while the
alignment between $\delta \boldsymbol{a}_l +
\boldsymbol{a}_{\mathcal{T}_{\overline{S}}}$ and
$-\boldsymbol{a}_{\mathit{\Pi}_{\overline{S}}}$ worsens with
decreasing $r_d$ (see figure \ref{fig:nsdDecompAlign}) presumably
because of direct dissipation and diffusion effects, so that $\delta
\boldsymbol{a}_l + \boldsymbol{a}_{\mathcal{T}_{\overline{S}}} \approx
0$ becomes a better approximation than equation \eqref{vortNew} at
$r_d < 0.5 \langle \lambda \rangle_{t}$. This observation is
consistent with our parallel observation that the magnitude of
$\boldsymbol{a}_{\mathcal{T}_{\overline{S}}}$ increases while the
magnitude of $\boldsymbol{a}_{\mathit{\Pi}_{\overline{S}}}$ decreases
with decreasing $r_d$ and that $\delta \boldsymbol{a}_{c_S}$ in
equation \eqref{NSD51} tends to be dominated by
$\boldsymbol{a}_{\mathcal{T}_{\overline{S}}}$ at the very smallest
scales.

On the other end of the spectrum, i.e. as the length scale $r_d$ grows
towards $\langle L \rangle_{t}$, the alignment between $\delta
\boldsymbol{a}_l$ and $-\boldsymbol{a}_{\mathcal{T}_{\overline{S}}}$ worsens while the
alignment between $\delta \boldsymbol{a}_l$ and
$-\boldsymbol{a}_{\mathit{\Pi}_{\overline{S}}}$ improves (see figure
\ref{fig:nsdDecompAlign}), both reaching a comparable level of
alignment/misalignment which contribute together to keep approximation
\eqref{vortNew} statistically well satisfied with 95\% alignment
between $\delta \boldsymbol{a}_l + \boldsymbol{a}_{\mathcal{T}_{\overline{S}}}$ and
$-\boldsymbol{a}_{\mathit{\Pi}_{\overline{S}}}$.

The strong anti-alignment between
$\boldsymbol{a}_{\mathcal{T}_{\overline{S}}}$ and $\delta
\boldsymbol{a}_l$, increasingly so at smaller $r_d$ (see figure
\ref{fig:nsdDecompAlign}) expresses the sweeping of the two-point
momentum difference $\delta \boldsymbol{u}$ at scales $r_d$ and
smaller by the mainly large scale velocity
$(\boldsymbol{u}^{+}+\boldsymbol{u}^{-})/2$. Note that this two-point
sweeping differs from anti-alignment between $\delta \boldsymbol{a}_l$
and $\delta \boldsymbol{a}_c$ for two reasons. Firstly, by using the
Helmholtz decomposition we have removed the pressure effect embodied
in the $\boldsymbol{a}_{c_I}$ contribution to $\boldsymbol{a}_c$ which
balances the pressure-gradient.  This was first understood in
\citet{Tsinober2001} in a one-point setting and is here extended to a
two-point setting. Secondly, $\delta \boldsymbol{a}_{c_S}$ is the sum
of an interspace transport
$\boldsymbol{a}_{\mathcal{T}_{\overline{S}}}$ and an interscale
transfer term $\boldsymbol{a}_{\mathit{\Pi}_{\overline{S}}}$ such that
the interpretation of two-point sweeping as anti-alignment between
$\boldsymbol{a}_{c_S}$ and $\boldsymbol{a}_l$ as sweeping cannot be
exactly accurate. The advection of $\delta \boldsymbol{u}$ by the
large scale velocity is attributable to
$\boldsymbol{a}_{\mathcal{T}_{\overline{S}}}$, and figure
\ref{fig:nsdDecompAlign} shows that the two-point sweeping
anti-alignment between $\delta \boldsymbol{a}_l$ and
$\boldsymbol{a}_{\mathcal{T}_{\overline{S}}}$ increases with
decreasing $r_d$.

The sweeping anti-alignment between $\delta \boldsymbol{a}_l$ and
$\boldsymbol{a}_{\mathcal{T}_{\overline{S}}}$ is by no means perfect
even if it reaches about 90\% accuracy at $r_{d} < \langle
\lambda\rangle_{t}$, as is clear from the similar magnitudes and very
strong alignment tendency between $\delta \boldsymbol{a}_l +
\boldsymbol{a}_{\mathcal{T}_{\overline{S}}}$ and
$-\boldsymbol{a}_{\mathit{\Pi}_{\overline{S}}}$ at scales
$|\boldsymbol{r}|\geq \langle \lambda\rangle_t$ (see figures
\ref{fig:decomposedNSD} and \ref{fig:nsdDecompAlign}). Note, in
passing, that the Lagrangian solenoidal acceleration $\delta
\boldsymbol{a}_l+\boldsymbol{a}_{\mathcal{T}_{\overline{S}}}$ and
$\boldsymbol{a}_{\mathit{\Pi}_{\overline{S}}}$ are both Galilean
invariant. Equation \eqref{vortNew} may be interpreted to mean that
the Lagrangian solenoidal acceleration of $\delta \boldsymbol{u}$
(which is actually solenoidal) moving with the mainly large scale
velocity $(\boldsymbol{u}^{+}+\boldsymbol{u}^{-})/2$, namely $\delta
\boldsymbol{a}_l+\boldsymbol{a}_{\mathcal{T}_{\overline{S}}}$, is
evolving in time and space in response to
$-\boldsymbol{a}_{\mathit{\Pi}_{\overline{S}}}$: when there is an
influx of momentum from larger scales
there is an increase in $\delta
\boldsymbol{a}_l+\boldsymbol{a}_{\mathcal{T}_{\overline{S}}}$ and
$\delta \boldsymbol{u}$ and vice versa.
\subsection{From NSD dynamics to KHMH dynamics in homogeneous/periodic turbulence} \label{subsec:khmh1}

The scale-by-scale evolution of $|\delta \boldsymbol{u}|^2$ locally in
space and time is governed by a KHMH equation. This makes KHMH
equations crucial tools for examining the turbulent energy
cascade. The original KHMH equation and the new solenoidal and
irrotational KHMH equations that we derive below are simply
projections of the corresponding NSD equations onto $2 \delta
\boldsymbol{u}$. Hence, KHMH dynamics depend on NSD dynamics and the
various NSD terms’ alignment or non-alignment tendencies with $2
\delta \boldsymbol{u}$. {In this subsection we present
  five KHMH results all clearly demarcated and identified in {\it
    italics}.}

By contracting the NSD equation \eqref{NSD1} with $2\delta \boldsymbol{u}$,
one obtains the KHMH equation \citep{HILL2002,Yasuda2018}:
\begin{multline}
\frac{\partial}{\partial t}|\delta \boldsymbol{u}|^2+\frac{u_k^{+}+u_k^{-}}{2}\frac{\partial}{\partial x_k}|\delta \boldsymbol{u}|^2
+ \frac{\partial}{\partial r_k}\big(\delta u_k|\delta \boldsymbol{u}|^2\big) =
-\frac{2}{\rho}\frac{\partial}{\partial x_k}\big(\delta u_k \delta p \big)
+ 2 \nu \frac{\partial^2}{\partial r_k^2}|\delta \boldsymbol{u}|^2 \\
+ \frac{\nu}{2}\frac{\partial^2}{\partial x_k^2}|\delta \boldsymbol{u}|^2
-\bigg[2 \nu\big(\frac{\partial u_i^{+}}{\partial x_k^{+}} \big)^2+ 2 \nu \big(\frac{\partial u_i^{-}}{\partial x_k^{-}} \big)^2 \bigg]
+2\delta u_k \delta f_k,   \label{KHMH_DNS}
\end{multline}
where no fluid velocity decomposition nor averaging operations have
been used. In line with the naming convention of \citet{Yasuda2018}
this equation can be written
\begin{equation}
	\mathcal{A}_t+\mathcal{T}+\mathit{\Pi} = \mathcal{T}_p+\mathcal{D}_{r,\nu}+\mathcal{D}_{x,\nu}-\mathcal{\epsilon}+\mathcal{I},  \label{khmhSymb}
\end{equation}
where the first, second and third terms on the left hand sides of
equations \eqref{KHMH_DNS} and \eqref{khmhSymb} correspond to each
other and so do the first, second, third, fourth and fifth terms on
the right hand sides. Preempting notation used further down in this
paper, equation \eqref{khmhSymb} is also written $\mathcal{A} =
\mathcal{T}_p + \mathcal{D} + \mathcal{I}$ or $\mathcal{A}_t +
\mathcal{A}_c = \mathcal{T}_p + \mathcal{D} + \mathcal{I}$ where
$\mathcal{A}_c \equiv \mathcal{T}+\mathit{\Pi}$, $\mathcal{A}\equiv
\mathcal{A}_t+\mathcal{A}_c$ and $\mathcal{D}\equiv
\mathcal{D}_{r,\nu}+\mathcal{D}_{x, \nu}-\mathcal{\epsilon}$.

To examine the KHMH dynamics in terms of irrotational and solenoidal
dynamics we contract the irrotational and solenoidal NSD equations
with $2 \delta \boldsymbol{u}$ to derive what we refer to as
irrotational and solenoidal KHMH equations. Each of the KHMH terms can
be subdivided into a contribution from the NSD irrotational part and a
contribution from the NSD solenoidal part of the respective term in
the NSD equation. A solenoidal KHMH term corresponding to a $\delta
\boldsymbol{q} (\boldsymbol{x},\boldsymbol{r},t)$ or $\boldsymbol{q}
(\boldsymbol{x},\boldsymbol{r},t)$ term in equation \eqref{NSDsolTr}
equals $\mathcal{Q}_{\overline{S}}=2\delta \boldsymbol{u} \cdot \delta
\boldsymbol{q}_{\overline{S}}$ or $\mathcal{Q}_{\overline{S}}=2\delta
\boldsymbol{u} \cdot \boldsymbol{q}_{\overline{S}}$, and an
irrotational KHMH term corresponding to a $\delta \boldsymbol{q}
(\boldsymbol{x},\boldsymbol{r},t)$ or $\boldsymbol{q}
(\boldsymbol{x},\boldsymbol{r},t)$ term in equation
\eqref{transportPressure} equals $\mathcal{Q}_{\overline{I}}=2\delta
\boldsymbol{u} \cdot \delta \boldsymbol{q}_{\overline{I}}$ or
$\mathcal{Q}_{\overline{I}}=2\delta \boldsymbol{u} \cdot
\boldsymbol{q}_{\overline{I}}$. With $\mathcal{Q} = 2\delta
\boldsymbol{u} \cdot \delta \boldsymbol{q}$ or $\mathcal{Q} = 2\delta
\boldsymbol{u} \cdot \boldsymbol{q}$, we have $\mathcal{Q}=
\mathcal{Q}_{\overline{I}} + \mathcal{Q}_{\overline{S}}$. The
irrotational and solenoidal KHMH equations for periodic/homogeneous
turbulence follow from equations \eqref{NSDsolTr} and
\eqref{transportPressure} respectively and read
\begin{align}
	\mathcal{A}_t  +\mathcal{T}_{\overline{S}} + \mathit{\Pi}_{\overline{S}} &= \mathcal{D}_{r,\nu} + \mathcal{D}_{x,\nu} - \mathcal{\epsilon} + \mathcal{I}  , \label{solKHMH2} \\
	\mathit{\Pi}_{\overline{I}} = \mathcal{T}_{\overline{I}} &= \frac{1}{2}\mathcal{T}_p,	 \label{irKHMH2}
\end{align}
where use has been made of the fact that the velocity and velocity
difference fields are solenoidal. {\it These two equations are our
  first KHMH result.}

Space-local changes in time of $|\delta \boldsymbol{u}|^2$, expressed
via $\mathcal{A}_t$, are only due to solenoidal KHMH dynamics in
equation \eqref{solKHMH2} which include interspace transport,
interscale transport, viscous and forcing effects. The irrotational
KHMH equation \eqref{irKHMH2} formulates how the imposition of
incompressibility by the pressure field affects interspace and
interscale dynamics and, in turn, energy cascade dynamics. Generalised
solenoidal and irrotational KHMH equations also valid for
non-periodic/non-homogeneous turbulence are given in appendix
\ref{sec:appC}.

We first consider the spatio-temporal average of these equations in
statistically steady forced periodic/homogeneous turbulence. As
$\langle \mathcal{T}_p \rangle=0$, we obtain from equation
\eqref{irKHMH2}, $\langle \mathit{\Pi}_{\overline{I}} \rangle= \langle
\mathcal{T}_{\overline{I}} \rangle=0$. As $\langle
\mathcal{T}_{\overline{S}}\rangle+\langle
\mathcal{T}_{\overline{I}}\rangle=\langle \mathcal{T}\rangle=0$, we
have $\langle \mathcal{T}_{\overline{S}}\rangle=0$, such that the
spatio-temporal average of \eqref{solKHMH2} reads
\begin{equation}
	\langle \mathit{\Pi}\rangle =  \langle \mathit{\Pi}_{\overline{S}}\rangle = \langle \mathcal{D}_{r,\nu} \rangle - \langle \mathcal{\epsilon} \rangle + \langle \mathcal{I} \rangle . \label{khmhAvg}
\end{equation}

\noindent If an intermediate inertial subrange of scales $\vert
\boldsymbol{r}\vert$ can be defined where viscous diffusion and
forcing are negligible, equation \eqref{khmhAvg} reduces to $\langle
\mathit{\Pi}_{\overline{S}}\rangle \approx - \langle
\mathcal{\epsilon} \rangle$ in that range. This
{theoretical} conclusion {(which is not
  part of our DNS study)} is the backbone of the
\citet{Kolmogorov1941c,Kolmogorov1941b,Kolmogorov1941a} theory for
high Reynolds number statistically homogeneous stationary small-scale
turbulence with the additional information that {\it the part of the
  average interscale transfer rate involved in Kolmogorov's
  equilibrium balance is the solenoidal interscale transfer rate
  only. This is our second KHMH result}. On average, there is a
cascade of turbulence energy from large to small scales where the rate
of interscale transfer is dominated by two-point vortex stretching
(see appendix \ref{sec:appC} for the relation between the solenoidal interscale
transfer and vortex stretching) and is equal to $- \langle
\mathcal{\epsilon} \rangle$ independently of $\vert
\boldsymbol{r}\vert$ over a range of scales where viscous diffusion
and forcing are negligible.

In this paper we
concentrate on the fluctuations around {the average
  picture described by the scale-by-scale equilibrium \eqref{khmhAvg}
  for any Reynolds number}. If we subtract the spatio-temporal average
solenoidal KHMH equation \eqref{khmhAvg} from the solenoidal KHMH
equation \eqref{solKHMH2} and use the generic notation
$\mathcal{Q}^{'}\equiv\mathcal{Q}-\langle \mathcal{Q}\rangle$, we
attain the fluctuating solenoidal KHMH equation
 \begin{equation}
 	\mathcal{A}_t +\mathcal{T}_{\overline{S}} +
        \mathit{\Pi}_{\overline{S}}^{'} = \mathcal{D}_{r,\nu}^{'} +
        \mathcal{D}_{x,\nu} - \mathcal{\epsilon}^{'} +
        \mathcal{I}^{'}. \label{khmhSolFluct}
 \end{equation}
This equation governs the fluctuations of the KHMH solenoidal dynamics
around its spatio-temporal average. Clearly, if these non-equilibrium
fluctuations are large relative to their average values, {the average picture expressed by equation
  \eqref{khmhAvg}} is not characteristic of the
{interscale transfer} dynamics. We now study the KHMH
fluctuations in statistically stationary periodic/homogeneous
turbulence on the basis of equations \eqref{irKHMH2} and
\eqref{khmhSolFluct}. {Concerning equation
  \eqref{irKHMH2}, note that $\mathit{\Pi}_{\overline{I}}^{'}=
  \mathit{\Pi}_{\overline{I}}$, $\mathcal{T}_{\overline{I}}^{'}=
  \mathcal{T}_{\overline{I}}$ and $\mathcal{T}_{p}^{'} =
  \mathcal{T}_{\overline{p}}$.}\par

\begin{figure}
	\centerline{\includegraphics{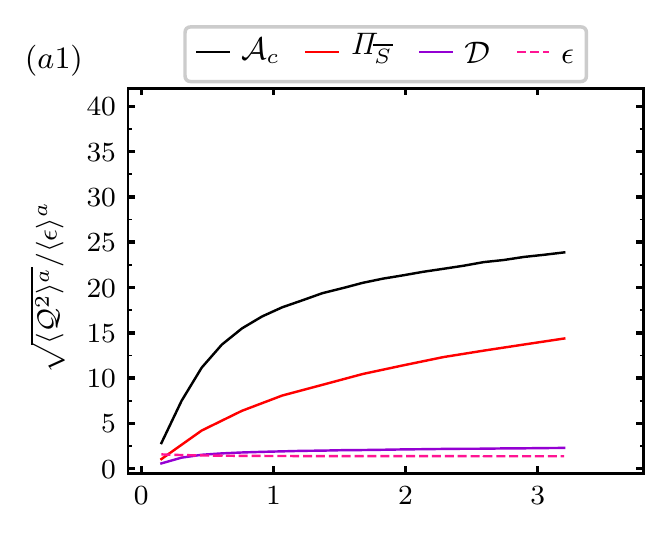} \hfill
          \includegraphics{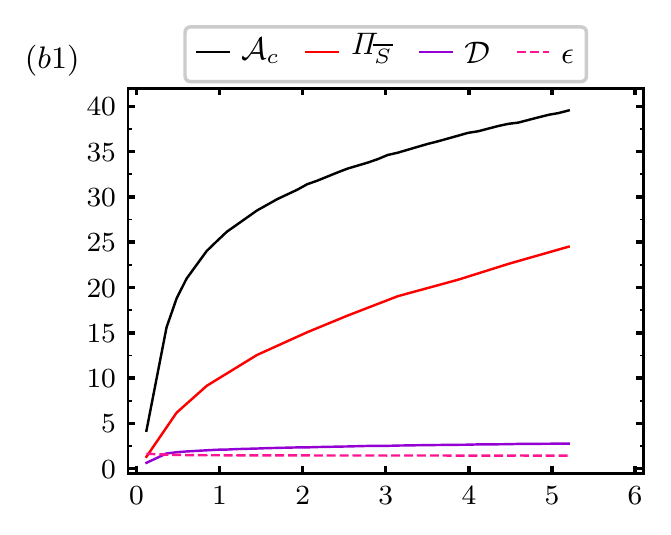}}
        \centerline{\includegraphics{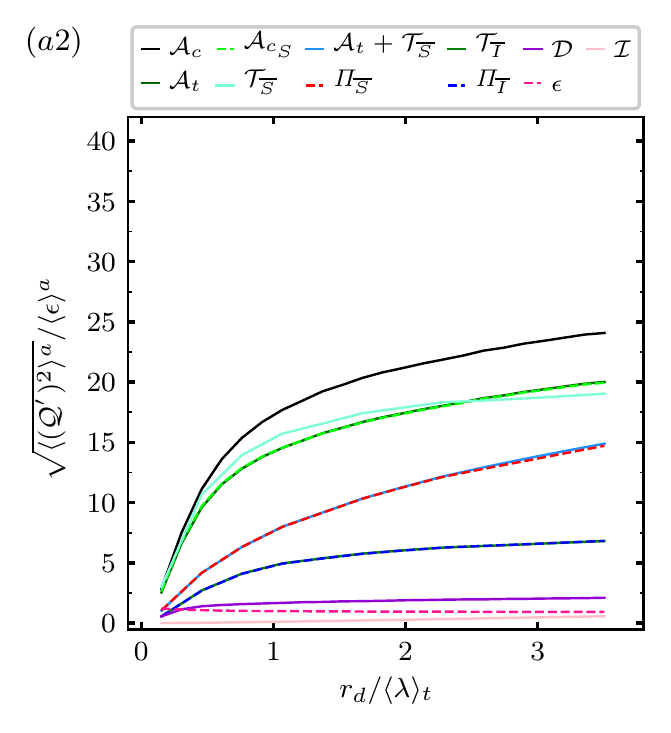} \hfill
          \includegraphics{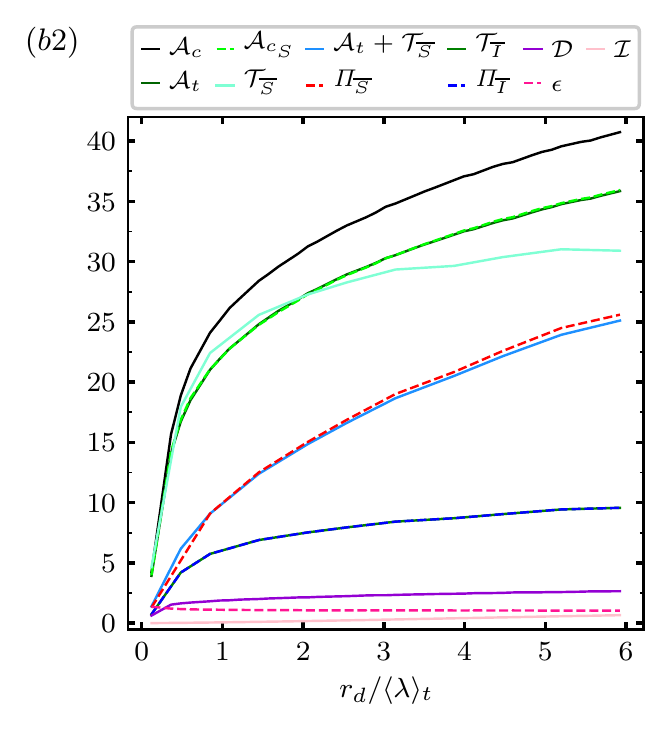}}
	\caption{($a1$, $b1$) KHMH average square magnitudes $\langle
          \mathcal{Q}^2\rangle^a$ and ($a2$, $b2$) KHMH average square
          fluctuating magnitudes $\langle
          (\mathcal{Q}^{'})^2\rangle^a$, where
          $\mathcal{Q}^{'}=\mathcal{Q}-\langle \mathcal{Q}\rangle$,
          for the KHMH terms $\mathcal{Q}$ listed above the figures
          and introduced in the third and fourth paragraph of
          \ref{subsec:khmh1}. All entries are normalised with $\langle
          \mathcal{\epsilon} \rangle^a$ (see equations
          \eqref{KHMH_DNS}-\eqref{khmhSymb}). {The
            following pairs of KHMH terms have overlapping magnitudes
            in $(a2,b2)$: $\mathcal{A}_t$ and $\mathcal{A}_{c_S}$;
            $\mathcal{A}_t+\mathcal{T}_{\overline{S}}$ and
            $\mathit{\Pi}_{\overline{S}}$;
            $\mathcal{T}_{\overline{I}}$ and
            $\mathit{\Pi}_{\overline{I}}$.}  ($a1$, $a2$) $\langle
          \Rey_\lambda\rangle_t = 112$, ($b1$,$b2$) $\langle
          \Rey_\lambda\rangle_t=174$.}
	\label{fig:khmhFluctMag}
\end{figure}

We start by determining the relative fluctuation magnitudes of the
spatio-temporal fluctuations of each term in the KHMH equations
\eqref{irKHMH2} and \eqref{khmhSolFluct}. These relative fluctuation
magnitudes can emulate those of respective terms in the NSD equations
under the following sufficient conditions: (i) the fluctuations are so
intense that they dwarf averages, so that $
\langle(\mathcal{Q}')^2\rangle \approx\langle \mathcal{Q}^2\rangle$;
(ii) the mean square of any KHMH term $\mathcal{Q} = 2\delta
\boldsymbol{u} \cdot \delta \boldsymbol{q}$ corresponding to a NSD
term $\delta \boldsymbol{q} (\boldsymbol{x},\boldsymbol{r},t)$
(equivalently $\mathcal{Q} = 2\delta \boldsymbol{u} \cdot
\boldsymbol{q}$ corresponding to
$\boldsymbol{q}(\boldsymbol{x},\boldsymbol{r},t)$)
{can be approximated as}
\begin{equation}
{\langle \mathcal{Q}^2 \rangle(\boldsymbol{r}) \approx
  4\langle |\delta \boldsymbol{u}|^2 \rangle \langle | \delta
  \boldsymbol{q}|^2 \rangle \langle \text{cos}^2(\theta_{q}) \rangle,}
          \label{Qcos}
\end{equation}
where the approximate equality results from a degree of decorrelation
and $\theta_q$ is the angle between $\delta
\boldsymbol{q}(\boldsymbol{x},\boldsymbol{r},t)$ (or
$\boldsymbol{q}(\boldsymbol{x},\boldsymbol{r},t)$) and $\delta
\boldsymbol{u}(\boldsymbol{x},\boldsymbol{r},t)$; (iii) $\langle
\text{cos}^2(\theta_{q}) \rangle$ is not very sensitive to the choice
of NSD term $\delta \boldsymbol{q}$ (or $\boldsymbol{q}$). Under these
conditions, we get
\begin{equation}
{\frac{\langle (2 \delta \boldsymbol{u} \cdot \delta
          \boldsymbol{q})^2\rangle (\boldsymbol{r})}{\langle (2\delta
          \boldsymbol{u} \cdot \delta \boldsymbol{w})^2 \rangle
          (\boldsymbol{r})} \approx \frac{\langle |\delta
          \boldsymbol{u}|^2\rangle \langle | \delta
          \boldsymbol{q}|^2\rangle \langle \text{cos}^2(\theta_q)
          \rangle (\boldsymbol{r})}{\langle|\delta
          \boldsymbol{u}|^2\rangle \langle |\delta \boldsymbol{w}|^2
          \rangle \langle \text{cos}^2(\theta_w)\rangle
          (\boldsymbol{r})} \approx \frac{\langle|\delta
          \boldsymbol{q}|^2\rangle (\boldsymbol{r})}{\langle|\delta
          \boldsymbol{w}|^2\rangle (\boldsymbol{r})}} , \label{magKH}
\end{equation}
which means that KHMH relative fluctuation magnitudes and NSD relative
fluctuation magnitudes are approximately
identical. {The first approximate equality in
  \eqref{magKH} follows directly from \eqref{Qcos} and the
  second approximate equality follows from hypothesis (iii) that
  $\text{cos}^2(\theta_q)$ and $\text{cos}^2(\theta_w)$ are about
  equal.}

{We test hypothesis (i) by comparing the plots in
  figure \ref{fig:khmhFluctMag}($a1$, $b1$) with those in figure
  \ref{fig:khmhFluctMag}($a2$, $b2$).} Figure
\ref{fig:khmhFluctMag}($a1$, $b1$) shows average magnitudes of KHMH
spatio-temporal fluctuations for terms with non-zero spatio-temporal
averages. Comparing with figure \ref{fig:khmhFluctMag}($a2$, $b2$), we
find $\langle(\mathcal{Q}')^2\rangle^a\approx\langle
\mathcal{Q}^2\rangle^a$, {i.e. hypothesis (i)}, for all
four terms plotted in figure \ref{fig:khmhFluctMag}($a1$, $b1$) at all
length scales $r_d$ considered. Note that this does not hold for
$\mathcal{D}_{r,\nu}^{'}$ and $\mathcal{I}^{'}$ which are the only
KHMH fluctuations such that $\sqrt{\langle
  (\mathcal{Q}')^2\rangle^a}/\langle\mathcal{\epsilon}\rangle^a$ is
smaller (in fact significantly smaller) than $1$ at all scales.
Figure \ref{fig:khmhFluctMag} makes it also clear that the magnitudes
of the fluctuations of all other KHMH terms (solenoidal and
irrotational) are much higher than those of the turbulence dissipation
at all scales $r_d > 0.5 \langle \lambda \rangle_{t}$, and more so for
the higher of the two Reynolds numbers. For scales $r_d \geq \langle
\lambda \rangle_t$, the largest average fluctuating magnitudes are
those of $\mathcal{A}_c^{'}$, followed closely by $\mathcal{A}_t$ and
$\mathcal{T}_{\overline{S}}$. Next come the magnitudes of
$\mathit{\Pi}_{\overline{S}}^{'}$ and
$\mathcal{A}_t+\mathcal{T}_{\overline{S}}$. Thereafter follow the
irrotational terms
$\mathit{\Pi}_{\overline{I}}=\mathcal{T}_{\overline{I}}$ $(=0.5
\mathcal{T}_{p})$ and finally the viscous, dissipative and forcing
terms $\mathcal{D}^{'}, \mathcal{\epsilon}^{'}$ and $\mathcal{I}^{'}$
in that order. {\it This order of fluctuations is our third KHMH
  result.}
An average description of the {interscale turbulent
  energy transfer} dynamics in terms of its spatio-temporal average
{cannot, therefore, be accurate.}
In order to characterise these dynamics, attention must be directed at
{most if not all KHMH term} fluctuations, and in fact
to much more than just the turbulence dissipation fluctuations given
that they are among the weakest.\par

\begin{figure}
	\centerline{\includegraphics{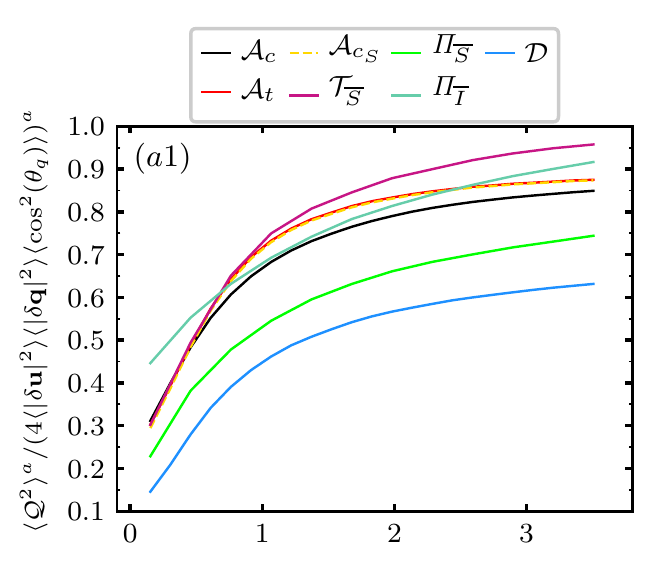} \hfill  \includegraphics{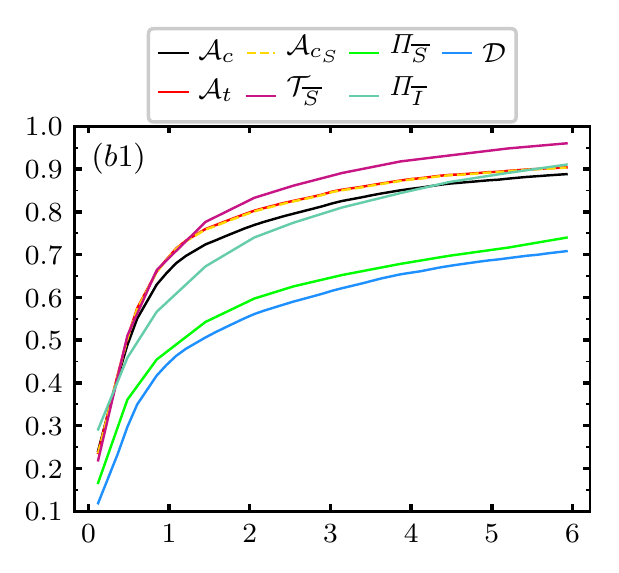}}
	\centerline{\includegraphics{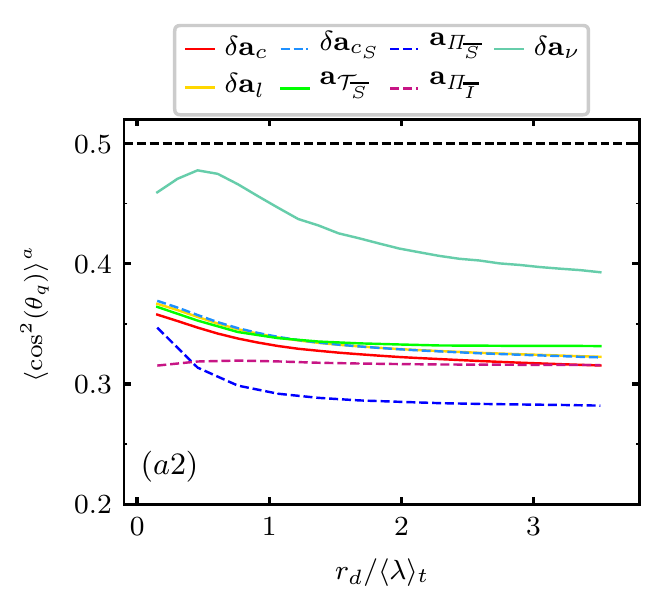} \hfill  \includegraphics{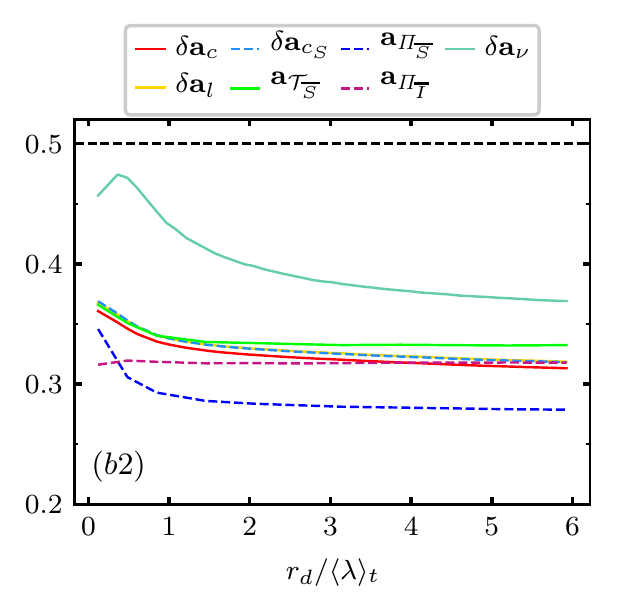}}
	\caption{Test of the assumptions (ii) and (iii) in the seventh
          paragraph of subsection \ref{subsec:khmh1} related to
          relations \eqref{Qcos}-\eqref{magKH} connecting NSD
          and KHMH relative magnitudes. ($a1$,$b1$) Test of assumption
          (ii) by taking the ratio of the left-hand and right-hand
          sides of \eqref{Qcos} for the KHMH terms
          $\mathcal{Q}$ listed above the figures.
          ($a2$,$b2$) test of assumption (iii) used in \eqref{magKH}
          by comparing the behaviour of $\langle \cos^2
          (\theta_q)\rangle^a$ for the various NSD terms listed above
          the figures. The black horizontal line $0.5$ corresponds to the
          value of $\langle \cos^2 (\theta_q)\rangle$ if $\theta_q$ is
          uniformly distributed. ($a1,a2$) $\langle
          \Rey_\lambda\rangle_t = 112$, ($b1,b2$) $\langle
          \Rey_\lambda\rangle_t=174$.}
	\label{fig:magCompCheck}
\end{figure}

Next, {we test hypothesis (ii) by testing} the validity
of \eqref{Qcos} and {hypothesis (iii) concerning}
approximately similar $\text{cos}^{2}(\theta_{q})$ behaviour for
different KHMH terms. {In figure
  \ref{fig:magCompCheck}($a1$, $b1$) we plot ratios of right hand
  sides to left hand sides of equation \eqref{Qcos} and see} that
\eqref{Qcos} is not valid, but that it is nevertheless about 65\% to
98\% accurate for $r_{d} \ge \langle \lambda \rangle_t$. Note that
\eqref{Qcos} might be sufficient but that it is by no means necessary
for the left-most and the right-most sides of \eqref{magKH} to
approximately balance. {In those cases where the
  variations between the ratios plotted in figure
  \ref{fig:magCompCheck}($a1$, $b1$) are not too large and the
  assumption of approximately similar $\text{cos}^{2}(\theta_{q})$ for
  different KHMH terms more or less holds, the left-most and the
  right-most sides of \eqref{magKH} can approximately balance}.

\begin{figure}
	\centerline{\includegraphics{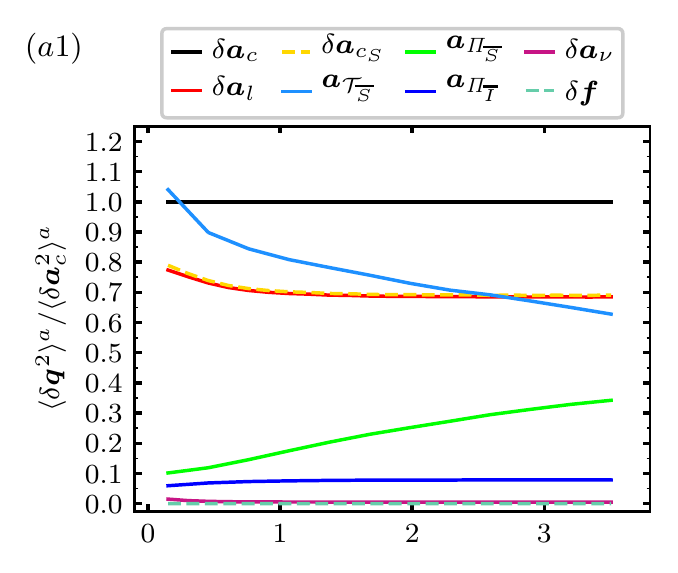} \hfill \includegraphics{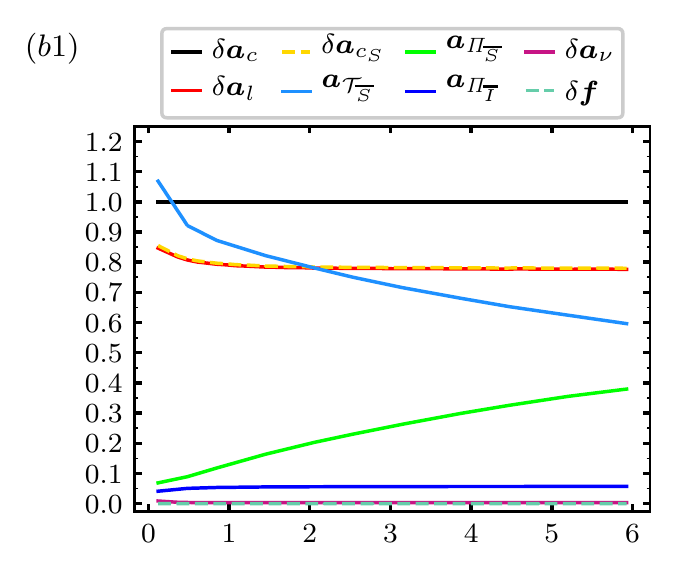}}
	\centerline{\includegraphics{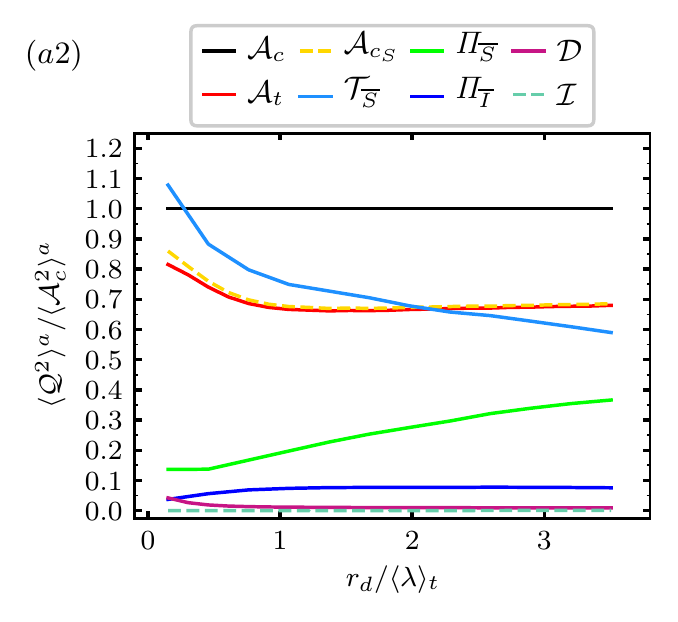} \hfill \includegraphics{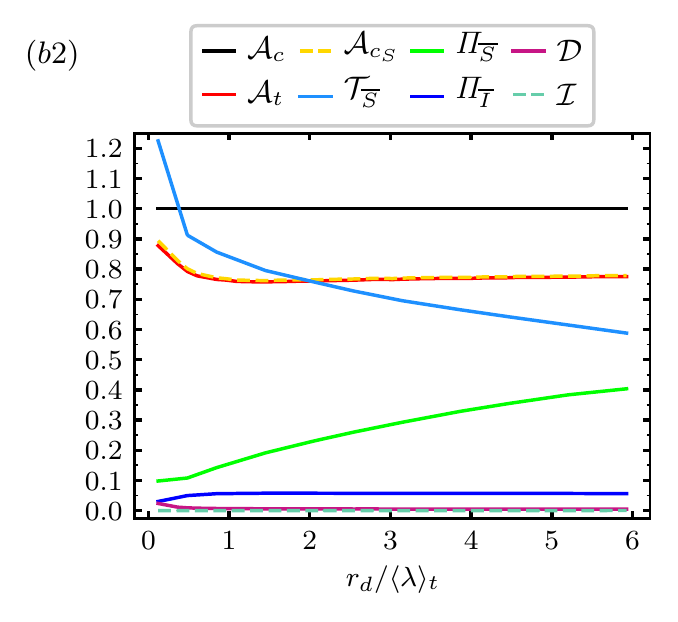}}
	\caption{NSD and KHMH relative average square magnitudes
          (which should be similar on the basis of \eqref{magKH}) for
          the terms listed above the figures: ($a1$) NSD and ($a2$)
          KHMH for $\langle \Rey_\lambda\rangle_t = 112$, ($b1$) NSD
          and ($b2$) KHMH for $\langle \Rey_\lambda\rangle_t=174$.}
	\label{fig:magComp}
\end{figure}

{Incidentally, figure \ref{fig:magCompCheck}($a2$,
  $b2$) also} shows that the angles $\theta_q$ are not random but that
they are more likely to be small rather than large in an approximately
similar way for all important NSD terms: $\text{cos}^2(\theta_{q})$
ranges between about 0.28 and 0.36 for all NSD terms (except the
viscous acceleration difference and the viscous force difference) at
all scales $r_d$. These values are much smaller than 0.5, the value
that $\text{cos}^2(\theta_{q})$ would have taken if the angles
$\theta_q$ were random. There is therefore an alignment tendency
between $\delta \boldsymbol{u}$ and NSD terms which is similar for all
the important NSD terms, thereby allowing the balance between the
left-most {(ratio of KHMH terms) and the right-most
  (ratio of NSD terms)} sides of \eqref{magKH} to approximately hold
as {seen by comparing the plots ($a1$)-($b1$) (mean square
  NSD terms) with the plots ($a2$)-($b2$) (mean square KHMH terms)} in
figure \ref{fig:magComp}.  (Note that the viscous term is bounded from
above, $\langle \mathcal{D}^2\rangle(\boldsymbol{r}) \leq 4 \langle
|\delta \boldsymbol{u}|^2 |\delta \boldsymbol{a}_{\nu}|^2 \rangle$,
which indicates limited magnitudes compared to the irrotational and
the dominant solenoidal terms because of the limited magnitude of
$\langle \delta \boldsymbol{a}_{\nu}^2 \rangle$. The limited
fluctuations of the viscous terms are clearly seen in figure
\ref{fig:khmhFluctMag}.)

Figure \ref{fig:magComp} does indeed confirm the close correspondence
between NSD and KHMH statistics which is a significant step further
from the correspondence reported earlier in this paper between NS and
NSD statistics. We can therefore use the approximate NSD relation
(\ref{vortNew}) to deduce the following approximate KHMH relation:
\begin{align}
	\mathcal{A}_t + \mathcal{T}_{\overline{S}} +
        \mathit{\Pi}_{\overline{S}}^{'} & \approx 0 , \label{solKHMH3}
\end{align}
understood in the sense that it holds in the majority of the domain
for the majority of the time but that there surely exist relatively
rare events within the flow where this approximate KHMH relation is violated.\par

{\it This approximate equation $\mathcal{A}_t +
  \mathcal{T}_{\overline{S}} + \mathit{\Pi}_{\overline{S}}^{'} \approx
  0$ can be considered to be our fourth KHMH result.} It is consistent
with the order of fluctuation magnitudes in figure \ref{fig:magComp}
which shows, in agreement with the NSD - KHMH correspondence just
established, that the largest fluctuating magnitudes are those of
$\mathcal{A}_c$, followed by the fluctuating magnitudes of
$\mathcal{T}_{\overline{S}}$, $\mathcal{A}_t$ and $\mathcal{A}_{c_S}$
($\mathcal{A}_{c_S} = \mathcal{T}_{\overline{S}} +
\mathit{\Pi}_{\overline{S}}$). Note though that there is a cross over
at about $r_{d} \approx 2 \langle \lambda \rangle_{t}$ for both
Reynolds numbers considered here between the fluctuation magnitudes of
$\mathcal{T}_{\overline{S}}$ and those of $\mathcal{A}_t$ and
$\mathcal{A}_{c_S}$ which are about equal to each other in agreement
with equation \eqref{solKHMH3}.

The fluctuation magnitudes of $\mathit{\Pi}_{\overline{S}}$ and
$\mathit{\Pi}_{\overline{I}}$ are both smaller than those just
mentioned, and those of $\mathit{\Pi}_{\overline{I}}$ are
significantly smaller than those of
$\mathit{\Pi}_{\overline{S}}$. Even smaller, are the fluctuation
magnitudes of $\mathcal{D}$ and $\mathcal{I}$, in that order. In
agreement with (\ref{magNSDD}), our third and fourth KHMH conclusions
incorporate the following:
\begin{equation}
\langle \mathcal{A}_{t}^{2} \rangle \approx \langle
\mathcal{A}_{c_S}^{2} \rangle \gg \langle \mathcal{T}_{p}^{2}\rangle =
4 \langle \mathit{\Pi}_{\overline{I}}^{2}\rangle = 4 \langle
\mathcal{T}_{\overline{I}}^{2}\rangle = \langle
\mathcal{A}_{c_I}^{2}\rangle \gg \langle \mathcal{D}^{2}\rangle \gg
\langle \mathcal{I}^{2}\rangle, \label{magIneqKHMH}
\end{equation}
where $\mathcal{A}_{c_I} = \mathcal{T}_{\overline{I}} +
\mathit{\Pi}_{\overline{I}}$.

An additional significant observation from figure \ref{fig:magComp} which
we can count as our {\it fifth KHMH result} is that, as $r_{d}$
decreases towards about $0.5 \langle \lambda \rangle_{t}$, the
fluctuation magnitude of $\mathcal{A}_{c_S} =
\mathcal{T}_{\overline{S}} + \mathit{\Pi}_{\overline{S}}$ remains
about constant but that of $\mathcal{T}_{\overline{S}}$ increases
while that of $\mathit{\Pi}_{\overline{S}}$ decreases. (At scale $r_d$
smaller than $0.5 \langle \lambda \rangle_{t}$, the fluctuation
magnitudes of both $\mathcal{A}_{c_S}$ and $\mathcal{T}_{\overline{S}}$
increase with diminishing $r_d$ whereas those of
$\mathit{\Pi}_{\overline{S}}$ remain about constant.) The convective
non-linearity is increasingly of the spatial transport type and
diminishingly of the interscale transfer type as the two-point
separation length decreases.

We now consider correlations between different intermediate and large
scale fluctuating KHMH terms in light of equations \eqref{irKHMH2} and
\eqref{solKHMH3}.
\section{Fluctuating KHMH dynamics in homogeneous/periodic turbulence}\label{sec:khmh2}

\subsection{Correlations} \label{subsec:corr}

\begin{figure}
	\centerline{\includegraphics{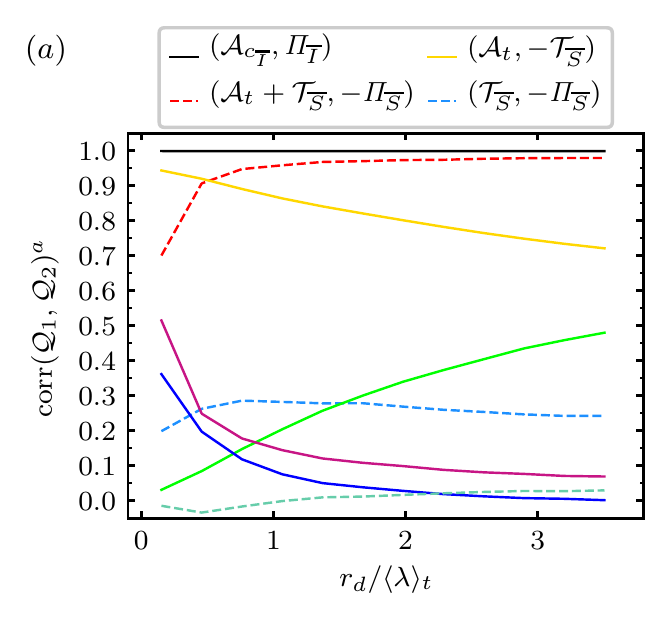} \hfill
          \includegraphics{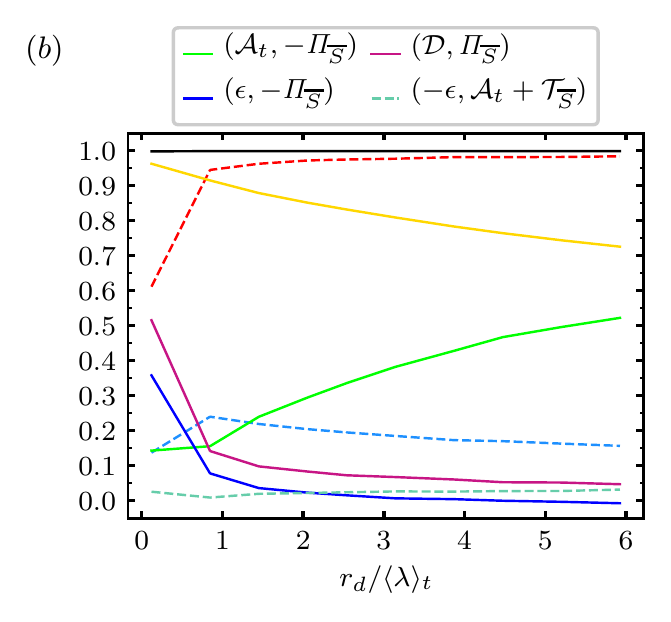}}
	\caption{Spherically averaged correlation coefficients between
          KHMH terms $(\mathcal{Q}_1, \mathcal{Q}_2)$ listed above
          {the plots $(a)$ and $(b)$. They are plotted as
            functions of scale $r_d$}. ($a$) $\langle
          \Rey_\lambda\rangle_t = 112$, ($b$) $\langle
          \Rey_\lambda\rangle_t = 174$.}
	\label{fig:khmhCorr}
\end{figure}

\begin{figure}
	\centerline{\includegraphics{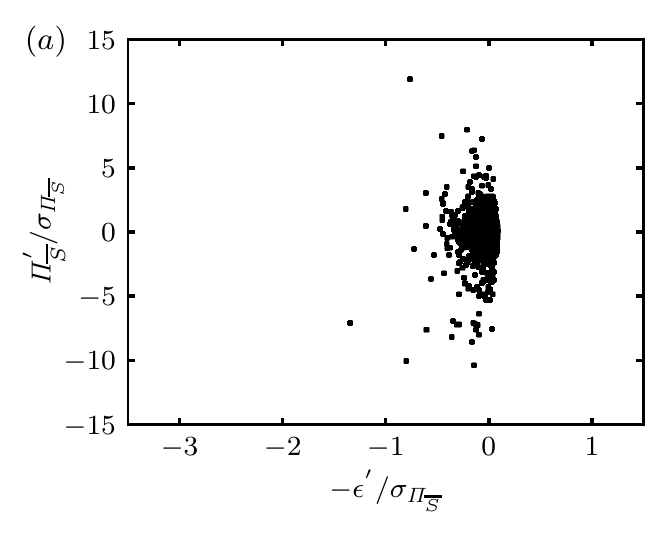} \hfill \includegraphics{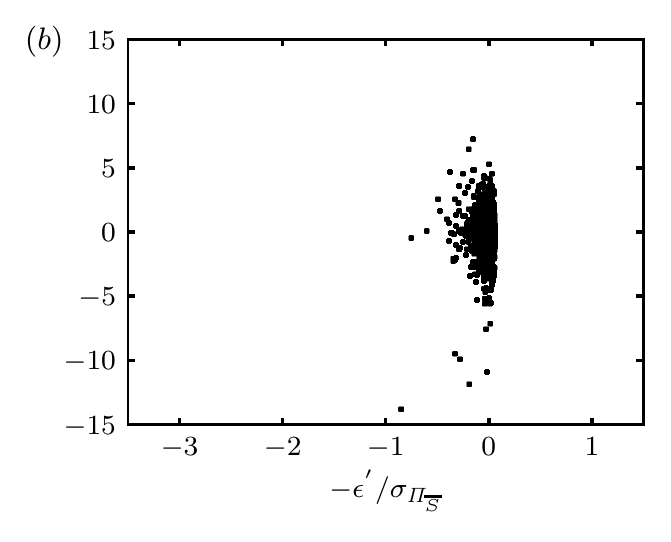}}
	\caption{Scatter plots of $\mathit{\Pi}_{\overline{S}}^{'}$ and $\-\mathcal{\epsilon}^{'}$ at random orientations $\boldsymbol{r}$ with $r_d/\langle \lambda\rangle_t=(1.45,3.1)$ for $(a,b)$, $\sigma_{\mathit{\Pi}_{\overline{S}}}$ is the standard deviation of $\mathit{\Pi}_{\overline{S}}$ and $\langle \Rey_\lambda\rangle_t=174$. }
	\label{fig:PisEpsScatter}
\end{figure}

We start this section by assessing the existence or non-existence of
local (in space and time) equilibrium between interscale transfer and
dissipation at some intermediate scales. {In figure
  \ref{fig:khmhCorr} we plot correlations between various KHMH
  terms. In particular, this figure} shows that the correlation
coefficient between $\mathit{\Pi}_{\overline{S}}^{'}$ and
$-\mathcal{\epsilon}^{'}$ lies well below $0.1$ for all scales $r_d
\geq \langle \lambda \rangle_t$. The scatter plots of these quantities
in figure \ref{fig:PisEpsScatter} confirm the absence of local
relation between interscale transfer rate and dissipation rate. For
example, for a given local/instantaneous dissipation fluctuation, the
corresponding local/instantaneous interscale transfer rate fluctuation
can be {close to} equally positive or negative. There
is no local equilibrium between these quantities as they fluctuate at
scales $r_d \geq \langle \lambda \rangle_t$. Such a correlation should
of course not necessarily be expected.
However, as $r_d$ decreases below $\langle \lambda\rangle_t$, the
correlations between $\mathit{\Pi}_{\overline{S}}^{'}$ and either
$-\mathcal{\epsilon}^{'}$ or $\mathcal{D}^{'}$ increase up to values
between about $0.3$ and about $0.5$. {This increased
  correlation may suggest} a feeble tendency towards
local/instantaneous equilibrium between interscale transfer rate and
dissipation rate at scales $r_d < \langle\lambda \rangle_t$. However,
these scales are strongly affected by direct viscous processes and can
therefore not be inertial range scales. \par

\begin{figure}
	\centerline{\includegraphics{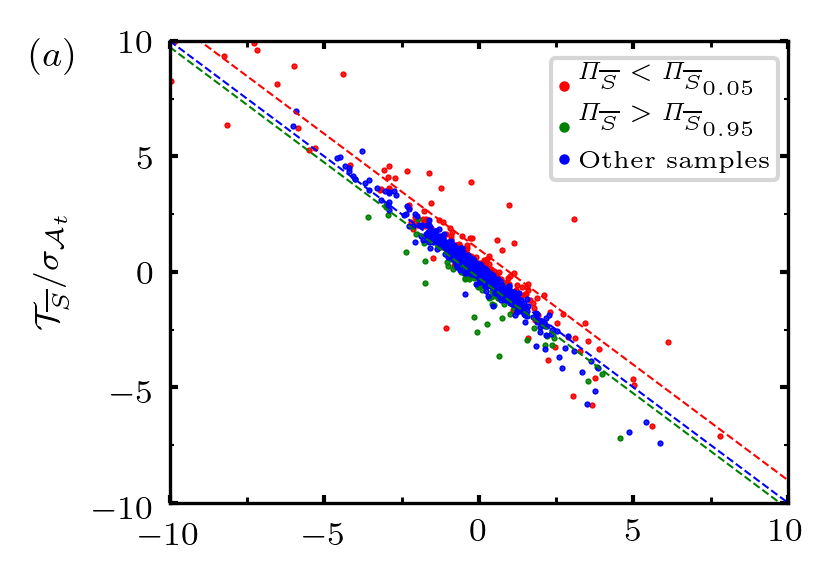} \hfill  \includegraphics{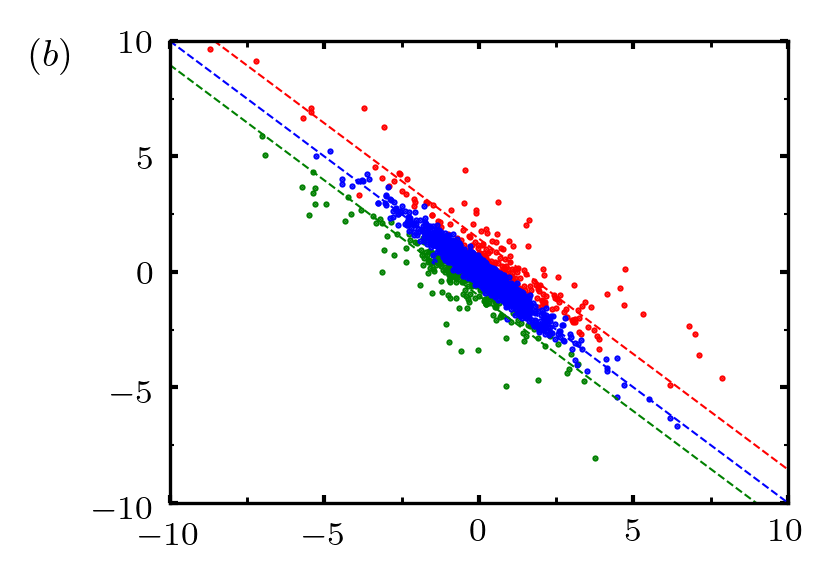}}
	\centerline{\includegraphics{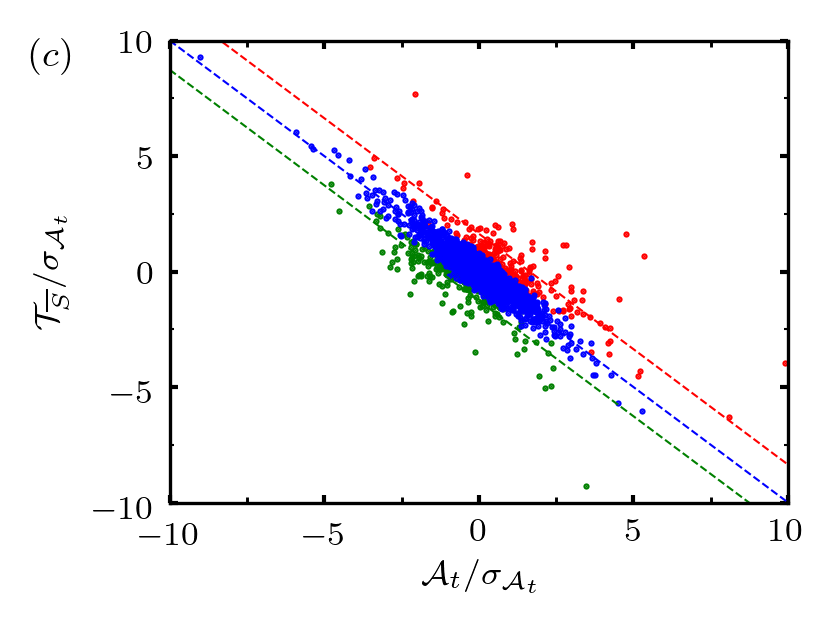} \hfill  \includegraphics{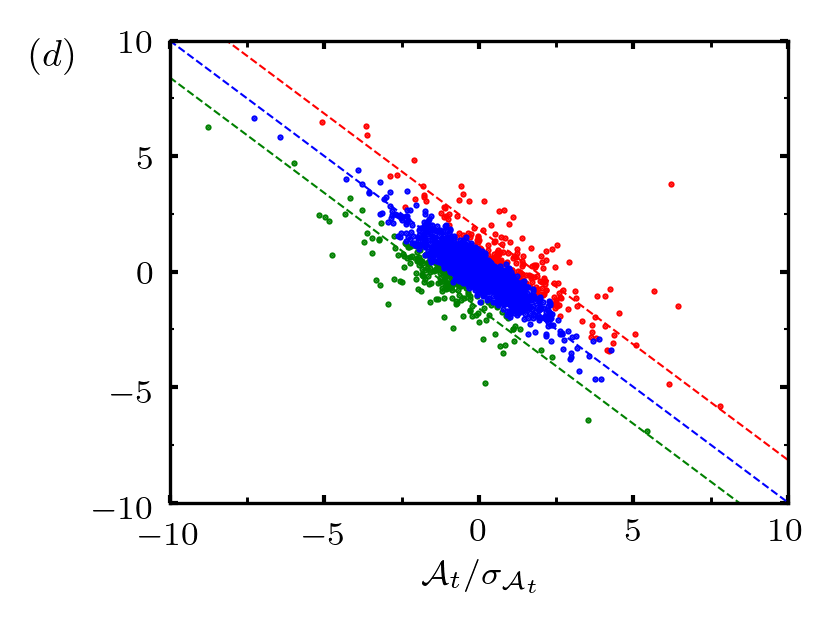}}
	\caption{Scatter plots of $\mathcal{A}_t$ and
          $\mathcal{T}_{\overline{S}}$ at random orientations
          $\boldsymbol{r}$ normalised by $\sigma_{\mathcal{A}_t}$ and
          $\sigma_{\mathcal{T}_{\overline{S}}}$, their respective
          standard deviations. $\mathit{\Pi}_{\overline{S}_{0.05}}$ is
          the value of $\mathit{\Pi}_{\overline{S}}$ at the respective
          $r_d$ for which $5\%$ of the samples are more negative than
          $\mathit{\Pi}_{\overline{S}_{0.05}}$ and
          $\mathit{\Pi}_{\overline{S}_{0.95}}$ is the value of
          $\mathit{\Pi}_{\overline{S}}$ for which $95\%$ of the
          samples are more positive than
          $\mathit{\Pi}_{\overline{S}_{0.95}}$. The events
          $\mathit{\Pi}_{\overline{S}}<\mathit{\Pi}_{\overline{S}_{0.05}}$
          and
          $\mathit{\Pi}_{\overline{S}}>\mathit{\Pi}_{\overline{S}_{0.95}}$
          are marked in red and green respectively, while the
          remaining events are marked in blue. {The red line marks
          $\mathcal{A}_t = -\mathcal{T}_{\overline{S}} - \langle
          \mathit{\Pi}_{\overline{S}} |
          \mathit{\Pi}_{\overline{S}}<\mathit{\Pi}_{\overline{S}_{0.05}}
          \rangle$, where $\langle \mathit{\Pi}_{\overline{S}} |
          \mathit{\Pi}_{\overline{S}}<\mathit{\Pi}_{\overline{S}_{0.05}}
          \rangle$ is the average value of $\mathit{\Pi}_{\overline{S}}$ conditioned on $\mathit{\Pi}_{\overline{S}}<\mathit{\Pi}_{\overline{S}_{0.05}}$. The green line marks $\mathcal{A}_t =
          -\mathcal{T}_{\overline{S}} - \langle
          \mathit{\Pi}_{\overline{S}} |
          \mathit{\Pi}_{\overline{S}}>\mathit{\Pi}_{\overline{S}_{0.95}}
          \rangle$ and the blue line marks $\mathcal{A}_t =
          -\mathcal{T}_{\overline{S}}$ (with all terms appropriately
          normalised with $\sigma_{\mathcal{A}_t}$ and
          $\sigma_{\mathcal{T}_{\overline{S}}}$)}. $r_d/\langle
          \lambda\rangle_t=(0.12,1.45,3.1,5.2)$ for $(a,b,c,d)$ and
          $\langle \Rey_\lambda\rangle_t=174$. }
	\label{fig:AtTsScatter}
\end{figure}

Following the question of local/instantaneous equilibrium, we now look
for local/instantaneous sweeping. Figure \ref{fig:khmhCorr} shows
strong anti-correlation between $\mathcal{A}_t$ and
$\mathcal{T}_{\overline{S}}$, increasingly so as $r_d$ decreases from
large to small scales. Along with the fifth KHMH result at the end of
the previous section (that the fluctuation magnitudes of
$\mathcal{A}_t$ and $\mathcal{T}_{\overline{S}}$ become increasingly
comparable as $r_d$ decreases), this anti-correlation tendency
suggests a tendency towards $\mathcal{A}_t +
\mathcal{T}_{\overline{S}} \approx 0$ at decreasing scales in
agreement with the concept of two-point sweeping introduced in section
\ref{subsec:nsd}. In other words, the sweeping of $|\delta
\boldsymbol{u}|^2$ by the mainly large scale advection velocity
$(\boldsymbol{u}^{+}+\boldsymbol{u}^{-})/2$ becomes increasingly
strong with decreasing $r_d$. The scatter plots of $\mathcal{A}_t$ and
$\mathcal{T}_{\overline{S}}$ in figure \ref{fig:AtTsScatter} make this
local/instantaneous two-point sweeping tendency with decreasing $r_d$
very evident, but also indicate that significant values of positive or
negative $\mathit{\Pi}_{\overline{S}}$ can cause increasing deviations
from $\mathcal{A}_t + \mathcal{T}_{\overline{S}}\approx 0$ as $r_d$
increases. Note $\mathcal{A}_t+\mathcal{T}_{\overline{S}}
+\mathit{\Pi}_{\overline{S}} \approx 0$ as indicated by the
correlation coefficients in figure \ref{fig:khmhCorr} between
$\mathcal{A}_t+\mathcal{T}_{\overline{S}}$ and
$-\mathit{\Pi}_{\overline{S}}$ (which exceed $0.95$ for $r_d\geq
\langle \lambda \rangle_t$ at our Reynolds numbers) and by their
overlapping fluctuation magnitudes in figure
\ref{fig:khmhFluctMag}($a2$,$b2$). The fluctuations of
$\mathit{\Pi}_{\overline{S}}$ increase in magnitude as $r_d$ increases
and so do high values of $\mathit{\Pi}_{\overline{S}}$ too. The
scatter plots in figure \ref{fig:AtTsScatter} highlight how the $5\%$
most negative $\mathit{\Pi}_{\overline{S}}$ events (values of
$\mathit{\Pi}_{\overline{S}}$ for which the probability that
$\mathit{\Pi}_{\overline{S}}$ is smaller than a negative value
$\mathit{\Pi}_{\overline{S}_{0.05}}$ is 0.05) and the $5\%$ most
positive $\mathit{\Pi}_{\overline{S}}$ events (values of
$\mathit{\Pi}_{\overline{S}}$ for which the probability that
$\mathit{\Pi}_{\overline{S}}$ is larger than a positive value
$\mathit{\Pi}_{\overline{S}_{0.95}}$ is also 0.05) cause significant
deviations from "perfect sweeping"
$\mathcal{A}_t=-\mathcal{T}_{\overline{S}}$, increasingly so for
increasing $r_d$, in agreement with
$\mathcal{A}_t+\mathcal{T}_{\overline{S}} +\mathit{\Pi}_{\overline{S}}
\approx 0$.

\begin{figure}
	\centerline{\includegraphics{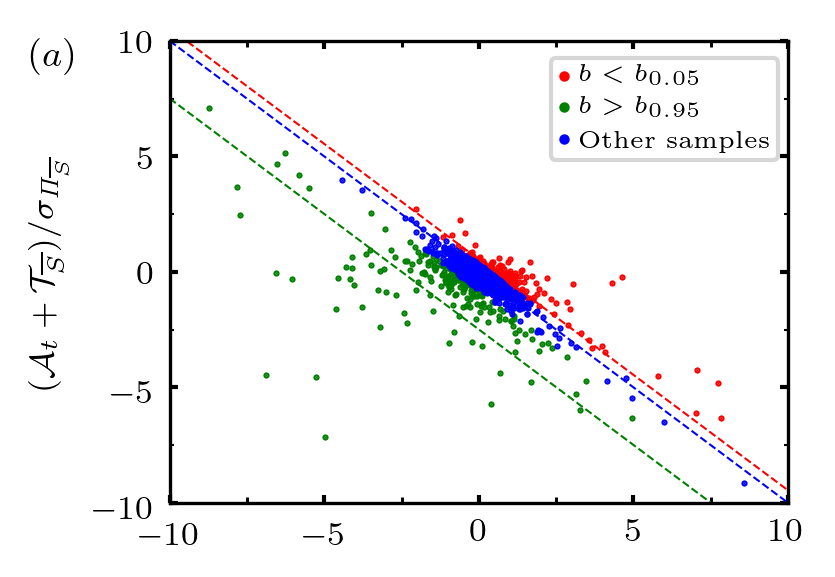} \hfill \includegraphics{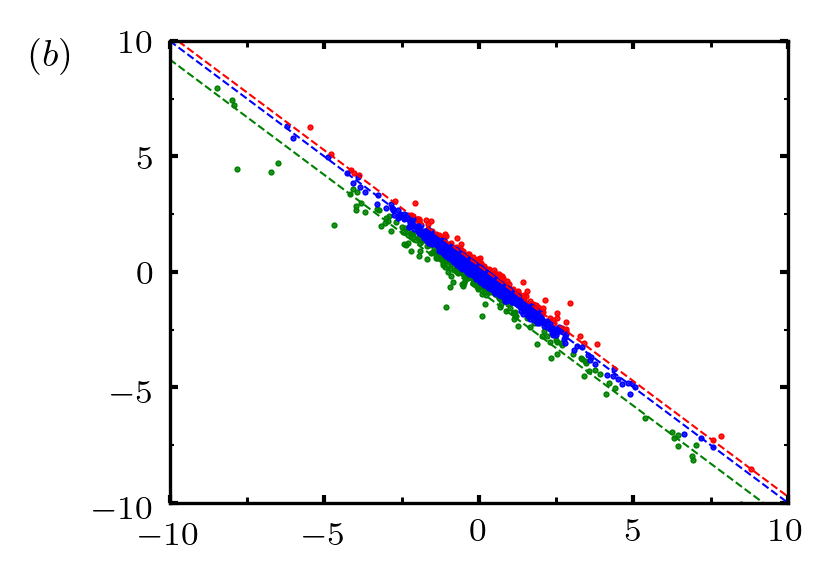}}
	\centerline{\includegraphics{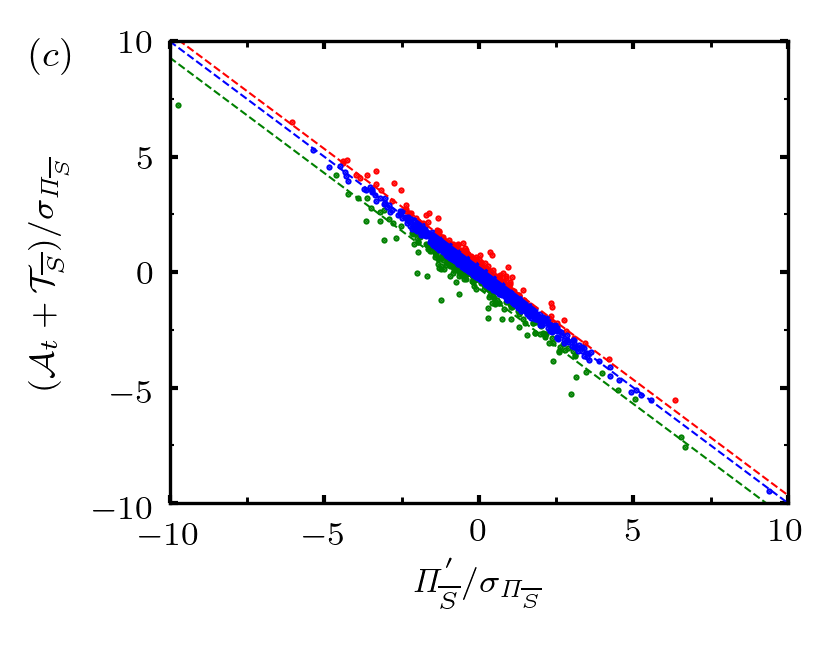} \hfill \includegraphics{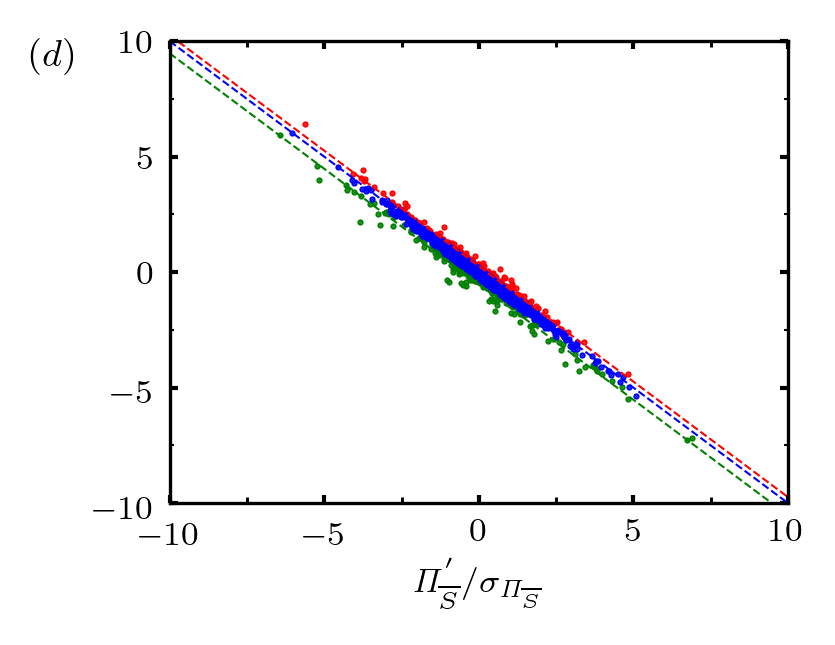}}
	\caption{Scatter plots of $\mathcal{A}_t+\mathcal{T}_{\overline{S}}$ and $\mathit{\Pi}_{\overline{S}}^{'}$ at random orientations $\boldsymbol{r}$. The residual $-b  \equiv \mathcal{A}_t+\mathcal{T}_{\overline{S}}+\mathit{\Pi}_{\overline{S}}^{'}$ and the values $b_{0.05}$ and $b_{0.95}$ are defined analogously as for $\mathit{\Pi}_{\overline{S}_{0.05}}$ and $\mathit{\Pi}_{\overline{S}_{0.95}}$ in the previous figure. The events $b<b_{0.05}$ and $b>b_{0.95}$ are marked in red and green respectively, while the remaining events are marked in blue. The red line marks $\mathcal{A}_t + \mathcal{T}_{\overline{S}}  =  - \mathit{\Pi}_{\overline{S}}^{'} - \langle b | b<b_{0.05} \rangle$, the green line $\mathcal{A}_t + \mathcal{T}_{\overline{S}} = - \mathit{\Pi}_{\overline{S}}^{'} - \langle b | b>b_{0.95} \rangle$ and the blue line $\mathcal{A}_t + \mathcal{T}_{\overline{S}} = -\mathit{\Pi}_{\overline{S}}^{'}$ (with all terms appropriately normalised with $\sigma_{\mathit{\Pi}_{\overline{S}}}$). $r_d/\langle \lambda\rangle_t=(0.12,1.45,3.1,5.2)$ for $(a,b,c,d)$ and $\langle \Rey_\lambda\rangle_t=174$.}
	\label{fig:AtTsPisScatter}
\end{figure}

The scatter plots in figure \ref{fig:AtTsPisScatter} show that it is
only in relatively rare circumstances that
$\mathcal{A}_t+\mathcal{T}_{\overline{S}} +\mathit{\Pi}_{\overline{S}}
\approx 0$ is significantly inaccurate for scales $r_d\geq \langle
\lambda \rangle_t$. Similarly to NSD dynamics,
$\mathcal{A}_t+\mathcal{T}_{\overline{S}}$ can be viewed as a
Lagrangian time-rate of change of $|\delta \boldsymbol{u}|^2$ moving
with $(\boldsymbol{u}^{+}+\boldsymbol{u}^{-})/2$. As more than average
$|\delta \boldsymbol{u}|^2$ is cascaded from larger to smaller scales
at a particular location $(\mathit{\Pi}_{\overline{S}}^{'}<0)$,
$\mathcal{A}_t+\mathcal{T}_{\overline{S}}$ increases; and as more than
average $|\delta \boldsymbol{u}|^2$ is inverse cascaded from smaller
to larger scales $(\mathit{\Pi}_{\overline{S}}^{'} > 0)$,
$\mathcal{A}_t+\mathcal{T}_{\overline{S}}$
decreases. $\mathit{\Pi}_{\overline{S}}^{'}$ is to a large extent
determined by $\boldsymbol{a}_{\mathit{\Pi}_{\overline{S}}}$ which, as
we show in appendix \ref{sec:appC}, is a non-local function in space
of the vortex stretching and compression dynamics determining the
two-point vorticity difference $\delta \boldsymbol{\omega}$.

A fairly complete way to summarise the details of the balance
$\mathcal{A}_t+\mathcal{T}_{\overline{S}} +\mathit{\Pi}_{\overline{S}}
\approx 0$ at scales $r_d\geq \langle \lambda \rangle_t$ is by noting
that, as $r_d$ decreases towards $\langle \lambda \rangle_t$, (i) the
fluctuation magnitude of $\mathcal{T}_{\overline{S}}$ tends to become
comparable to that of $\mathcal{A}_t$ while that of
$\mathit{\Pi}_{\overline{S}}$ decreases by comparison, (ii) the
correlation coefficient between $\mathcal{A}_t$ and
$-\mathcal{T}_{\overline{S}}$ increases towards $0.9$, and also (iii)
(not mentioned till now but evident in figure \ref{fig:khmhCorr}) the
correlation coefficient between $\mathcal{A}_t$ and
$-\mathit{\Pi}_{\overline{S}}$ decreases towards values below $0.2$.

\subsection{Conditional correlations} \label{subsec:condcorr}

\begin{figure}
	\centerline{\includegraphics{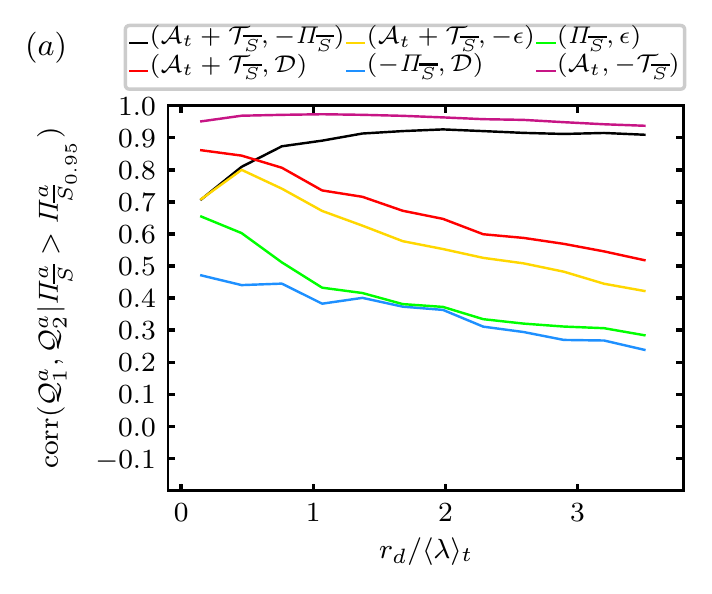} \hfill \includegraphics{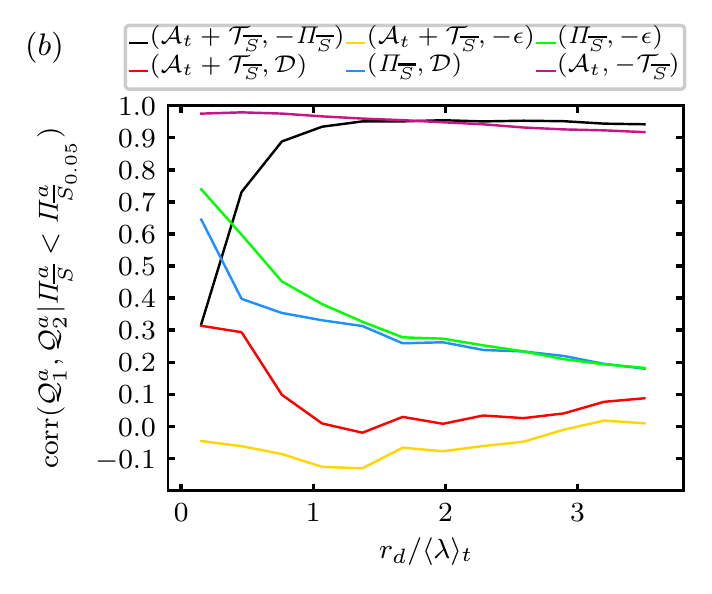}}
	\caption{($a$) Correlation coefficients among the $5\%$
          strongest spherically averaged backward interscale transfer
          events
          $\mathit{\Pi}_{\overline{S}}^a>\mathit{\Pi}_{\overline{S}_{0.95}}^a$
          for KHMH terms $(\mathcal{Q}_1^a,\mathcal{Q}_2^a)$ listed on
          top of the figure. ($b$) Correlation coefficients among the
          $5\%$ strongest spherically averaged forward interscale
          transfer events
          $\mathit{\Pi}_{\overline{S}}^a<\mathit{\Pi}_{\overline{S}_{0.05}}^a$
          for KHMH terms $(\mathcal{Q}_1^a,\mathcal{Q}_2^a)$ listed on
          top of the figure. $\langle
          \Rey_\lambda\rangle_t=112$. (Corresponding plots for
          $\langle \Rey_\lambda\rangle_t=174$ are omitted because they
          are very similar.)}
	\label{fig:extremeCorr}
\end{figure}

At scales $r_d$ below $\langle \lambda \rangle_t$, the relation
$\mathcal{A}_t+\mathcal{T}_{\overline{S}} +\mathit{\Pi}_{\overline{S}}
\approx 0$ becomes less accurate as the correlation coefficient
between $\mathcal{A}_t+\mathcal{T}_{\overline{S}}$ and
$-\mathit{\Pi}_{\overline{S}}$ drops from $0.95$ to $0.7$ with
decreasing $r_d$, reflecting the increase of correlation between
$\mathcal{\epsilon}$ and $-\mathit{\Pi}_{\overline{S}}$ and the even
higher increase towards values close to $0.5$ of the correlation
coefficient between $\mathcal{D}$ and
$\mathit{\Pi}_{\overline{S}}$. This increase of correlation appears to
reflect the impact of relatively rare yet intense local/instantaneous
occurances of interscale transfer rate as shown in figure
\ref{fig:extremeCorr} {where we plot correlations
  conditional on relatively rare interscale events where the
  magnitudes of the spherically-averaged interscale transfer rates are
  higher than 95\% of all interscale transfer rates of same sign
  (positive for backward and negative for forward transfer) in our
  overall spatio-temporal sample}. This impact is highest at scales
smaller than $\langle \lambda \rangle_t$ where the correlation
coefficient conditioned on intense forward or backward interscale
transfer rate events of $\pm\mathit{\Pi}_{\overline{S}}$ and either
$\mathcal{\epsilon}$ or $\mathcal{D}$ can be as high as $0.7$
($+\mathit{\Pi}_{\overline{S}}$ in the case of backward events and
$-\mathit{\Pi}_{\overline{S}}$ in the case of forward events which
causes significantly higher correlations between
$\mathcal{A}_t+\mathcal{T}_{\overline{S}}$ and either
$-\mathcal{\epsilon}$ or $\mathcal{D}$ in the case of backward events
than in the case of forward events as seen in figure
\ref{fig:extremeCorr}). However, the impact of such
{relatively rare} events is also manifest at scales
larger than $\langle \lambda \rangle_t$ (see figure
\ref{fig:extremeCorr}) where the conditioned correlation coefficient
is significantly higher than the unconditioned one in figure
\ref{fig:khmhCorr}. Interestingly, conditioning on these
{relatively rare} events does not change the
correlation coefficients of $\mathcal{A}_t+\mathcal{T}_{\overline{S}}$
with $-\mathit{\Pi}_{\overline{S}}^{'}$ except at scales $r_d$ smaller
than $\langle \lambda \rangle_t$ where, consistently with the
increased conditioned correlations between
$-\mathit{\Pi}_{\overline{S}}$ and $\mathcal{D}$, they are smaller
than the unconditional correlation coefficients of
$\mathcal{A}_t+\mathcal{T}_{\overline{S}}$ with
$-\mathit{\Pi}_{\overline{S}}^{'}$, particularly at relatively rare
forward interscale events where this conditional correlation drops to
values close to $0.3$ at scales well below $\langle \lambda
\rangle_t$.

\begin{figure}
	\centerline{\includegraphics{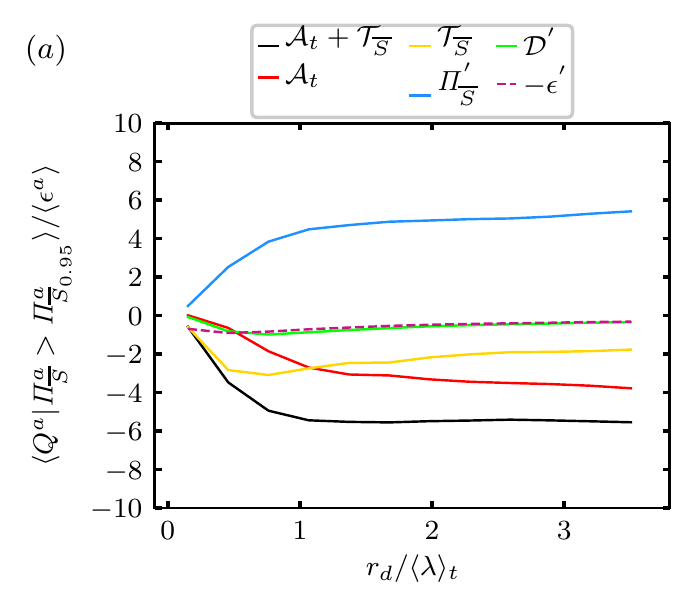} \hfill \includegraphics{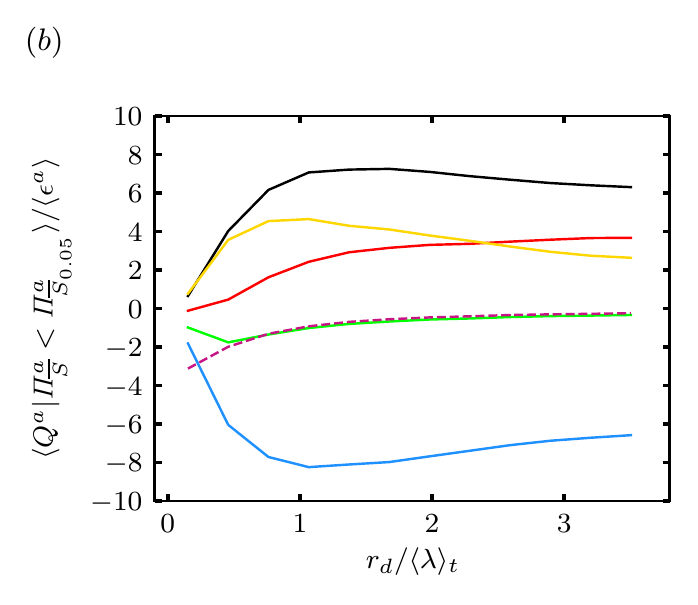}}
	\caption{($a$) Spatio-temporal averages of KHMH terms
          $\mathcal{Q}^a$ conditioned on the $5\%$ strongest
          spherically averaged backward ($a$) and forward $(b)$
          interscale transfer events. The KHMH terms are listed above
          figure $(a)$ and $\langle
          \Rey_\lambda\rangle_t=112$.(Corresponding plots for $\langle
          \Rey_\lambda\rangle_t=174$ are omitted because they are very
          similar.)}
	\label{fig:condAvg}
\end{figure}

Given that {our relatively rare intense} interscale
transfer rates can be the seat of some correlation between
$\mathit{\Pi}_{\overline{S}}$ and either $-\mathcal{\epsilon}$ or
$\mathcal{D}$ particularly for $r_d < \langle \lambda \rangle_t$ , and
given that $\mathcal{A}_t+\mathcal{T}_{\overline{S}} \approx 0$ is a
good approximation at scales smaller than $\langle \lambda \rangle_t$,
do we have approximate two-point sweeping and approximate equilibrium
$\mathit{\Pi}_{\overline{S}} \approx \mathcal{D}$ if we condition on
{relatively rare} forward or backward interscale
transfer rate events? In fact the conditional correlations between
$\mathcal{A}_t$ and $-\mathcal{T}_{\overline{S}}$ are very high (close
to and above 0.95) at all scales (see figure \ref{fig:extremeCorr}),
higher than the corresponding unconditional correlations. However, the
conditional averages of $\mathcal{A}_t$ and
$-\mathcal{T}_{\overline{S}}$ shown in figure \ref{fig:condAvg} are
also significantly different at all scales, implying that these strong
conditional correlations do not actually amount to two-point sweeping
at {relatively rare} forward and backward
events. Furthermore, if we condition on high negative/positive values
of $\mathit{\Pi}_{\overline{S}}$, the averages of both $\mathcal{A}_t$
and $\mathcal{T}_{\overline{S}}$ are positive/negative (figure
\ref{fig:condAvg}), even though these conditional averages do tend to
$0$ as $r_d$ tends to $0$. This has two implications. (i) It implies
that, even though $\mathcal{A}_t$ and $-\mathcal{T}_{\overline{S}}$
are very well correlated at these {relatively rare}
events, $\mathcal{A}_t + \mathcal{T}_{\overline{S}}$ fluctuates around
a constant $C$ where $C>0$ if we condition the fluctuations on
{relatively rare} negative
$\mathit{\Pi}_{\overline{S}}$ but $C<0$ if we condition them on
{relatively rare} positive
$\mathit{\Pi}_{\overline{S}}$ ($C=0$ if we do not condition). This
amounts to a systematic deviation on the average from two-point
sweeping even though the strong correlation between the high magnitude
fluctuations of $\mathcal{A}_t$ and $-\mathcal{T}_{\overline{S}}$
point at a tendency towards sweeping which is frustrated by the
presence of the comparatively low non-zero local
$\mathit{\Pi}_{\overline{S}}$. Given equation \eqref{khmhSolFluct},
the presence of this non-zero constant $C$ (clearly non-zero for all
scales, and non-zero but tending towards zero as $r_d$ tends to $0$
well below $\langle \lambda \rangle_t$) means that the equilibrium
$\mathit{\Pi}_{\overline{S}} \approx \mathcal{D}$ for scales smaller
than $\langle \lambda \rangle_t$ does not hold either, even at scales
smaller than $\langle \lambda \rangle_t$ where the conditional
correlation between $\mathit{\Pi}_{\overline{S}}$ and $\mathcal{D}$ is
significant. In fact, figure \ref{fig:condAvg} shows that the
conditional averages of $\mathit{\Pi}_{\overline{S}}'$ are much larger
than those of both $\mathcal{D}'$ and $-\mathcal{\epsilon}^{'}$; they
are much closer to those of $\mathcal{A}_t +
\mathcal{T}_{\overline{S}}$.

\begin{figure}
	\centerline{\includegraphics{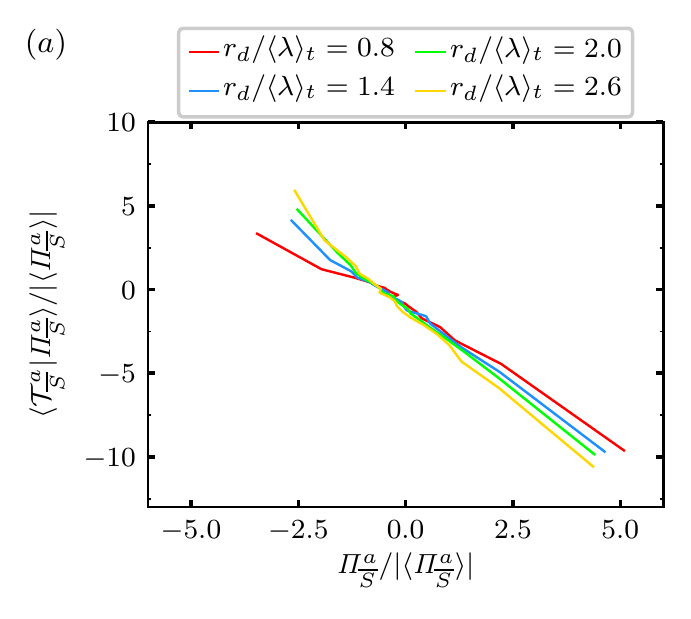} \hfill \includegraphics{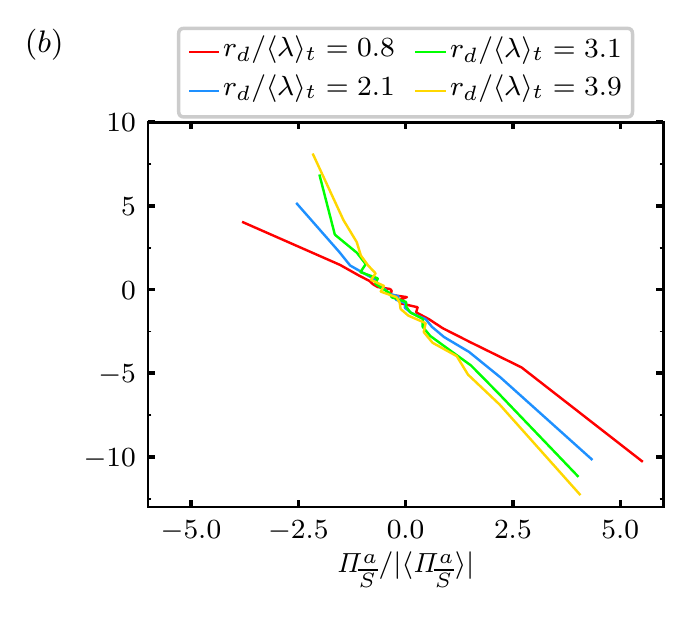}}
	\caption{{Spatio-temporal averages of
            $\mathcal{T}_{\overline{S}}^a$ across scales $r_d$
            conditioned on $\mathit{\Pi}_{\overline{S}}^a$ being
            within a certain range of $\mathit{\Pi}_{\overline{S}}^a$
            values and we consider $20$ such ranges of increasing
            values of $\mathit{\Pi}_{\overline{S}}^a$: the $5\%$
            smallest/most negative $\mathit{\Pi}_{\overline{S}}^a$,
            the $5\%$ to $10\%$ smallest/most negative
            $\mathit{\Pi}_{\overline{S}}^a$ values and so on until the
            $5\%$ largest/most positive
            $\mathit{\Pi}_{\overline{S}}^a$ values.} ($a$) $\langle
          \Rey \rangle_t=112$, ($b$) $\langle \Rey \rangle_t=174$.}
	\label{fig:condAvgTsPis}
\end{figure}

(ii) The second implication of the conditional signs of
$\mathcal{T}_{\overline{S}}$ is the existence of a relation between
conditional average of solenoidal interspace transfer rate
$\mathcal{T}_{\overline{S}}$ and the solenoidal interspace transfer
rate $\mathit{\Pi}_{\overline{S}}$ on which the average is
conditioned: when one is positive/negative the other is
negative/positive, and we also find that their absolute magnitudes
increase together (see figure \ref{fig:condAvgTsPis}). This is an observation which may prove important in the future for both subgrid scale modeling and the
detailed study of the very smallest scales of turbulence fluctuations.

In conclusion, $\mathit{\Pi}_{\overline{S}}$ does not fluctuate with
neither $-\mathcal{\epsilon}$ nor $\mathcal{D}$. Instead,
$\mathit{\Pi}_{\overline{S}}$ and
$\mathcal{A}_t+\mathcal{T}_{\overline{S}}$ fluctuate together at all
scales, in particular scales larger than $\langle \lambda \rangle_t$,
and even at {relatively rare} interscale transfer
events. At scales smaller than $\langle \lambda \rangle_t$, we have a
general tendency towards two-point sweeping if we do not condition on
particular events. At {our relatively rare} interscale
transfer events this correlation tendency (now conditional) is in fact
amplified but there is nevertheless a systematic average deviation
from two-point sweeping consistent with the absence of equilibrium
$\mathit{\Pi}_{\overline{S}} \approx \mathcal{D}$ at these
events. Finally, a relation exists between interspace and interscale
transfer rates because the average interspace transfer rate
conditioned on positive/negative values of interscale transfer rate is
negative/positive. It must be stressed, however, that this relation
does not imply an anticorrelation between interscale and interspace
transport rates. The unconditioned correlation coefficients between
$-\mathit{\Pi}_{\overline{S}}$ and $\mathcal{T}_{\overline{S}}$ are
around $0.2$ (see figure \ref{fig:khmhCorr}), and we checked that this
$0.2$ correlation does not change significantly if we condition on
relatively rare intense occurances of interscale transfer rate.
\section{Inhomogeneity contribution to interscale transfer}\label{sec:khmh3}
\subsection{Average values and PDFs} \label{subsec:avgandPDF}

The decomposition $\mathit{\Pi} = \mathit{\Pi}_{\overline{I}}
+\mathit{\Pi}_{\overline{S}}$ helped us distinguish between the
solenoidal vortex stretching/compression and the pressure-related
aspects of the interscale transfer. As recently shown by
\citet{AlvesPortela2020}, the interscale transfer rate $\mathit{\Pi}$
can also be decomposed in a way which brings out the fact that it has
a direct inhomogeneity contribution to it. This last part of the
present study is an examination of the decomposition introduced by
\citet{AlvesPortela2020} which is
$\mathit{\Pi}=\mathit{\Pi}_{I}+\mathit{\Pi}_H$ where
\begin{align}
	\mathit{\Pi}_I &=  \ \ \frac{1}{2} \delta u_i \frac{\partial}{\partial x_i }(u_k^{+}u_k^{+}-u_k^{-}u_k^{-}) \label{piI}, \\
	\mathit{\Pi}_H &= -2 \delta u_i \frac{\partial}{\partial r_i}  (u_k^{-}u_k^{+}).\label{piH}
\end{align}
\noindent $\mathit{\Pi}_I$ can be locally/instantaneously non-zero
only in the presence of a local/instantaneous inhomogeneity. However,
it averages to zero, i.e. $\langle \mathit{\Pi}_I \rangle=0$, in
periodic/statistically homogeneous turbulence. Note that
$\mathit{\Pi}=\mathit{\Pi}_{I}=\mathit{\Pi}_{H}=0$ at
$\boldsymbol{r}=\boldsymbol{0}$.
With $\boldsymbol{r}$-orientation-averaging, the decomposition
$\mathit{\Pi}^{a}=\mathit{\Pi}_{I}^{a}+\mathit{\Pi}_{H}^{a}$ is unique
in the sense that any potentially suitable (e.g. such that it equals
$0$ at $\boldsymbol{r}=\boldsymbol{0}$) $\boldsymbol{x}$-gradient term
added to $\mathit{\Pi}_{I}$
vanishes after $\boldsymbol{r}$-orientation-averaging (see
\citet{AlvesPortela2020}).

An equivalent expression for $\mathit{\Pi}_I$ which immediately
reveals where the decomposition $\mathit{\Pi} = \mathit{\Pi}_I +
\mathit{\Pi}_H$ comes from is $\mathit{\Pi}_I = \delta u_i
\frac{\partial}{\partial r_i }(u_k^{+}u_k^{+}+u_k^{-}u_k^{-})$. Given
that the total interscale transfer rate is $\mathit{\Pi} = \delta u_i
\frac{\partial}{\partial r_i }(\delta u_k \delta u_k)$, the
$\mathit{\Pi}_I$ part of the interscale transfer concerns the
transfered energy differences coming mostly from differences between
velocity amplitudes, i.e. local/instantaneous inhomogeneities of
``turbulence intensity'' in the flow; the $\mathit{\Pi}_H$ part of the
interscale transfer concerns transfered energy differences coming
mostly from differences between velocity orientations. Consistently
with its link to local/instantaneous non-homogeneity, $\mathit{\Pi}_I$
can be written in the form (\eqref{piI}) making it clear that
$\mathit{\Pi}_I$ is zero where and when fluctuating velocity
magnitudes are locally uniform.

In comparing the decompositions $\mathit{\Pi} =
\mathit{\Pi}_{\overline{S}} +\mathit{\Pi}_{\overline{I}}$ and
$\mathit{\Pi}=\mathit{\Pi}_{I}+\mathit{\Pi}_H$, it is worth noting
that $\mathit{\Pi}_{I}= \mathit{\Pi}_{I_{\overline{I}}}$ given that
$\mathit{\Pi}_{I_{\overline{S}}} = 0$ from its centroid gradient form
(see equation \eqref{piI}).
It therefore follows that
\begin{align}
	\mathit{\Pi}_{\overline{S}} &=
        \mathit{\Pi}_{H_{\overline{S}}} \label{pihs}, \\ \mathit{\Pi}_{\overline{I}}
        &= \mathit{\Pi}_{I} + \mathit{\Pi}_{H_{\overline{I}}}. \label{pihi}
\end{align}
The inhomogeneity-based interscale transfer rate influences only the
irrotational part of the total interscale transfer rate whereas
$\mathit{\Pi}_{H}$ influences both the irrotational and the solenoidal
parts. As $\langle \mathit{\Pi}_I \rangle=0$ and $\langle
\mathit{\Pi}_{\overline{I}} \rangle =0$, it follows that $\langle
\mathit{\Pi}_{H_{\overline{I}}}\rangle = 0$. More to the point,
$\langle \mathit{\Pi}_{\overline{S}} \rangle$ equals $\langle
\mathit{\Pi}_{H_{\overline{S}}}\rangle$ and so equation
\eqref{khmhAvg} reduces to
\begin{equation}
	\langle \mathit{\Pi}\rangle = \langle
        \mathit{\Pi}_{H_{\overline{S}}}\rangle = \langle
        \mathcal{D}_{r,\nu} \rangle - \langle \mathcal{\epsilon}
        \rangle + \langle \mathcal{I} \rangle .
\end{equation}
The part of the interscale transfer rate which is present in the
average interscale transfer/cascade dynamics is in fact
$\mathit{\Pi}_{H_{\overline{S}}}$.

\begin{figure}
	\centerline{\includegraphics{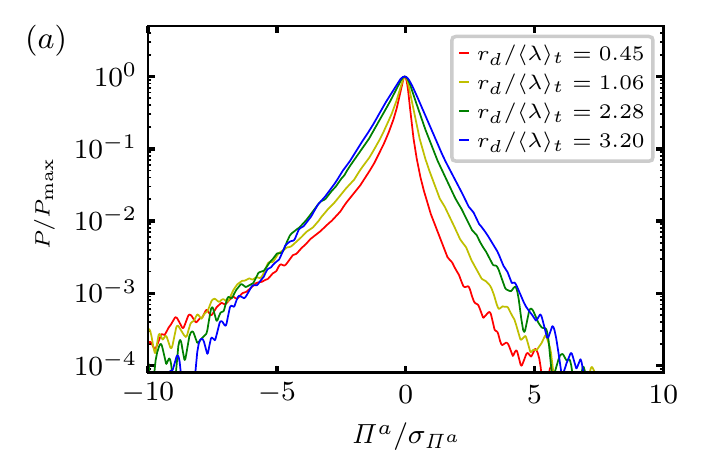} \hfill  \includegraphics{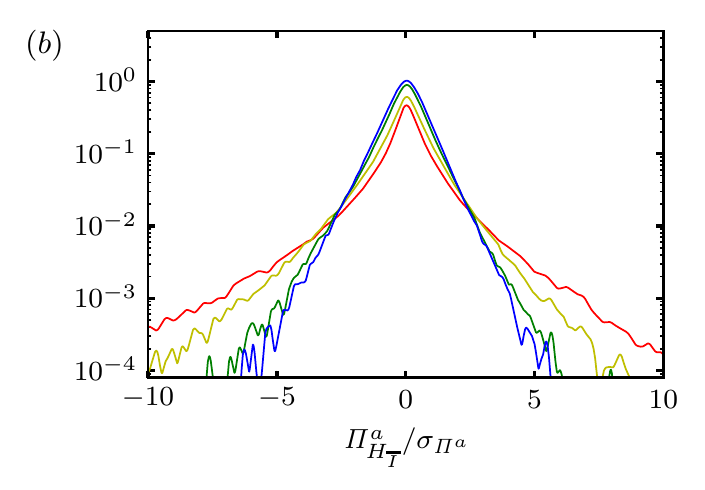}}
	\centerline{\includegraphics{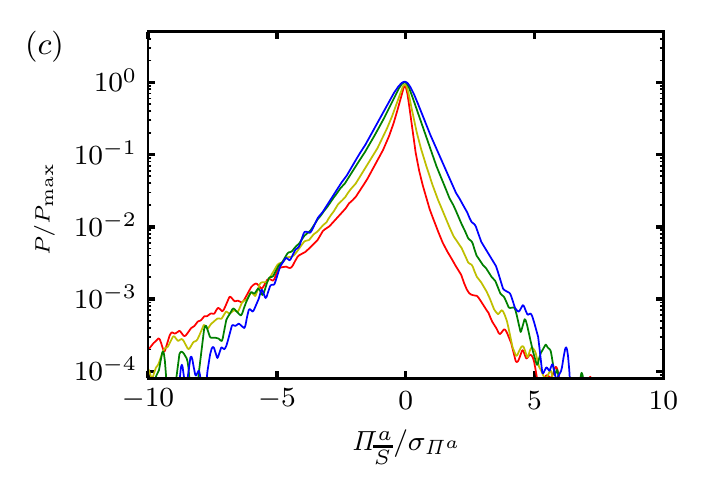} \hfill  \includegraphics{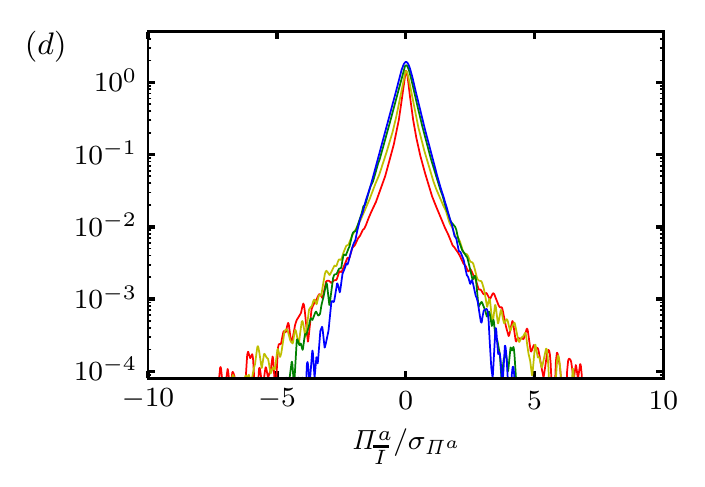}}
	\centerline{\includegraphics{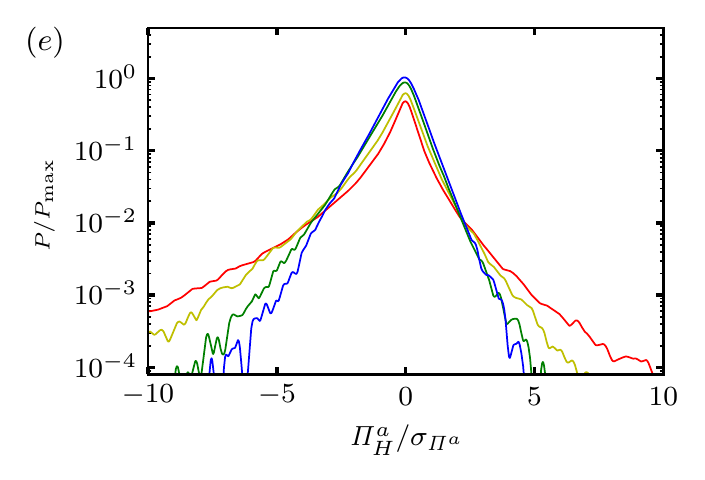} \hfill \includegraphics{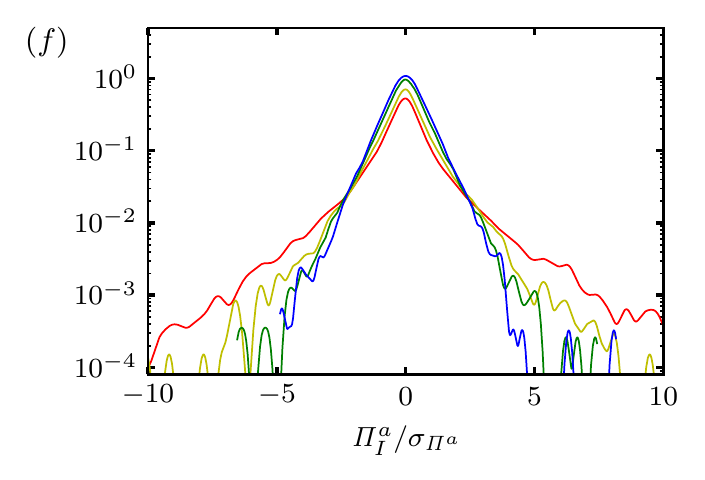}}
	\caption{($a,b,c,d,e,f$) PDFs of $\mathit{\Pi}$ decompositions $(\mathit{\Pi},\mathit{\Pi}_{H_{\overline{I}}},\mathit{\Pi}_{\overline{S}},\mathit{\Pi}_{\overline{I}},\mathit{\Pi}_H,\mathit{\Pi}_I)$ at $\langle \Rey_\lambda\rangle_t=112$. $\sigma_{\mathit{\Pi}^a}$ denotes the standard deviation of $\mathit{\Pi}^a$ and $P_{\text{max}}$ denotes the maximum value of the PDF of $\mathit{\Pi}^a$. The inhomogeneity and homogeneity interscale transfer rates $\mathit{\Pi}_I$ and $\mathit{\Pi}_H$ are defined in equations \eqref{piI}-\eqref{piH} and the irrotational part of the homogeneity interscale transfer rate $\mathit{\Pi}_{H_{\overline{I}}}$ in equation \eqref{pihi}.}
	\label{fig:piPDF}
\end{figure}

\begin{figure}
	\centerline{\includegraphics{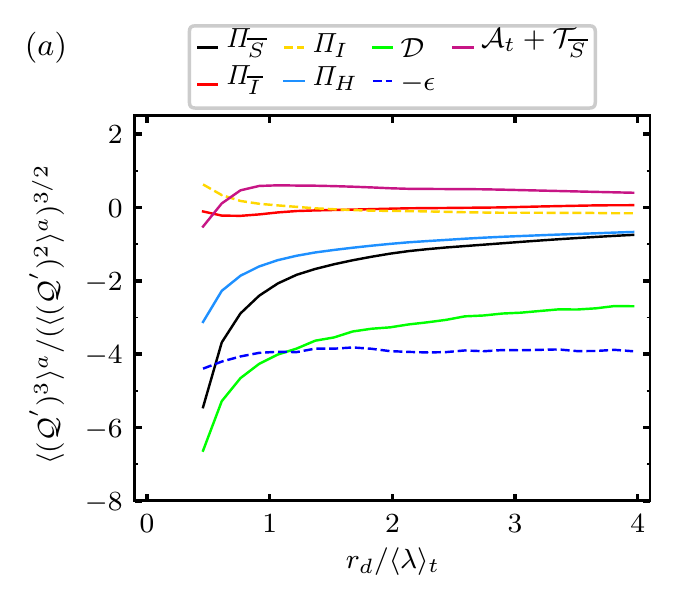}	\hfill			  \includegraphics{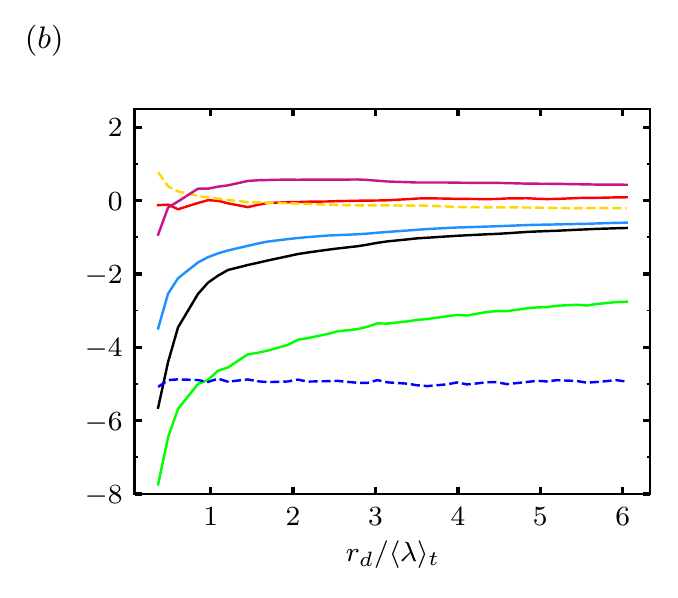}}
	\caption{Skewness factors for KHMH terms $\mathcal{Q}$ listed on top of $(a)$: ($a$) $\langle \Rey_\lambda\rangle_t=112$, ($b$) $\langle \Rey_\lambda\rangle_t=174$.}
	\label{fig:skewFac}
\end{figure}

Given that the average interscale transfer is controlled by
$\mathit{\Pi}_{H_{\overline{S}}} = \mathit{\Pi}_{\overline{S}}$, it is
worth asking whether the well-known negative skewness of the PDF of
$\mathit{\Pi}^{a}$ (e.g. see \citet{Yasuda2018} and references
therein) is also present in the PDF of
$\mathit{\Pi}_{\overline{S}}^{a}$ or/and whether it is spread across
different terms of our two interscale transfer rate decompositions. In
figure \ref{fig:piPDF} we plot the PDFs of $\mathit{\Pi}^{a}$ and of
the different $\boldsymbol{r}$-orientation-averaged terms in the
decompositions of $\mathit{\Pi}$ that we use. It is clear that the
PDFs of $\mathit{\Pi}$ and $\mathit{\Pi}_{\overline{S}}$ are nearly
identical whilst the PDFs of $\mathit{\Pi}_H$ are different though
also negatively skewed. The PDFs of $\mathit{\Pi}_{H_{\overline{I}}}$,
$\mathit{\Pi}_{\overline{I}}$ and $\mathit{\Pi}_{I}$ are not
significantly skewed.
In figure \ref{fig:skewFac} we plot the skewnes factors of the various
interscale transfer terms as well as some other KHMH terms. The
inhomogeneity interscale transfer $\mathit{\Pi}_{I}$ has close to zero
skewness across scales. Both $\mathit{\Pi}_{\overline{S}}$ and
$\mathit{\Pi}_{H}$ are negatively skewed, the former more so than the
latter. Given equations \eqref{pihs}-\eqref{pihi} and $\mathit{\Pi}_H=
\mathit{\Pi}_{\overline{S}}+\mathit{\Pi}_{H_{\overline{I}}}$, this
difference in skewness factors is due to the irrotational part of
$\mathit{\Pi}_H$ which is not significantly skewed and reduces the
skewness of $\mathit{\Pi}_H$ relative to that of
$\mathit{\Pi}_{\overline{S}}$. All in all, the skewness towards
forward rather than inverse interscale transfers is present in its
homogeneous and solenoidal components but is absent in its
non-homogeneous and irrotational parts. \par

Figure \ref{fig:skewFac} also shows that
$\mathcal{A}_t+\mathcal{T}_{\overline{S}}$ is slightly positively
skewed with flatness factors of approximately $0.5$ at scales $r_d
\geq \langle \lambda \rangle_t$ and close to 0 or below at scales
below $\langle \lambda \rangle_t$. The skewness factor of
$-\mathit{\Pi}_{\overline{S}}$ with which
$\mathcal{A}_t+\mathcal{T}_{\overline{S}}$ is very well correlated (as
we have seen in the previous section) is about the same at scales
close to the integral scale but steadily increases to values well
above $0.5$ as $r$ decreases, reaching nearly $6.0$ at scales close to
$0.5\langle \lambda \rangle_t$. This is a concrete illustration of the
fact already mentioned earlier in this paper that
$\mathcal{A}_t+\mathcal{T}_{\overline{S}} \approx
-\mathit{\Pi}_{\overline{S}}$ is a very good approximation for most
locations and most times but not all. Given the very significantly
increased correlation/anti-correlation of
$\mathit{\Pi}_{\overline{S}}$ with both $\mathcal{D}$ and
$\mathcal{\epsilon}$ at relatively {intense}
forward/inverse interscale transfer events and with decreasing scale
$r_d$, it is natural to expect the skewness factor of
$\mathit{\Pi}_{\overline{S}}$ to veer towards the skewness factors of
$\mathcal{D}$ and $-\mathcal{\epsilon}$ which, as can be seen in
figure \ref{fig:skewFac}, are highly negative with values between
$-3.0$ and $-7.0$.

\begin{figure}
	\centerline{\includegraphics{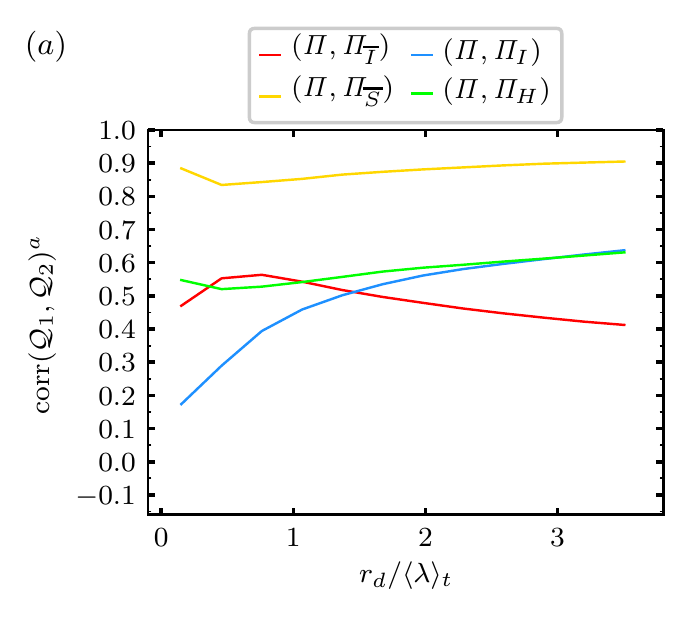} \hfill \includegraphics{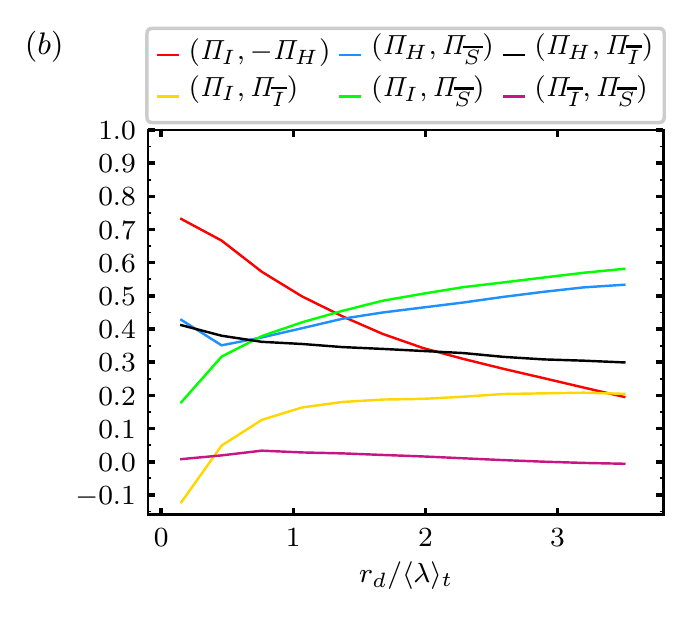}}
	\caption{Correlation coefficients between various
          $\mathit{\Pi}$ decompositions
          $(\mathcal{Q}_1,\mathcal{Q}_2)$ listed on top of the figures
          at $\langle \Rey_\lambda\rangle_t=112$. (Corresponding plots
          for $\langle \Rey_\lambda\rangle_t=174$ are omitted because
          they are very similar.)}
	\label{fig:corrPiIH}
\end{figure}

\subsection{Correlations} \label{subsec:corrPI}

We now consider the local/instantaneous relations between the various
interscale transfer terms in terms of correlation coefficients plotted
in figure \ref{fig:corrPiIH}$a$. First, note the very strong
correlation between $\mathit{\Pi}$ and $\mathit{\Pi}_{\overline{S}}$
and the moderate correlation between $\mathit{\Pi}$ and
$\mathit{\Pi}_{\overline{I}}$. Even though $\mathit{\Pi}$ and
$\mathit{\Pi}_{\overline{S}}$ are highly correlated, we cannot ignore
$\mathit{\Pi}_{\overline{I}}$ and cannot write $\mathit{\Pi} \approx
\mathit{\Pi}_{\overline{S}}$. As seen earlier in the paper, we cannot
ignore $\mathit{\Pi}_{\overline{I}}$ because it is the part of the
interscale transfer which balances the pressure term, but we have also
seen that the fluctuation magnitude of $\mathit{\Pi}_{\overline{S}}$
is significantly higher than the fluctuation magnitude of
$\mathit{\Pi}_{\overline{I}}$. However, even if smaller, the
fluctuation magnitude of $\mathit{\Pi}_{\overline{I}}$ is not
neglible. There is no correlation between
$\mathit{\Pi}_{\overline{S}}$ and $\mathit{\Pi}_{\overline{I}}$ (see
figure \ref{fig:corrPiIH}b), and so $\mathit{\Pi}$ correlates with
both $\mathit{\Pi}_{\overline{S}}$ (strongly) and
$\mathit{\Pi}_{\overline{I}}$ (moderately) for different independent
reasons. $\mathit{\Pi}$ feels the influence of solenoidal vortex
stretching/compression via $\mathit{\Pi}_{\overline{S}}$ and the
influence of pressure fluctuations via $\mathit{\Pi}_{\overline{I}}$,
the former influencing $\mathit{\Pi}$ more than the latter.

Figure \ref{fig:corrPiIH}a also shows significantly smaller
correlations between $\mathit{\Pi}$ and $\mathit{\Pi}_{H}$ than
between $\mathit{\Pi}$ and $\mathit{\Pi}_{\overline{S}}$. This must be
due to a decorrelating effect of $\mathit{\Pi}_{H_{\overline{I}}}$ as
$\mathit{\Pi}_{H}=\mathit{\Pi}_{\overline{S}}+\mathit{\Pi}_{H_{\overline{I}}}$.
The correlations between $\mathit{\Pi}$ and $\mathit{\Pi}_{I}$ are
even smaller at the smaller scales but at integral size scales these
correlations are equal to those between $\mathit{\Pi}$ and
$\mathit{\Pi}_{H}$ (figure \ref{fig:corrPiIH}a).

Figure \ref{fig:corrPiIH}b reveals a strong anti-correlation between
$\mathit{\Pi}_I$ and $\mathit{\Pi}_{H}$ at the small scales and a weak
one at the large scales. As the scales decrease, the interscale
transfers of fluctuating velocity differences caused by
local/instantaneous non-homogeneities and the interscale transfers of
fluctuating velocity differences caused by orientation differences get
progressively more anti-correlated. This anti-correlation tendency
results in $\mathit{\Pi}_{H}$ and $\mathit{\Pi}_{I}$ having larger
fluctuation magnitudes than $\mathit{\Pi}$ at smaller scales, in
particular scales smaller than $\langle \lambda\rangle_t$ (verified
with our DNS data but not shown here for economy of space).

The other significant correlations revealed in figure
\ref{fig:corrPiIH}b are those between $\mathit{\Pi}_H$ and
$\mathit{\Pi}_{\overline{S}}$ and those between $\mathit{\Pi}_I$ and
$\mathit{\Pi}_{\overline{S}}$, particularly as $r_d$ increases from
around/below $\langle \lambda\rangle_t$ to the integral length
scale. These correlations relate to the very stong correlations
between $\mathit{\Pi}$ and $\mathit{\Pi}_{\overline{S}}$ but are
weaker. One can imagine that $\mathit{\Pi}_{\overline{S}}$ correlates
with $\mathit{\Pi}_H$ sometimes and with $\mathit{\Pi}_{I}$ some other
times, but not too often with both given that $\mathit{\Pi}_{I}$ and
$\mathit{\Pi}_H$ tend to be anti-correlated, and that this happens in
a way subjected to a continuously strong correlation between
$\mathit{\Pi} = \mathit{\Pi}_{H} + \mathit{\Pi}_{I}$ and
$\mathit{\Pi}_{\overline{S}}$.

\begin{figure}
	\centerline{\includegraphics{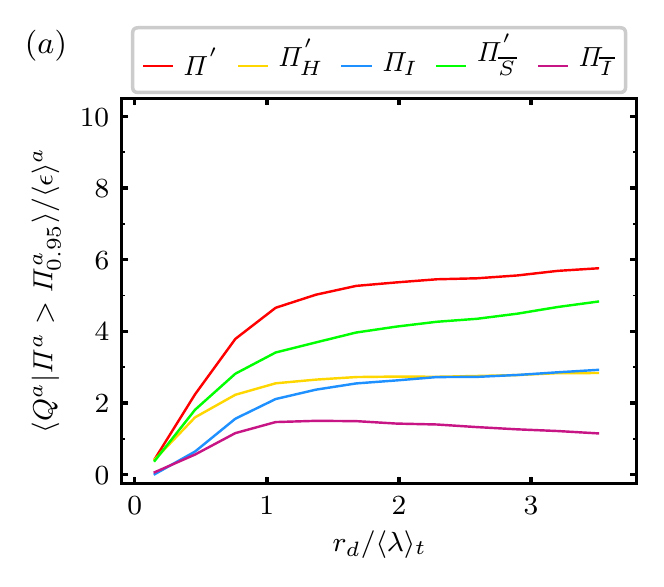} \hfill \includegraphics{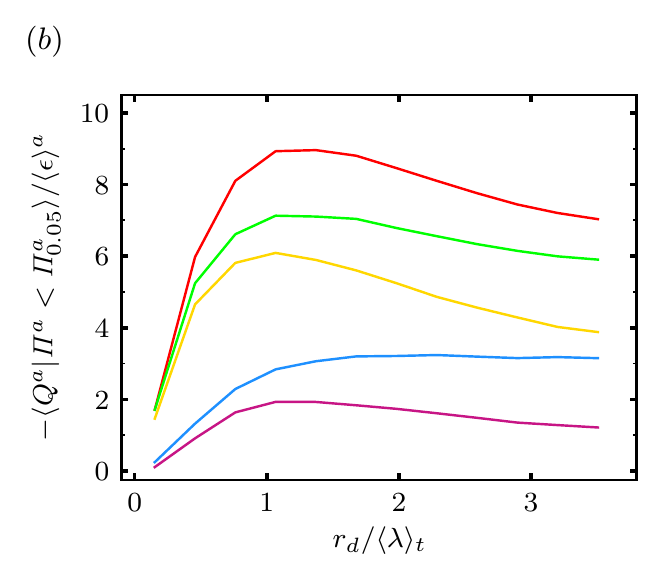}}
	\caption{Average values of $\mathit{\Pi}$ decompositions
          conditioned on ($a$) intense backward events, ($b$) intense
          forward events at $\langle \Rey_\lambda\rangle_t=112$. The
          top of $(a)$ lists the $\mathit{\Pi}$
          decompositions. (Corresponding plots for $\langle
          \Rey_\lambda\rangle_t=174$ are omitted because they are very
          similar.)}
	\label{fig:extremeMag}
\end{figure}

We finally consider in figure \ref{fig:extremeMag} the average
contributions of the various $\mathit{\Pi}$-decomposition terms
conditional on {relatively rare intense}
$\mathit{\Pi}$-events.  We calculate averages conditioned on $5\%$
most negative (forward transfer) $\mathit{\Pi}_{\overline{S}}$ events
(values of $\mathit{\Pi}_{\overline{S}}$ for which the probability
that $\mathit{\Pi}_{\overline{S}}$ is smaller than a negative value
$\mathit{\Pi}_{\overline{S}_{0.05}}$ is 0.05) and on $5\%$ most
positive $\mathit{\Pi}_{\overline{S}}$ (inverse transfer) events
(values of $\mathit{\Pi}_{\overline{S}}$ for which the probability
that $\mathit{\Pi}_{\overline{S}}$ is larger than a positive value
$\mathit{\Pi}_{\overline{S}_{0.95}}$ is also 0.05). All these averages
tend to $0$ as $r_d$ tends to $0$ below $\langle
\lambda\rangle_t$. The largest such conditional averages are those of
$\mathit{\Pi}'$ followed by those of
$\mathit{\Pi}_{\overline{S}}^{'}$. This is the forward-skewed part of
the interscale transfer (in terms of PDFs) and it is dominant at both
forward and backward {intense} interscale transfer
events. The weakest such conditional averages are those of
$\mathit{\Pi}_{\overline{I}}$ for all $r_d$ and both forward and
inverse extreme interscale transfer events. This is consistent with
our observation in section 3.4 that the unconditional fluctuation
magnitude of $\mathit{\Pi}_{\overline{I}}$ is smaller that the
unconditional fluctuation magnitudes of $\mathit{\Pi}$ followed by
those of $\mathit{\Pi}_{\overline{S}}$.

The most interesting point to notice in figure \ref{fig:extremeMag},
however, is the difference between conditional averages of
$\mathit{\Pi}_{H}^{'}$ and $\mathit{\Pi}_{I}$ when conditioned on {intense} foward or {intense} inverse
interscale transfer events. Whilst the conditional averages of these
two quantities are about the same at {intense} inverse
events, they differ substantially at forward transfer events where the
conditional average of $-\mathit{\Pi}_{H}^{'}$ is substantially higher
that the conditional average of $-\mathit{\Pi}_{I}$ except close to
the integral length-scale.
\section{Conclusions} \label{sec:conc}

The balance between space-time-averaged interscale energy transfer
rate on the one hand and space-time-averaged viscous diffusion,
turbulence dissipation rate {and power input on} the
other does not represent in any way the actual energy transfer
dynamics in statistically stationary homogeneous/periodic
turbulence. In this paper we have studied the fluctuations of
two-point acceleration terms in the NSD equation and their relation to
the various terms of the KHMH equation and we now give a
{summary of results in eleven points}.

{1.} The various corresponding terms in
{the NSD and KHMH} equations behave similarly relative
to each other because the two-point velocity difference has a similar
tendency of alignment with each one of the acceleration terms of the
NSD equation.

{2.} The terms in the two-point energy balance which
fluctuate with the highest magnitudes are $\mathcal{A}_{c}^{'}$
followed closely by the time-derivative term $\mathcal{A}_t$ and the
solenoidal interspace transfer rate $\mathcal{T}_{\overline{S}}$. The
fluctuation intensity of $\mathcal{A}_t + \mathcal{T}_{\overline{S}}$
is much reduced by comparison to both these terms (two-point sweeping)
and is comparable to the fluctuation intensity of the solenoidal
interscale transfer rate. {The solenoidal interscale
  transfer rate, which averages according to equation \eqref{khmhAvg},
  does not fluctuate with viscous diffusion and/or turbulence
  dissipation with which it is negligibly correlated at scales larger
  than $\langle \lambda \rangle_{t}$ and rather weakly correlated at
  scales smaller than $\langle \lambda \rangle_{t}$. Its fluctuation
  magnitude is also significantly larger than that of
  $\mathcal{D}_{r,\nu}$, $-\mathcal{\epsilon}$ and $\mathcal{I}$ at
  all scales. Instead, the solenoidal interscale transfer rate
  fluctuates with $\mathcal{A}_t + \mathcal{T}_{\overline{S}}$ with
  which it is extremely well correlated at length scales larger than
  $\langle \lambda \rangle_{t}$ and very significantly correlated at
  length scales smaller than $\langle \lambda \rangle_{t}$.}

{3.} In fact, for scales larger than $\langle \lambda
\rangle_{t}$, the relation
\begin{align}
	\mathcal{A}_t + \mathcal{T}_{\overline{S}} +
        \mathit{\Pi}_{\overline{S}}^{'} & \approx 0,
\end{align}
is a good approximation for most times and most locations in the
flow. $\mathcal{A}_t+\mathcal{T}_{\overline{S}}$ can be viewed as a
Lagrangian time-rate of change of $|\delta \boldsymbol{u}|^2$ moving
with $(\boldsymbol{u}^{+}+\boldsymbol{u}^{-})/2$. As more than average
$|\delta \boldsymbol{u}|^2$ is cascaded from larger to smaller scales
at a particular location $(\mathit{\Pi}_{\overline{S}}^{'}<0)$,
$\mathcal{A}_t+\mathcal{T}_{\overline{S}}$ increases; and as more than
average $|\delta \boldsymbol{u}|^2$ is inverse cascaded from smaller
to larger scales $(\mathit{\Pi}_{\overline{S}}^{'} > 0)$,
$\mathcal{A}_t+\mathcal{T}_{\overline{S}}$ decreases. The relatively
rare space-time events which do not comply with this relation are
responsible for the different skewness factors of the PDFs of
$\mathcal{A}_t + \mathcal{T}_{\overline{S}}$ (small, mostly positive,
skewness factor) and of $\mathit{\Pi}_{\overline{S}}^{'}$ (negative
skewness factor reaching increasingly large negative values with
decreasing scale).

{4.} As the length scale (i.e. two point separation
length) decreases, the correlation between $\mathcal{A}_t$ and
$-\mathcal{T}_{\overline{S}}$ increases and so do their fluctuation
magnitudes relative to the fluctuation magnitude of
$\mathit{\Pi}_{\overline{S}}^{'}$ which reaches to be an order of
magnitude smaller by comparison. In this limit, the correlation
between $\mathcal{A}_t$ and $-\mathit{\Pi}_{\overline{S}}$
decreases. At length scales smaller than $\langle \lambda \rangle_{t}$
the correlation between $\mathcal{A}_t$ and
$-\mathcal{T}_{\overline{S}}$ is extremely good indicating a tendency
towards two-point sweeping. However, the correlation between
$\mathcal{A}_t + \mathcal{T}_{\overline{S}}$ and
$\mathit{\Pi}_{\overline{S}}^{'}$ remains strong even if reduced from
its near perfect values at length scales larger than $\langle \lambda
\rangle_{t}$ and there remains a small difference of fluctuation
magnitudes between $\mathcal{A}_t$ and $\mathcal{T}_{\overline{S}}$
which is mostly related to the small fluctuation magnitude of
$\mathit{\Pi}_{\overline{S}}^{'}$. At the other end of the length
scale range, i.e. as the length scale tends towards the integral scale
and larger, the fluctuation magnitudes of $\mathcal{T}_{\overline{S}}$
and $\mathit{\Pi}_{\overline{S}}^{'}$ tend to become the same.

{5.} The irrotational part of the interscale transfer
rate has zero spatio-temporal average but is exactly equal to the
irrotational part of the interspace transfer rate and half the
two-point pressure work term in the KHMH equation. A complete dynamic
picture of the interscale transfer rate needs to also take this into
account, even though the fluctuation magnitudes of these irrotational
terms are smaller than the ones of the terms discussed in the previous
paragraph. In fact, the exact relation $\mathit{\Pi}_{\overline{I}} =
\mathcal{T}_{\overline{I}} = \frac{1}{2}\mathcal{T}_p$ explains the
significant correlation between interscale transfer rate
$\mathit{\Pi}$ and $\mathcal{T}_p$ reported by \citet{Yasuda2018}.

{6.} The increase towards small correlations at length
scales below $\langle \lambda \rangle_{t}$ between
$\mathit{\Pi}_{\overline{S}}$ and both $\mathcal{D}_{r,\nu}$ and
$-\mathcal{\epsilon}$ is accountable to the significant correlations
between these terms at these {viscous} scales when
conditioned on relatively rare {intense}
$\mathit{\Pi}_{\overline{S}}$ events, both forward cascading events
with negative values of $\mathit{\Pi}_{\overline{S}}$
{of high magnitude} and backward cascading events with
positive values of $\mathit{\Pi}_{\overline{S}}$ {of
  high magnitude}. The choice of $\mathit{\Pi}_{\overline{S}}$ to
identify {relatively rare intense} events is predicated
on the fact that the PDFs of $\mathit{\Pi}_{\overline{S}}$ are
negatively skewed similarly to the PDFs of $\mathit{\Pi}$, whereas the
PDFs of $\mathit{\Pi}_{\overline{I}}$ are not. The solenoidal part of
the interscale transfer rate derives from the integrated two-point
vorticity equation and includes non-local vortex
stretching/compression effects at all scales whereas the irrotational
part of the interscale transfer rate derives from the integrated
Poisson equation for two-point pressure fluctuations.

{7.} At these {relatively rare intense}
interscale transfer rate events, the tendency for two-point sweeping
may appear increased because of the extremely good conditional
correlation between $\mathcal{A}_t$ and $-\mathcal{T}_{\overline{S}}$
at all length-scales, however $\mathcal{A}_t$ and
$-\mathcal{T}_{\overline{S}}$ have also very significantly different
average values given the high absolute values of
$\mathit{\Pi}_{\overline{S}}$ at these {relatively}
rare interscale transfer events. This implies that there is neither
local/instantaneous sweeping nor local/instantaneous balance between
$\mathit{\Pi}_{\overline{S}}$ and $\mathcal{D}$ or
$\mathit{\Pi}_{\overline{S}}$ and $-\mathcal{\epsilon}$ at these
relatively rare intense events, a conclusion confirmed by the
observation that the conditional averages and the conditional
fluctuation magnitudes of $\mathit{\Pi}_{\overline{S}}$ are much
higher than those of $\mathcal{D}$ and $-\mathcal{\epsilon}$ in
absolute values.

{8.} Another property of these
{relatively rare intense} solenoidal interscale
transfer rate events is that the conditional averages of solenoidal
interscale and interspace transfer rates have opposite signs when
sampling on these events. There is therefore a relation between them
which may however be {concealed} by the fact that the
fluctuation magnitudes of the interspace transport rate are higher
than those of the interscale transfer rate.

{9.} We have also considered the decomposition into
homogeneous and inhomogeneous interscale transfer rates recently
introduced by \citet{AlvesPortela2020} and have studied their
fluctuations in statistically stationary homogeneous turulence. The
PDFs of the homogeneous interscale transfer rate are skewed towards
forward cascade events whereas the PDFs of the inhomogeneous
interscale transfer rate are not significantly skewed. However, the
skewness factors of the PDFs of the homogeneous interscale transfer
rate are not as high as those of both the full and the solenoidal
interscale transfer rates.  Relating to this, $\mathit{\Pi}$ is highly
correlated with $\mathit{\Pi}_{\overline{S}}$ more than with
$\mathit{\Pi}_{\overline{I}}$, $\mathit{\Pi}_{H}$ and
$\mathit{\Pi}_{I}$ with all of which $\mathit{\Pi}$ is, nevertheless,
significantly correlated.

{10.} There is an increasing correlation between
$\mathit{\Pi}_{I}$ and $-\mathit{\Pi}_{H}$ as the length-scale
decreases, in particular below $\langle \lambda \rangle_t$ where it
reaches values above 0.6. The interscale transfer of velocity
difference energy caused by local inhomogeneities in fluctuating
velocity magnitudes tends to cancel the interscale transfer of
fluctuating velocity difference energy caused by misalignments between
the two neighboring fluctuating velocities, in particular at length
scales below $\langle \lambda \rangle_t$. As a result, the fluctuation
magnitudes of $\mathit{\Pi}$ are smaller than those of both
$\mathit{\Pi}_{I}$ and $-\mathit{\Pi}_{H}$.

{11.} Finally, the decomposition $\mathit{\Pi} =
\mathit{\Pi}_{I} + \mathit{\Pi}_{H}$ can be used to physically
distinguish between {intense} forward and {intense} inverse interscale transfer events. The
averages of $\mathit{\Pi}_{H}^{'}$ and $\mathit{\Pi}_{I}$ when
conditioned on {intense} inverse interscale transfer
events are about the same, but they differ substantially when
conditioned on {intense} forward interscale transfer
events where the conditional average of $-\mathit{\Pi}_{H}^{'}$ is
substantially higher that the conditional average of
$-\mathit{\Pi}_{I}$ except close to the integral length-scale.

Future subgrid scale models for Large Eddy Simulations (LES) which are
dynamic reduced order approaches to turbulent flows and their
fluctuating large scales cannot rely on average cascade phenomenology
describing spatio-temporal averages and should benefit from detailed
descriptions of the fluctuating dynamics of interscale and interspace
energy transfers such as the one presented in this
paper. {Whilst LES models based on local equilibrium
  such as the Smagorinsky model can reproduce structure function
  exponents and correlations between velocity increments and
  subgrid-scale energy transfers as shown by \citet{Linkmann2018},
  \citet{Dairay2017} have found that the Smagorinsky model is unable
  to ensure regularisation and is very sensitive to numerical
  errors. Furthermore, the recent review by \citet{Moser2021} makes it
  clear that the need for new subgrid models which can faithfully
  operate with coarse resolutions remains unanswered.} The results in
the present paper suggest that LES models based on local equilibrium
(e.g. the Smagorinsky model) {cannot be fully suitable
  for calculating fluctuations in subgrid stresses, a weakness which
  may become increasingly evident with coarser resolution}. On the
other hand, the good correlations between subgrid stresses from
similarity models \citep{Bardina1980,Cimarelli2019} and subgrid
stresses from DNS suggest that these models might indeed approximate
(unawarely) at least some of the cascade dynamics reported in this
paper, for example the fact that $\mathcal{A}_t +
\mathcal{T}_{\overline{S}} + \mathit{\Pi}_{\overline{S}}^{'} \approx
0$ holds in most of the flow most of the time. {This
  relation incorporates both forward and backward interscale
  transfers, yet a recent work by \citet{Vela-Martin2022} argues that
  backscatter represents spatial fluxes and can therefore be
  ignored. It is not yet clear how such a claim can be understood in
  the context of the present paper's results. Some new questions are
  therefore now raised concerning LES subgrid stess modeling which
  also need to be addressed in future work.} \\

\backsection[Acknowledgements]{We thank Professor S. Goto for
allowing us to use his parallelised pseudo-spectral DNS code for
periodic turbulence.}

\backsection[Funding]{HSL and JCV acknowledge support from EPSRC
award number EP/L016230/1. JCV also acknowledges the Chair of
Excellence CoPreFlo funded by I-SITE-ULNE (grant
no. R-TALENT-19-001-VASSILICOS); MEL (grant no. CONVENTION-219-ESR-06)
and Region Hauts de France (grant no. 20003862).}

\backsection[Declaration of Interests]{The authors report no
conflict of interest.}

\appendix				

\section{The Helmholtz decomposition in Fourier space} \label{sec:appA}

{In this appendix we list the Helmholtz decomposition for periodic fields and note how this decomposition relates to the more general solution to the Helmholtz decomposition in the case of incompressible fields and fields which can be written as gradients of scalar fields.}\par 

Let $\boldsymbol{q}(\boldsymbol{x},t)$ be a periodic,
twice continuously differentiable $3$D vector field with the Helmholtz
decomposition
$\boldsymbol{q}(\boldsymbol{x},t)=\boldsymbol{q}_I(\boldsymbol{x},t)+\boldsymbol{q}_S(\boldsymbol{x},t)$,
where $\boldsymbol{q}_I(\boldsymbol{x},t)=-\nabla_{\boldsymbol{x}}
\phi(\boldsymbol{x},t)$, $\boldsymbol{q}_S(\boldsymbol{x},t)=
\nabla_{\boldsymbol{x}} \times \boldsymbol{B}(\boldsymbol{x},t)$. The
scalar and vector potentials $\phi$ and $\boldsymbol{B}$ are unique
within constants when $\nabla_{\boldsymbol{x}} \cdot \boldsymbol{q}$
and $\nabla_{\boldsymbol{x}} \times \boldsymbol{q}$ are known in the
domain and $\boldsymbol{q}$ is known at the boundary
\citep{Bhatia2013}. $\boldsymbol{q}(\boldsymbol{x},t)$ has the
Fourier representation
$\widehat{\boldsymbol{q}}(\boldsymbol{k},t)$, which can be decomposed
into a component parallel to $\boldsymbol{k}$ (the longitudinal
$\widehat{\boldsymbol{q}}^{L}$) and transverse to $\boldsymbol{k}$
(the transverse $\widehat{\boldsymbol{q}}^{T}$)
\begin{equation}
	  \widehat{\boldsymbol{q}}^{L}(\boldsymbol{k},t) = \frac{\boldsymbol{k} [\widehat{\boldsymbol{q}}(\boldsymbol{k},t)\cdot \boldsymbol{k}]}{k^2} \ , \quad \widehat{\boldsymbol{q}}^{T}(\boldsymbol{k},t) = \widehat{\boldsymbol{q}}(\boldsymbol{k},t)-\widehat{\boldsymbol{q}}^{L}(\boldsymbol{k},t) \ . \label{decompQ}
\end{equation}

\noindent {It can be easily shown (see e.g. \citet{Stewart2012}) that the irrotational part of $\boldsymbol{q}$ equals its longitudinal part $\boldsymbol{q}_I=\boldsymbol{q}^{L}$ and that the solenoidal part of $\boldsymbol{q}$ equals its transverse part $\boldsymbol{q}_S=\boldsymbol{q}^{T}$. Hence, \eqref{decompQ} provides the Fourier representation of the Helmholtz decomposition of $\boldsymbol{q}$.}

{The Helmholtz decomposition can also be written for very general boundary conditions as \citep{Sprossig2010}}
\begin{align}
	\boldsymbol{q}_{IV}(\boldsymbol{x},t) &= \frac{1}{4\upi} \int_V d\boldsymbol{y} \frac{\boldsymbol{x}-\boldsymbol{y}}{|\boldsymbol{x}-\boldsymbol{y}|^3} [\nabla_{\boldsymbol{y}} \cdot \boldsymbol{q}(\boldsymbol{y},t)] ,  \label{irotV} \\
	\boldsymbol{q}_{IB}(\boldsymbol{x},t) &= -\frac{1}{4\upi}\int_S dS_{\boldsymbol{y}}  \frac{\boldsymbol{x}-\boldsymbol{y}}{|\boldsymbol{x}-\boldsymbol{y}|^3}  [\widehat{\boldsymbol{n}}_{\boldsymbol{y}} \cdot \boldsymbol{q}(\boldsymbol{y},t)], \label{irotB} \\
	\boldsymbol{q}_{SV}(\boldsymbol{x},t) &= -\frac{1}{4\upi} \int_V d\boldsymbol{y} \frac{\boldsymbol{x}-\boldsymbol{y}}{|\boldsymbol{x}-\boldsymbol{y}|^3} \times [\nabla_{\boldsymbol{y}} \times \boldsymbol{q}(\boldsymbol{y},t)], \label{solV} \\
	\boldsymbol{q}_{SB}(\boldsymbol{x},t) &= \frac{1}{4\upi}\int_S dS_{\boldsymbol{y}}  \frac{\boldsymbol{x}-\boldsymbol{y}}{|\boldsymbol{x}-\boldsymbol{y}|^3} \times [\widehat{\boldsymbol{n}}_{\boldsymbol{y}} \times \boldsymbol{q}(\boldsymbol{y},t)]. \label{solB}
\end{align}
where $\boldsymbol{q}_I=\boldsymbol{q}_{IV}+\boldsymbol{q}_{IB}$, $\boldsymbol{q}_S=\boldsymbol{q}_{SV}+\boldsymbol{q}_{SB}$ and $\widehat{\boldsymbol{n}}_{\boldsymbol{y}}$ denotes the unit surface normal at $\boldsymbol{y}$ and $dS_{\boldsymbol{y}}$ is the differential surface element at $\boldsymbol{y}$. For periodic vector fields $\boldsymbol{q}(\boldsymbol{x},t)$ that are incompressible or that can be written as the gradient of a scalar, this solution simplifies. In the case of a field $\boldsymbol{q}(\boldsymbol{x},t)$ which is incompressible $\nabla_{\boldsymbol{x}} \cdot \boldsymbol{q}(\boldsymbol{x},t)=0$, it can be shown that $ \widehat{\boldsymbol{q}}(\boldsymbol{k},t) \cdot \boldsymbol{k}=0$ for every $\boldsymbol{k}$ \citep{Pope2000}. By inspection of \eqref{decompQ}, it is clear that this condition yields $\widehat{\boldsymbol{q}}^{L}(\boldsymbol{k},t)=0$ for every $\boldsymbol{k}$ such that $\widehat{\boldsymbol{q}}(\boldsymbol{k},t) = \widehat{\boldsymbol{q}}(\boldsymbol{k},t)^{T}$. By applying the Fourier transform to this relation and apply $\boldsymbol{q}^{T}(\boldsymbol{x},t)=\boldsymbol{q}_S(\boldsymbol{x},t)$ from above, we have that $\boldsymbol{q}(\boldsymbol{x},t)=\boldsymbol{q}_S(\boldsymbol{x},t)$ for incompressible periodic vector fields. In the case of $\boldsymbol{q}(\boldsymbol{x},t)=\nabla_{\boldsymbol{x}} \psi(\boldsymbol{x,t})$, where $\psi(\boldsymbol{x},t)$ is some scalar field, it can be shown that $\widehat{\boldsymbol{q}}(\boldsymbol{k},t)=i\boldsymbol{k}\widehat{\psi}(\boldsymbol{k},t)$ \citep{Pope2000}. If we insert this expression into the definition of $\widehat{\boldsymbol{q}}^{L}(\boldsymbol{k},t)$, it follows that $\widehat{\boldsymbol{q}}(\boldsymbol{k},t)=\widehat{\boldsymbol{q}}^{L}(\boldsymbol{k},t)$, which implies that $\boldsymbol{q}(\boldsymbol{x},t) = \boldsymbol{q}_{I}(\boldsymbol{x},t)$. If these properties are combined with equations \eqref{irotV}-\eqref{solB}, we have that a periodic incompressible vector field will have $\boldsymbol{q}_{IB}=\boldsymbol{q}_{IV}=0$ and that a periodic vector field that can be written as a gradient of a scalar field has $ \boldsymbol{q}_{SB}= \boldsymbol{q}_{SV}=0$. \par

\section{Irrotational and solenoidal NSD tranport terms in Fourier space} \label{sec:appB}

We start this appendix with demonstrating that $\delta \boldsymbol{q}_{I}=\delta \boldsymbol{q}_{\overline{I}}$ and $\delta \boldsymbol{q}_{S}=\delta \boldsymbol{q}_{\overline{S}}$ for a periodic vector field $\boldsymbol{q}$ {(see the second pararaph of section \ref{subsec:nsdTwo})}. The field $\boldsymbol{q}$ has the Fourier representation
\begin{equation}
	\boldsymbol{q}(\boldsymbol{x},t) = \sum_{\boldsymbol{k}} \widehat{\boldsymbol{q}}(\boldsymbol{k},t) e^{i\boldsymbol{k}\cdot \boldsymbol{x}},
\end{equation}
\noindent with the shifted fields
\begin{align}
	\boldsymbol{q}^{+}(\boldsymbol{x},\boldsymbol{r},t)=\boldsymbol{q}(\boldsymbol{x}+\boldsymbol{r}/2,t) &= \sum_{\boldsymbol{k}} \widehat{\boldsymbol{q}}(\boldsymbol{k},t) e^{i\boldsymbol{k}\cdot (\boldsymbol{x}+\boldsymbol{r}/2)}, \\
	\boldsymbol{q}^{-}(\boldsymbol{x},\boldsymbol{r},t)=\boldsymbol{q}(\boldsymbol{x}-\boldsymbol{r}/2,t) &= \sum_{\boldsymbol{k}} \widehat{\boldsymbol{q}}(\boldsymbol{k},t) e^{i\boldsymbol{k}\cdot (\boldsymbol{x}-\boldsymbol{r}/2)},
\end{align}
which have the Fourier coefficients
\begin{align}
	\widehat{\boldsymbol{q}^{+}}(\boldsymbol{k},\boldsymbol{r},t)&= \widehat{\boldsymbol{q}}(\boldsymbol{k},t)e^{i\boldsymbol{k}\cdot \boldsymbol{r}/2}, \\
	\widehat{\boldsymbol{q}^{-}}(\boldsymbol{k},\boldsymbol{r},t)&= \widehat{\boldsymbol{q}}(\boldsymbol{k},t)e^{-i\boldsymbol{k}\cdot \boldsymbol{r}/2}.
\end{align}
From the definition of the irrotational part of a vector field in \eqref{decompQ}, it follows
\begin{align}
	\delta \boldsymbol{q}_I(\boldsymbol{x},\boldsymbol{r},t) &= \boldsymbol{q}_{I}^{+}(\boldsymbol{x},\boldsymbol{r},t)-\boldsymbol{q}_{I}^{-}(\boldsymbol{x},\boldsymbol{r},t) , \\
									&= \sum_{\boldsymbol{k}} [\widehat{\boldsymbol{q}^{+}_I}(\boldsymbol{k},\boldsymbol{r},t)-\widehat{\boldsymbol{q}^{-}_I}(\boldsymbol{k},\boldsymbol{r},t)] e^{i\boldsymbol{k}\cdot \boldsymbol{x}} , \\
									&= \sum_{\boldsymbol{k}} \frac{\boldsymbol{k}}{k^2} [\widehat{\boldsymbol{q}}(\boldsymbol{k},t)\cdot\boldsymbol{k}](e^{i\boldsymbol{k}\cdot\boldsymbol{r}/2}-e^{-i\boldsymbol{k}\cdot\boldsymbol{r}/2})e^{i\boldsymbol{k}\cdot \boldsymbol{x}} , \\
									&= \sum_{\boldsymbol{k}} \frac{\boldsymbol{k}}{k^2} [\widehat{\boldsymbol{q}}(\boldsymbol{k},t)\cdot\boldsymbol{k}]2i \sin (\boldsymbol{k}\cdot \boldsymbol{r}/2) e^{i\boldsymbol{k}\cdot \boldsymbol{x}}.
\end{align}

\noindent Similarly, we can write
\begin{align}
	\delta \boldsymbol{q}(\boldsymbol{x},\boldsymbol{r},t) &= \boldsymbol{q}^{+}(\boldsymbol{x},\boldsymbol{r},t)-\boldsymbol{q}^{-}(\boldsymbol{x},\boldsymbol{r},t), \\
								  &= \sum_{\boldsymbol{k}} \widehat{\boldsymbol{q}}(\boldsymbol{k},t) 2i \sin (\boldsymbol{k}\cdot \boldsymbol{r}/2) e^{i\boldsymbol{k}\cdot\boldsymbol{x}},
\end{align}
\noindent and then calculate its irrotational centroid part
\begin{equation}
	\delta \boldsymbol{q}_{\overline{I}}(\boldsymbol{x},\boldsymbol{r},t) = \sum_{\boldsymbol{k}} \frac{\boldsymbol{k}}{k^2}[\widehat{\boldsymbol{q}}(\boldsymbol{k},t) \cdot \boldsymbol{k}] 2i \sin (\boldsymbol{k}\cdot \boldsymbol{r}/2) e^{i\boldsymbol{k}\cdot\boldsymbol{x}},
\end{equation}
\noindent which shows that $\delta \boldsymbol{q}_I(\boldsymbol{x},\boldsymbol{r},t) = \delta \boldsymbol{q}_{\overline{I}}(\boldsymbol{x},\boldsymbol{r},t)$. By combining this with $\delta \boldsymbol{q} = \delta \boldsymbol{q}_{I}+\delta \boldsymbol{q}_{S}=\delta \boldsymbol{q}_{\overline{I}}+\delta \boldsymbol{q}_{\overline{S}}$, we have also $\delta \boldsymbol{q}_S(\boldsymbol{x},\boldsymbol{r},t) = \delta \boldsymbol{q}_{\overline{S}}(\boldsymbol{x},\boldsymbol{r},t)$, which is what we wanted to show.

Next we demonstrate that $\boldsymbol{a}_{\mathit{\Pi}_{\overline{I}}}(\boldsymbol{k},\boldsymbol{r},t) = \boldsymbol{a}_{\mathcal{T}_{\overline{I}}}(\boldsymbol{x},\boldsymbol{r},t)$ in {homogeneous/periodic turbulence}. We list the following expressions for the vectors and tensors related to these two terms
\begin{align}
	\widehat{\delta u_j}(\boldsymbol{k},\boldsymbol{r},t) &= 2i \sin (\boldsymbol{k} \cdot \boldsymbol{r}/2) \widehat{u}_j(\boldsymbol{k},t), \label{FourTransport1}  \\
	 \widehat{(u_j^{+}+u_j^{-})/2}(\boldsymbol{k},\boldsymbol{r},t)&= \cos (\boldsymbol{k} \cdot \boldsymbol{r}/2) \widehat{u}_j(\boldsymbol{k},t), \label{FourTransport2} \\
	\widehat{\frac{\partial \delta u_i}{\partial r_j}}(\boldsymbol{k},\boldsymbol{r},t) &= ik_j  \cos (\boldsymbol{k} \cdot \boldsymbol{r}/2) \widehat{u}_i(\boldsymbol{k},t), \label{FourTransport3} \\ \widehat{\frac{\partial \delta u_i}{\partial x_j}}(\boldsymbol{k},\boldsymbol{r},t) &= -2k_j \sin (\boldsymbol{k}\cdot \boldsymbol{r}/2) \widehat{u}_i(\boldsymbol{k},t)  \label{FourTransport4}.
\end{align}

\noindent By use of these equations, we have that the Fourier coefficients of the transport terms read
\begin{align}
	\widehat{\boldsymbol{a}_{\mathcal{T}}}(\boldsymbol{k},\boldsymbol{r},t) = \sum_{\boldsymbol{k}=\boldsymbol{k}^{'}+\boldsymbol{k}^{''}} - 2 \sin (\boldsymbol{k}^{''}\cdot \boldsymbol{r}/2) \cos (\boldsymbol{k}^{'}\cdot \boldsymbol{r}/2) \widehat{u}_j(\boldsymbol{k}^{'})k_j^{''}\widehat{\boldsymbol{u}}(\boldsymbol{k}^{''}) , \\
	\widehat{\boldsymbol{a}_{\mathit{\Pi}}}(\boldsymbol{k},\boldsymbol{r},t) = \sum_{\boldsymbol{k}=\boldsymbol{k}^{'}+\boldsymbol{k}^{''}} - 2 \sin (\boldsymbol{k}^{'}\cdot \boldsymbol{r}/2) \cos (\boldsymbol{k}^{''}\cdot \boldsymbol{r}/2) \widehat{u}_j(\boldsymbol{k}^{'})k_j^{''}\widehat{\boldsymbol{u}}(\boldsymbol{k}^{''}) .
\end{align}
Their irrotational parts are given per \eqref{decompQ}
\begin{align}
	\widehat{\boldsymbol{a}_{\mathcal{T}_{\overline{I}}}}(\boldsymbol{k},\boldsymbol{r},t) = -\frac{\boldsymbol{k}}{k^2} \sum_{\boldsymbol{k}=\boldsymbol{k}^{'}+\boldsymbol{k}^{''}}  2 \sin (\boldsymbol{k}^{''}\cdot \boldsymbol{r}/2) \cos (\boldsymbol{k}^{'}\cdot \boldsymbol{r}/2) \widehat{u}_j(\boldsymbol{k}^{'})k_j^{''}\widehat{u}_l(\boldsymbol{k}^{''})k^{'}_l , \\
	\widehat{\boldsymbol{a}_{\mathit{\Pi}_{\overline{I}}}}(\boldsymbol{k},\boldsymbol{r},t) = -\frac{\boldsymbol{k}}{k^2} \sum_{\boldsymbol{k}=\boldsymbol{k}^{'}+\boldsymbol{k}^{''}}  2 \sin (\boldsymbol{k}^{'}\cdot \boldsymbol{r}/2) \cos (\boldsymbol{k}^{''}\cdot \boldsymbol{r}/2) \widehat{u}_j(\boldsymbol{k}^{'})k_j^{''}\widehat{u}_l(\boldsymbol{k}^{''})k^{'}_l .
\end{align}
If we employ the trigonometric identity $\sin x \cos y = \frac{1}{2}[\sin (x+y)+\sin (x-y)]$, we get
\begin{align}
	\widehat{\boldsymbol{a}_{\mathcal{T}_{\overline{I}}}}(\boldsymbol{k},\boldsymbol{r},t) = -\frac{\boldsymbol{k}}{k^2} \sum_{\boldsymbol{k}=\boldsymbol{k}^{'}+\boldsymbol{k}^{''}} [\sin (\boldsymbol{k}\cdot \boldsymbol{r}/2)+ \sin (\boldsymbol{k}^{''}\cdot \boldsymbol{r}/2-\boldsymbol{k}^{'}\cdot \boldsymbol{r}/2)] \widehat{u}_j(\boldsymbol{k}^{'})k_j^{''}\widehat{u}_l(\boldsymbol{k}^{''})k^{'}_l , \\
	\widehat{\boldsymbol{a}_{\mathit{\Pi}_{\overline{I}}}}(\boldsymbol{k},\boldsymbol{r},t) = -\frac{\boldsymbol{k}}{k^2} \sum_{\boldsymbol{k}=\boldsymbol{k}^{'}+\boldsymbol{k}^{''}}  [\sin (\boldsymbol{k}\cdot \boldsymbol{r}/2)- \sin (\boldsymbol{k}^{''}\cdot \boldsymbol{r}/2-\boldsymbol{k}^{'}\cdot \boldsymbol{r}/2)] \widehat{u}_j(\boldsymbol{k}^{'})k_j^{''}\widehat{u}_l(\boldsymbol{k}^{''})k^{'}_l .
\end{align}
Consider the term $\sin (\boldsymbol{k}^{''}\cdot \boldsymbol{r}/2-\boldsymbol{k}^{'}\cdot \boldsymbol{r}/2) \widehat{u}_j(\boldsymbol{k}^{'})k_j^{''}\widehat{u}_l(\boldsymbol{k}^{''})k^{'}_l$. If one adds this term with the wavenumber triad $\boldsymbol{k}^{'}=\boldsymbol{k}^{a}$ and $\boldsymbol{k}^{''}=\boldsymbol{k}^{b} \neq \boldsymbol{k}^{a}$ with the same term with the wavenumber triad $\boldsymbol{k}^{'}=\boldsymbol{k}^{b}$ and $\boldsymbol{k}^{''}=\boldsymbol{k}^{a}$ the result is zero. Furthermore, in the case of $\boldsymbol{k}^{a}=\boldsymbol{k}^{b}$ this term is zero per incompressibility. This yields that this term does not contribute instantaneously in the above expressions such that we attain {the final result (see section \ref{subsec:nsdTwo} and equation \eqref{transportPressure})}
\begin{equation}
	\widehat{\boldsymbol{a}_{\mathcal{T}_{\overline{I}}}}(\boldsymbol{k},\boldsymbol{r},t) = \widehat{\boldsymbol{a}_{\mathit{\Pi}_{\overline{I}}}}(\boldsymbol{k},\boldsymbol{r},t) = -\frac{\boldsymbol{k}}{k^2}  \sin (\boldsymbol{k}\cdot \boldsymbol{r}/2) \sum_{\boldsymbol{k}=\boldsymbol{k}^{'}+\boldsymbol{k}^{''}}  \widehat{u}_j(\boldsymbol{k}^{'})k_j^{''}\widehat{u}_l(\boldsymbol{k}^{''})k^{'}_l.
\end{equation}

\section{Irrotational and solenoidal dynamics in non-homogeneous turbulence} \label{sec:appC}

Here we deduce the generalised Tsinober equations and the irrotational
and solenoidal NSD and KHMH equations applicable to non-homogeneous
turbulence. Consider the twice continously differentiable vector field
vector field $\boldsymbol{q}(\boldsymbol{x},t)$ defined on a domain $V
\subseteq \mathbb{R}^3$ with the bounding surface $S$. This field can
be uniquely decomposed into the irrotational and solenoidal vector
fields
\begin{equation}
	\boldsymbol{q}(\boldsymbol{x},t) = \boldsymbol{q}_I(\boldsymbol{x},t)+\boldsymbol{q}_S(\boldsymbol{x},t) = -\nabla_{\boldsymbol{x}} \boldsymbol{\phi}(\boldsymbol{x},t) + \nabla_{\boldsymbol{x}} \times \boldsymbol{B}(\boldsymbol{x},t), \label{helm0}
\end{equation}
\noindent The solution to this problem under very general conditions \citep{Sprossig2010} is $\boldsymbol{q}_I=\boldsymbol{q}_{IV}+\boldsymbol{q}_{IB}$ and $\boldsymbol{q}_S=\boldsymbol{q}_{SV}+\boldsymbol{q}_{SB}$, where the solenoidal and irrotational volume and boundary terms are given in equations \eqref{irotV}-\eqref{solB}.

Consider an incompressible fluid that satisfies the incompressible vorticity equation
\begin{equation}
	\nabla_{\boldsymbol{y}} \times \big( \frac{\partial \boldsymbol{u}}{\partial  t} + \boldsymbol{u}\cdot \nabla_{\boldsymbol{y}}\boldsymbol{u} -\nu \nabla_{\boldsymbol{y}}^2 \boldsymbol{u} -  \boldsymbol{f} \big) = 0  .
\end{equation}
By comparing this equation with \eqref{solV}, it is clear that the vorticity equation can be used to derive an evolution equation for the solenoidal volume parts of the NS terms. We can apply the following operator to this equation
\begin{equation}
	- \frac{1}{4\upi} \int_V d\boldsymbol{y} \frac{\boldsymbol{x}-\boldsymbol{y}}{|\boldsymbol{x}-\boldsymbol{y}|^3} \times \Big[ \nabla_{\boldsymbol{y}} \times \big( \frac{\partial \boldsymbol{u}}{\partial  t} + (\boldsymbol{u}\cdot \nabla_{\boldsymbol{y}})\boldsymbol{u} -\nu \nabla_{\boldsymbol{y}}^2 \boldsymbol{u} -  \boldsymbol{f} \big) \Big] = 0  ,
\end{equation}
\noindent and use \eqref{solV} to rewrite this equation as
\begin{equation}
	(\frac{\partial \boldsymbol{u}}{\partial  t})_{SV}+ (\boldsymbol{u}\cdot \nabla_{\boldsymbol{x}}\boldsymbol{u})_{SV}= (\nu \nabla_{\boldsymbol{x}}^2 \boldsymbol{u})_{SV}+\boldsymbol{f}_{SV} . \label{solVolEq}
\end{equation}
\noindent We can in a similar manner obtain the evolution equation for the irrotational volume NS terms from the Poisson equation for pressure
\begin{equation}
	\frac{1}{4\upi} \int_V d\boldsymbol{y} \frac{\boldsymbol{x}-\boldsymbol{y}}{|\boldsymbol{x}-\boldsymbol{y}|^3} \Big[ \nabla_{\boldsymbol{y}} \cdot \big( \boldsymbol{u}\cdot \nabla_{\boldsymbol{y}}\boldsymbol{u} +\frac{1}{\rho}\nabla_{\boldsymbol{y}}p - \boldsymbol{f} \big) \Big] = 0 ,
\end{equation}
which yields
\begin{equation}
	(\boldsymbol{u}\cdot \nabla_{\boldsymbol{x}}\boldsymbol{u})_{IV} = (-\frac{1}{\rho}\nabla_{\boldsymbol{x}}p)_{IV} + \boldsymbol{f}_{IV} , \label{irotVolEq}
\end{equation}

\noindent The equations \eqref{solVolEq} and \eqref{irotVolEq} state that in all incompressible turbulent flows the solenoidal accelerations from volume contributions balance with solenoidal forces from volume contributions and irrotational accelerations from volume contributions balance with irrotational forces from volume contributions. The former can be viewed as an integrated vorticity equation which dictates a part of the solenoidal NS dynamics, while the latter equation as an integrated pressure Poisson equation which dictates a part of the irrotational NS dynamics. Due to the non-local character of the solenoidal and irrotational volume terms, we reformulate these equations in terms of full NS term minus boundary terms. E.g., for the time-derivative $(\frac{\partial \boldsymbol{u}}{\partial  t})_{SV}=\frac{\partial \boldsymbol{u}}{\partial  t}-(\frac{\partial \boldsymbol{u}}{\partial  t})_{IB}-(\frac{\partial \boldsymbol{u}}{\partial  t})_{SB}$. The irrotational volume component (see \eqref{irotV}) involves an integral of the divergence of the respective term ($\nabla_{\boldsymbol{y}} \cdot \boldsymbol{q}(\boldsymbol{y})$). Thus, due to incompressibility, the time derivative and viscous terms have zero volume irrotational components, $(\frac{\partial \boldsymbol{u}}{\partial t})_{IV}=(\nu \nabla_{\boldsymbol{x}}^2 \boldsymbol{u})_{IV}=0$. The solenoidal volume component (see \eqref{solV}) involves an integral of the curl of the respective term, and as the curl of the pressure gradient equals zero, this term will have a zero solenoidal volume component, $(-\frac{1}{\rho}\nabla_{\boldsymbol{x}}p)_{SV}=0$. We rewrite the solenoidal volume terms in equation \eqref{solVolEq} in terms of combinations of full terms and boundary terms to obtain
\begin{multline}
	\frac{\partial \boldsymbol{u}}{\partial  t}+((\boldsymbol{u}\cdot \nabla_{\boldsymbol{x}})\boldsymbol{u})_S = \nu \nabla_{\boldsymbol{x}}^2 \boldsymbol{u}+\boldsymbol{f}_{S} + \\
	(\frac{\partial \boldsymbol{u}}{\partial  t})_{IB} - (\nu \nabla_{\boldsymbol{x}}^2 \boldsymbol{u})_{IB}+(\frac{\partial \boldsymbol{u}}{\partial  t})_{SB}+((\boldsymbol{u}\cdot \nabla_{\boldsymbol{x}})\boldsymbol{u})_{SB} - (\nu \nabla_{\boldsymbol{x}}^2 \boldsymbol{u})_{SB}-\boldsymbol{f}_{SB}  \ ,
\end{multline}
\noindent where the sum of the four rightmost terms on the RHS equals $(-\frac{1}{\rho}\nabla_{\boldsymbol{x}}p)_{SB}$ as the NS equations are satisfied at the boundary. By using this simplification and writing out all the boundary terms, we arrive at
\begin{multline}
\frac{\partial \boldsymbol{u}}{\partial t}+((\boldsymbol{u}\cdot \nabla_{\boldsymbol{x}})\boldsymbol{u})_{S} = \nu\nabla_{\boldsymbol{x}}^2 \boldsymbol{u} + \boldsymbol{f}_{S} \\ -\frac{1}{4\upi} \int_S dS_{\boldsymbol{y}} \frac{\boldsymbol{x}-\boldsymbol{y}}{|\boldsymbol{x}-\boldsymbol{y}|^3}[\widehat{\boldsymbol{n}}_{\boldsymbol{y}} \cdot (\frac{\partial \boldsymbol{u}}{\partial t}-\nu \nabla_{\boldsymbol{y}}^{2}\boldsymbol{u})] - \frac{1}{4\upi} \int_S dS_{\boldsymbol{y}}  \frac{\boldsymbol{x}-\boldsymbol{y}}{|\boldsymbol{x}-\boldsymbol{y}|^3} \times [\widehat{\boldsymbol{n}}_{\boldsymbol{y}} \times \nabla_{\boldsymbol{y}} \frac{1}{\rho}p] . \label{rot0}
\end{multline}

\noindent By rewriting the irrotational volume components in equation \eqref{irotVolEq} in terms of the full terms and the boundary terms, we have
\begin{equation}
	((\boldsymbol{u}\cdot \nabla_{\boldsymbol{x}})\boldsymbol{u})_{I} = -\frac{1}{\rho}\nabla_{\boldsymbol{x}}p + \boldsymbol{f}_{I} + ((\boldsymbol{u}\cdot \nabla_{\boldsymbol{x}})\boldsymbol{u})_{IB} - (-\frac{1}{\rho}\nabla_{\boldsymbol{x}}p)_{IB} -\boldsymbol{f}_{IB} - (-\frac{1}{\rho}\nabla_{\boldsymbol{x}}p)_{SB}  ,
\end{equation}
\noindent where the sum of the irrotational boundary terms equals $-(\frac{\partial \boldsymbol{u}}{\partial t})_{IB}+(\nu\nabla_{\boldsymbol{x}}^2 \boldsymbol{u})_{IB}$ by the NS equations at the boundary. If we use this relation and write out all boundary terms, we have
\begin{multline}
	((\boldsymbol{u}\cdot \nabla_{\boldsymbol{x}})\boldsymbol{u})_{I} = -\frac{1}{\rho}\nabla_{\boldsymbol{x}} p+\boldsymbol{f}_{I} \\
	+\frac{1}{4\upi} \int_S dS_{\boldsymbol{y}} \frac{\boldsymbol{x}-\boldsymbol{y}}{|\boldsymbol{x}-\boldsymbol{y}|^3} [\widehat{\boldsymbol{n}}_{\boldsymbol{y}} \cdot (\frac{\partial \boldsymbol{u}}{\partial t}-\nu \nabla_{\boldsymbol{y}}^{2}\boldsymbol{u})]+ \frac{1}{4\upi} \int_S dS_{\boldsymbol{y}}  \frac{\boldsymbol{x}-\boldsymbol{y}}{|\boldsymbol{x}-\boldsymbol{y}|^3} \times [\widehat{\boldsymbol{n}}_{\boldsymbol{y}} \times \nabla_{\boldsymbol{y}} \frac{1}{\rho}p] . \label{poisson0}
\end{multline}

The equations \eqref{rot0} and \eqref{poisson0} are generalisations of
equations \eqref{rotHIT1}-\eqref{poissonHIT1} for homogeneous/periodic
turbulence and these equations are valid for all incompressible
turbulent flows. The difference from homogeneous/periodic turbulence
is the collection of boundary terms
\begin{align}
\boldsymbol{R}(\boldsymbol{x},t) &\equiv \frac{1}{4\upi} \int_S dS_{\boldsymbol{y}} \frac{\boldsymbol{x}-\boldsymbol{y}}{|\boldsymbol{x}-\boldsymbol{y}|^3} [\boldsymbol{n}_{\boldsymbol{y}} \cdot (\frac{\partial \boldsymbol{u}}{\partial t}-\nu \nabla_{\boldsymbol{y}}^{2}\boldsymbol{u})]+ \frac{1}{4\upi} \int_S dS_{\boldsymbol{y}}  \frac{\boldsymbol{x}-\boldsymbol{y}}{|\boldsymbol{x}-\boldsymbol{y}|^3} \times [\boldsymbol{n}_{\boldsymbol{y}} \times \nabla_{\boldsymbol{y}} \frac{1}{\rho}p] , \label{boundCol} \\
&= - (\boldsymbol{a}_l)_{IB} + (\boldsymbol{a}_{\nu})_{IB} - (\boldsymbol{a}_p)_{SB} ,
\end{align}

which yields the final expressions for the general irrotational and
solenoidal NS equations
\begin{align}
	\frac{\partial \boldsymbol{u}}{\partial t}+((\boldsymbol{u}\cdot \nabla_{\boldsymbol{x}})\boldsymbol{u})_{S} &= \nu\nabla_{\boldsymbol{x}}^2 \boldsymbol{u} + \boldsymbol{f}_{S} -\boldsymbol{R}(\boldsymbol{x},t) \label{nsSolFinal}\\
	((\boldsymbol{u}\cdot \nabla_{\boldsymbol{x}})\boldsymbol{u})_{I} &= -\frac{1}{\rho}\nabla_{\boldsymbol{x}} p+\boldsymbol{f}_{I} +\boldsymbol{R}(\boldsymbol{x},t) \label{nsIrrFinal}
\end{align}

In homogeneous/periodic turbulence all the boundary terms in
$\boldsymbol{R}(\boldsymbol{x},t)$ equal zero individually (see the
last parapgraph of \ref{sec:appA}), such that we recover equations
\eqref{rotHIT1}-\eqref{poissonHIT1}. In general, the boundary terms
will be non-zero and differ in different flows. E.g., at a solid wall
the boundary term from the time-derivative will vanish because of
no-slip and the NS equations at the wall can be used to rewrite the
boundary terms as a non-local function of the pressure gradient only.

The NSD irrotational and solenoidal equations in general turbulent flows are obtained by subtracting the solenoidal and irrotational NS equations \eqref{nsSolFinal}-\eqref{nsIrrFinal} at $\boldsymbol{x}-\boldsymbol{r}/2$ from the same equations at $\boldsymbol{x}+\boldsymbol{r}/2$
\begin{align}
	\frac{\partial \delta \boldsymbol{u}}{\partial t}+\delta \boldsymbol{a}_{c_S} &= \delta \boldsymbol{a}_\nu +\delta \boldsymbol{f}_S  - \delta \boldsymbol{R} ,  \label{NSD12} \\
	\delta \boldsymbol{a}_{c_I} &=  -\frac{1}{\rho}\nabla_{\boldsymbol{x}}\delta p + \delta \boldsymbol{f}_I + \delta \boldsymbol{R} , \label{NSD13}
\end{align}

\noindent The rephrasing of the irrotational and solenoidal NSD
equations in terms of the interscale and interspace transport terms
can also be performed for non-homogeneous turbulence. We derive the
centroid irrotational and solenoidal NSD equations similarly as for
the NS irrotational and solenoidal equations by starting with the NSD
equation \eqref{NSD1}. This yields the equations
\begin{align}
\delta \boldsymbol{a}_{l} +
\boldsymbol{a}_{\mathcal{T}_{\overline{S}}} +
\boldsymbol{a}_{\mathit{\Pi}_{\overline{S}}} &= \delta
\boldsymbol{a}_{\nu} + \delta \boldsymbol{f}_{\overline{S}} -
\overline{\boldsymbol{R}} , \label{TsolNSD}
\\ \boldsymbol{a}_{\mathcal{T}_{\overline{I}}} +
\boldsymbol{a}_{\mathit{\Pi}_{\overline{I}}} &= \delta
\boldsymbol{a}_p + \delta \boldsymbol{f}_{\overline{I}} +
\overline{\boldsymbol{R}} , \label{TirrNSD}
\end{align}
\noindent where
\begin{align}
	\overline{\boldsymbol{R}}(\boldsymbol{x},\boldsymbol{r},t) &\equiv \frac{1}{4\upi} \int_S dS_{\boldsymbol{y}} \frac{\boldsymbol{x}-\boldsymbol{y}}{|\boldsymbol{x}-\boldsymbol{y}|^{3}}[\widehat{\boldsymbol{n}}_{\boldsymbol{y}} \cdot (\delta \boldsymbol{a}_l - \delta \boldsymbol{a}_{\nu})] -\frac{1}{4\upi} \int_S dS_{\boldsymbol{y}} \frac{\boldsymbol{x}-\boldsymbol{y}}{|\boldsymbol{x}-\boldsymbol{y}|^{3}} \times [\widehat{\boldsymbol{n}}_{\boldsymbol{y}} \times \delta \boldsymbol{a}_{p}],\\
	&= - (\delta \boldsymbol{a}_l)_{IB} + (\delta \boldsymbol{a}_{\nu})_{IB} - (\delta \boldsymbol{a}_p)_{SB} .
\end{align}

\noindent These boundary terms are individually equal to zero in
homogeneous/periodic turbulence for the analogue reason as for the NS
dynamics. Regarding the irrotational dynamics, in general,
$\boldsymbol{a}_{\mathcal{T}_{\overline{I}}} \neq
\boldsymbol{a}_{\mathit{\Pi}_{\overline{I}}}$, but the irrotational
volume terms are always equal,
$(\boldsymbol{a}_{\mathcal{T}})_{\overline{IV}}=(\boldsymbol{a}_{\mathit{\Pi}})_{\overline{IV}}$
from equation \eqref{irotV} and
\begin{equation}
	\nabla_{\boldsymbol{x}} \cdot \boldsymbol{a}_{\mathit{\Pi}} =  \nabla_{\boldsymbol{x}} \cdot \boldsymbol{a}_{\mathcal{T}} = \frac{1}{2}\big( \frac{\partial u_k^{+}}{\partial x^{+}_{i}}\frac{\partial u_i^{+}}{\partial x^{+}_{k}}-\frac{\partial u_k^{-}}{\partial x^{-}_{i}}\frac{\partial u_i^{-}}{\partial x^{-}_{k}} \big) .
\end{equation}

The solenoidal interscale transfer term $\boldsymbol{a}_{\mathit{\Pi}_{\overline{S}}}$ in non-homogeneous turbulence can be written as
\begin{multline}
	\boldsymbol{a}_{\mathit{\Pi}_{\overline{S}}}(\boldsymbol{x},\boldsymbol{r},t) = -\frac{1}{4\upi} \int_V d\boldsymbol{y} \frac{\boldsymbol{x}-\boldsymbol{y}}{|\boldsymbol{x}-\boldsymbol{y}|^3} \times [\nabla_{\boldsymbol{y}} \times \boldsymbol{a}_{\mathit{\Pi}}(\boldsymbol{y},\boldsymbol{r}, t)] + \\ \frac{1}{4\upi}\int_S dS_{\boldsymbol{y}}  \frac{\boldsymbol{x}-\boldsymbol{y}}{|\boldsymbol{x}-\boldsymbol{y}|^3} \times [\widehat{\boldsymbol{n}}_{\boldsymbol{y}} \times  \boldsymbol{a}_{\mathit{\Pi}}(\boldsymbol{y},\boldsymbol{r}, t)],
\end{multline}
\noindent where the surface integral is of smaller order of magnitude
than the volume integral away from boundaries and increasingly so with
increasing $\langle \Rey_\lambda \rangle_t$ (verified in our periodic
DNS). Hence, for a qualitative interpretation of
$\boldsymbol{a}_{\mathit{\Pi}_{\overline{S}}}$, we consider
$\boldsymbol{a}_{\mathit{\Pi}_{\overline{S}}} \approx
\boldsymbol{a}_{\mathit{\Pi}_{\overline{SV}}}$ with
\begin{equation}
	(\nabla_{\boldsymbol{x}} \times \boldsymbol{a}_{\mathit{\Pi}})_i =  \delta u_k \frac{\partial \delta \omega_i}{\partial r_k} - \frac{\delta \omega_k}{2} \frac{s_{ij}^{+}+s_{ij}^{-}}{2}-\frac{\omega_k^{+}+\omega_k^{-}}{4} \delta s_{ij} + \frac{\epsilon_{ijk}}{2}[\frac{\partial u_l^{+}}{\partial x_j^{+}}\frac{\partial u_k^{-}}{\partial x_l^{-}} -  \frac{\partial u_l^{-}}{\partial x_j^{-}}\frac{\partial u_k^{+}}{\partial x_l^{+}}], \label{vortNonLoc}
\end{equation}
where $s_{ij}$ is the strain-rate tensor and $\epsilon_{ijk}$ is the Levi-Civita tensor. This set of terms constitutes a part of the non-linear term in the the evolution equation for the vorticity difference $\delta \boldsymbol{\omega}(\boldsymbol{x},\boldsymbol{r},t)$, i.e. vorticity at scales $|\boldsymbol{r}|$ and smaller, as $\nabla_{\boldsymbol{x}} \times \delta \boldsymbol{a}_{c}= \nabla_{\boldsymbol{x}} \times (\boldsymbol{a}_{\mathit{\Pi}} + \boldsymbol{a}_{\mathcal{T}})$. If one contracts \eqref{vortNonLoc} with $2\delta \boldsymbol{\omega}$, the RHS corresponds to non-linear terms which determine the evolution of the enstrophy $|\delta \boldsymbol{\omega}|^2$  at scales smaller or comparable to $|\boldsymbol{r}|$. We interpret the first term on the RHS in \eqref{vortNonLoc} as vorticity interscale transfer. By the connection to $|\delta \boldsymbol{\omega}|^2$, we interpret the second and third terms as related to the enstrophy production/destruction at scales smaller or comparable to $|\boldsymbol{r}|$ due to interactions between the vorticity and strain fields. These three terms justify the interpretation of $\boldsymbol{a}_{\mathit{\Pi}_{\overline{S}}}$ being related non-locally in space to vortex stretching and compression dynamics. The last term in \eqref{vortNonLoc} appears in $\nabla_{\boldsymbol{x}} \times \boldsymbol{a}_{\mathcal{T}_{\overline{SV}}}$ with a negative sign such that these terms cancel.

The exact solenoidal and irrotational KHMH equations follows from contracting equations \eqref{TsolNSD}-\eqref{TirrNSD} with $2\delta \boldsymbol{u}$
\begin{align}
		\mathcal{A}_t + \mathcal{T}_{\overline{S}} + \mathit{\Pi}_{\overline{S}} &= \mathcal{D}_{r,\nu} + \mathcal{D}_{X,\nu} - \mathcal{\epsilon} +I_{\overline{S}} - 2\delta \boldsymbol{u} \cdot\overline{\boldsymbol{R}} ,  \label{TsolKHMH} \\
		\mathcal{T}_{\overline{I}} + \mathit{\Pi}_{\overline{I}} &= \mathcal{T}_p + I_{\overline{I}} + 2\delta \boldsymbol{u} \cdot \overline{\boldsymbol{R}} , \label{TirrKHMH}
\end{align}
where $\mathcal{T}_{\overline{IV}}=\mathit{\Pi}_{\overline{IV}}$. This
shows that the solenoidal and irrotational KHMH equations can be
extended to non-homogeneous turbulence. In contrast to
homogeneous/periodic turbulence, in general boundary terms couple the
irrotational and solenoidal dynamics.

\bibliographystyle{jfm}
\bibliography{jfm}

\end{document}